\documentclass[acmtog, authorversion]{acmart}
\acmSubmissionID{papers\_1830}

\usepackage[ruled]{algorithm2e} 
\usepackage{listings}
\usepackage{makecell}
\usepackage{amsmath}
\usepackage{amsthm}
\usepackage{multirow}
\usepackage{colortbl}

\usepackage{bm}
\usepackage{subcaption}
\usepackage{float}

\newcommand{\ifcommentsenabled}[1]{}
\newcommand{\ifACM}[2]{#1}

\definecolor{tab_color}{rgb}{0.45,0.0,0.65}

\definecolor{felix_color}{rgb}{.6,.4,.05}
\definecolor{michael_color}{rgb}{0,0.35,0}
\definecolor{lukas_color}{rgb}{0.35,0.6,0.6}
\definecolor{thomas_color}{rgb}{0,0,0.85}
\definecolor{thomas_color2}{rgb}{0.1,0.2,0.55}
\definecolor{markus_color}{rgb}{0,0.35,0.35}
\definecolor{bernhard_color}{rgb}{0.35,0.35,0}

\newcommand{\lukas}[1]{\ifcommentsenabled{\textcolor{lukas_color}{Lukas: #1}}}

\newcommand{\markus}[1]{\ifcommentsenabled{\textcolor{markus_color}{Markus: #1}}}

\newcommand{\appref}[1]{App. \ref*{#1}\xspace}

\newcommand{\ie}{\textit{i.e.}\xspace}
\newcommand{\eg}{\textit{e.g.}\xspace}
\newcommand{\cf}{cf.\xspace}

\newcommand{\mb}[1]{\mathbf{#1}}
\newcommand{\R}[0]{\mathbb{R}}
\newcommand{\opa}[0]{o}
\newcommand{\teval}[0]{t_{\text{eval}}}

\newcommand{\gof}{GOF\xspace}

\newcommand{\gs}{3DGS\xspace}
\newcommand{\mue}{\bm{\mu}}

\newcommand{\rayd}[0]{t_{\mb{r}}}
\newcommand{\FLIP}{\protect\reflectbox{F}LIP\xspace}

\newcommand{\near}[0]{\tau_n}
\newcommand{\far}[0]{\tau_f}

\definecolor{edited_color}{rgb}{.8,.15,.15}
\newcommand{\new}[1]{#1} 

\definecolor{revised_color}{rgb}{.15,.15,.8}
\newcommand{\revised}[2]{#1} 

\SetAlFnt{\small}
\SetAlCapFnt{\small}
\SetAlCapNameFnt{\small}
\SetAlCapHSkip{0pt}

\copyrightyear{2025}
\acmYear{2025}
\setcopyright{cc}
\setcctype{by}
\acmConference[SA Conference Papers '25]{SIGGRAPH Asia 2025 Conference Papers}{December 15--18, 2025}{Hong Kong, Hong Kong}
\acmBooktitle{SIGGRAPH Asia 2025 Conference Papers (SA Conference Papers '25), December 15--18, 2025, Hong Kong, Hong Kong}\acmDOI{10.1145/3757377.3763933}
\acmISBN{979-8-4007-2137-3/2025/12}

\begin{teaserfigure}
\footnotesize\sffamily
\renewcommand{\arraystretch}{1.1}%
\resizebox{.99\linewidth}{!}{
\begin{tabular}{lccccc}
\multicolumn{1}{c}{Ground Truth} & \multicolumn{2}{c}{Ours} & \multicolumn{2}{c}{Gaussian Opacity Fields} 
\\[-0.25mm]
\makecell{\includegraphics[width=0.20\linewidth]{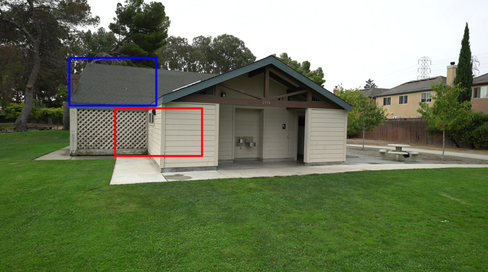}} 
&
\makecell{\includegraphics[width=0.20\linewidth]{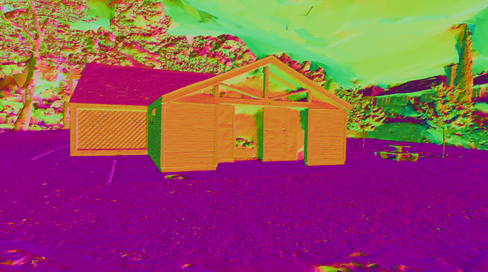}} 
&
\makecell{
\fcolorbox{red}{white}{\includegraphics[width=0.095\linewidth]{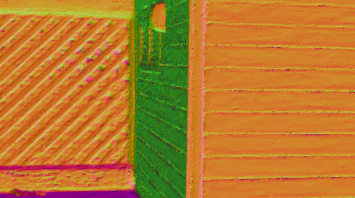}}\\[-0.25mm]
\fcolorbox{blue}{white}{\includegraphics[width=0.095\linewidth]{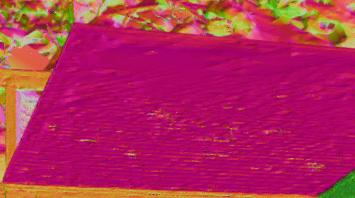}}} 
&
\makecell{\includegraphics[width=0.20\linewidth]{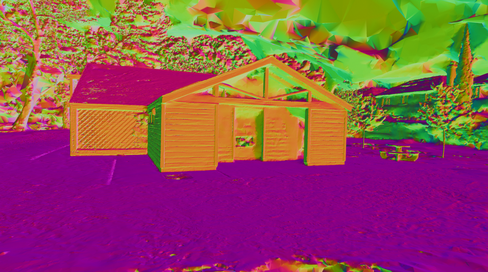}} 
&
\makecell{
\fcolorbox{red}{white}{\includegraphics[width=0.095\linewidth]{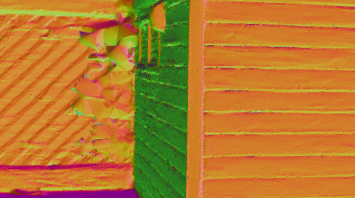}}\\[-0.25mm]
\fcolorbox{blue}{white}{\includegraphics[width=0.095\linewidth]{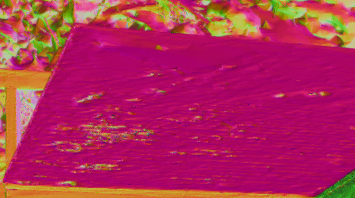}}} 
\\[-0.5mm]
Precision/Recall/F1-Score\textsuperscript{$\uparrow$} & \multicolumn{2}{c}{\textbf{0.589}/\textbf{0.490}/\textbf{0.535}} & \multicolumn{2}{c}{0.522/0.451/0.484}
\\[-0.5mm]
Optimization/Meshing (minutes)& \multicolumn{2}{c}{\textbf{14.4}/\textbf{3.1}} & \multicolumn{2}{c}{40.1/30.3} 
\\
\makecell{\includegraphics[width=0.20\linewidth]{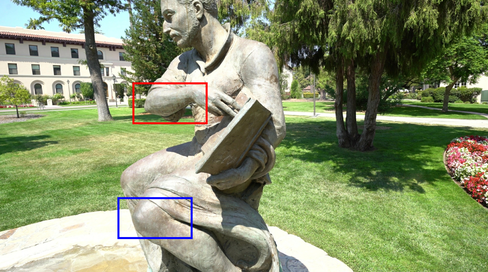}} 
&
\makecell{\includegraphics[width=0.20\linewidth]{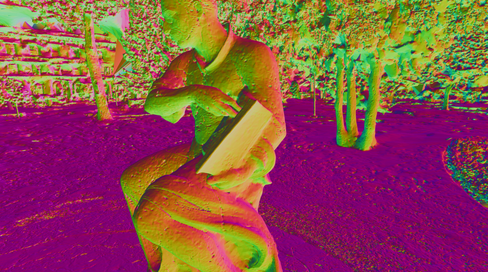}} 
&
\makecell{
\fcolorbox{red}{white}{\includegraphics[width=0.095\linewidth]{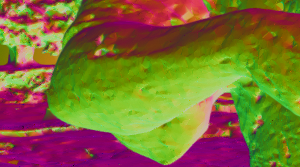}}\\[-0.25mm]
\fcolorbox{blue}{white}{\includegraphics[width=0.095\linewidth]{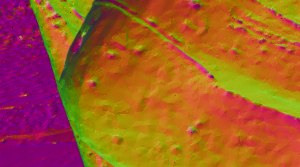}}} 
&
\makecell{\includegraphics[width=0.20\linewidth]{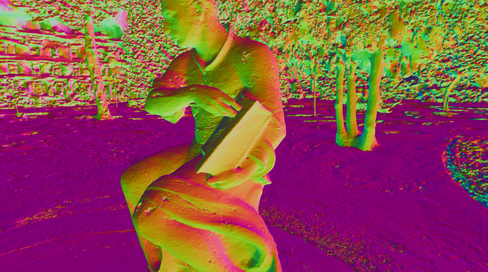}} 
&
\makecell{
\fcolorbox{red}{white}{\includegraphics[width=0.095\linewidth]{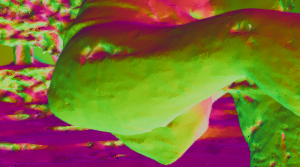}}\\[-0.25mm]
\fcolorbox{blue}{white}{\includegraphics[width=0.095\linewidth]{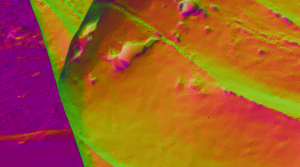}}} 
\\[-0.5mm]
Precision/Recall/F1-Score\textsuperscript{$\uparrow$} & 
\multicolumn{2}{c}{\textbf{0.768}/\textbf{0.707}/\textbf{0.736}} & 
\multicolumn{2}{c}{0.725/0.631/0.674}
\\[-0.5mm]
Optimization/Meshing (minutes) & 
\multicolumn{2}{c}{\textbf{23.7}/\textbf{5.2}} & 
\multicolumn{2}{c}{50.6/15.6} 
\\
\end{tabular}
}
\caption{
We introduce Sorted Opacity Fields (SOF), which allow for swift extraction of high-quality unbounded meshes. 
Compared to current state-of-the-art Gaussian Opacity Fields \cite{yu2024gof}, our meshes are more detailed and exhibit fewer artifacts.
In addition, our approach accelerates optimization by over $3\times$ and meshing by up to an order of magnitude.
}
\label{fig:teaser}
\end{teaserfigure}

\begin{document}

\title{SOF: Sorted Opacity Fields for Fast Unbounded Surface Reconstruction}


\author{Lukas Radl}
\affiliation{%
 \institution{Graz University of Technology}
 \country{Austria}}
\email{lukas.radl@tugraz.at}

\author{Felix Windisch}
\affiliation{%
 \institution{Graz University of Technology}
 \country{Austria}}
\email{felix.windisch@tugraz.at}

\author{Thomas Deixelberger}
\email{thomas.deixelberger@gmail.com}
\affiliation{%
 \institution{Huawei Technologies}
 \country{Austria}}
 
\author{Jozef Hladky}
\email{jozef.hladky@huawei.com}
\affiliation{%
 \institution{Huawei Technologies}
 \country{Switzerland}}

\author{Michael Steiner}
\email{michael.steiner@tugraz.at}
\affiliation{%
 \institution{Graz University of Technology}
 \country{Austria}}

\author{Dieter Schmalstieg}
\affiliation{%
 \institution{Graz University of Technology}
 \country{Austria}
}
\affiliation{%
 \institution{University of Stuttgart}
 \country{Germany}}
\email{dieter.schmalstieg@visus.uni-stuttgart.de}

\author{Markus Steinberger}
\affiliation{%
 \institution{Graz University of Technology}
 \country{Austria}
}
\affiliation{%
 \institution{Huawei Technologies}
 \country{Austria}
}
\email{steinberger@tugraz.at}

\renewcommand{\shortauthors}{Radl, et al.}

\begin{abstract}
Recent advances in 3D Gaussian representations have significantly improved the quality and efficiency of image-based scene reconstruction. 
Their explicit nature facilitates real-time rendering and fast optimization, yet extracting accurate surfaces---particularly in large-scale, unbounded environments---remains a difficult task. 
Many existing methods rely on approximate depth estimates and global sorting heuristics, which can introduce artifacts and limit the fidelity of the reconstructed mesh.
In this paper, we present Sorted Opacity Fields (SOF), a method designed to recover detailed surfaces from 3D Gaussians with both speed and precision. 
Our approach improves upon prior work by introducing hierarchical resorting and a robust formulation of Gaussian depth, which better aligns with the level-set. 
To enhance mesh quality, we incorporate a level-set regularizer operating on the opacity field and introduce losses that encourage geometrically-consistent primitive shapes. 
In addition, we develop a parallelized Marching Tetrahedra algorithm tailored to our opacity formulation, reducing meshing time by up to an order of magnitude.
As demonstrated by our quantitative evaluation, SOF achieves higher reconstruction accuracy while cutting total processing time by more than a factor of three.
These results mark a step forward in turning efficient Gaussian-based rendering into equally efficient geometry extraction.
\end{abstract}

%
%

\ccsdesc[500]{Computing methodologies~Artificial intelligence}
\ccsdesc[300]{Computing methodologies~Graphics systems and interfaces}

%
%

\keywords{Novel View Synthesis, Geometry Reconstruction, Rasterization}

\maketitle

\section{Introduction}
\label{sec:intro}
Reconstruction of environments based on captured images alone has received widespread research attention in previous years, benefiting from recent advances in 3D scene representations and potential applications in telepresence, mixed- and virtual reality.
These methods either use implicit functions, explicit, optimizable primitives, or a combination of the aforementioned to enable learning-based scene optimization, which are subsequently used for novel view synthesis.
Particularly noteworthy are Neural Radiance Fields (NeRF) \cite{Mildenhall2020NeRF} and 3D Gaussian Splatting (3DGS) \cite{kerbl20233dgs}.
While both methods build upon differentiable volume rendering, NeRF leverages large MLPs, evaluated at many sample positions in 3D space.
On the other hand, 3DGS optimizes explicit 3D Gaussians, which drastically reduces computational cost due to the complete removal of redundant computations in empty space.

Since the advent of NeRF, surface reconstruction from implicit 3D scene representations has been investigated in various works, using signed distance fields \cite{yariv2021volsdf, yariv2023baked, Yu2022MonoSDF} or occupancy networks \cite{Oechsle2021unisurf}.
However, with the research community swiftly switching to 3DGS due to its rasterization-based rendering speed, a plethora of works have investigated conversion from trained point clouds to meshes.
Most of these works either align 3D Gaussians to surfaces \cite{guedon2023sugar, yu2024gof} or work with Gaussian disks directly \cite{huang20242dgs, Dai2024GaussianSurfels}.
While the aforementioned works significantly improve processing time whilst mostly retaining reconstruction quality, virtually all current surface reconstruction methods still rely on TSDF fusion \cite{curless1996volumetric}, and are thus unable to extract detailed meshes for background regions.

\begin{figure}[!h]
  \tiny\sffamily
\setlength{\tabcolsep}{1pt}%
\setlength{\fboxsep}{0pt}%
\setlength{\fboxrule}{0.25pt}%
\renewcommand{\arraystretch}{1.1}%
\resizebox{.99\linewidth}{!}{
\begin{tabular}{ccc}
Scene & Ours & \gof
\\[-0.25mm]
\makecell{\includegraphics[width=0.20\linewidth]{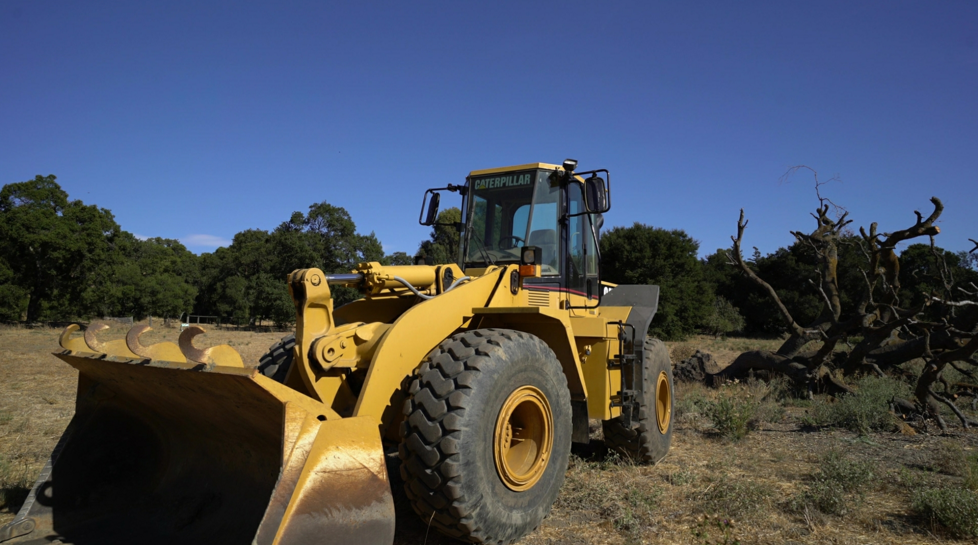}} 
&
\makecell{\includegraphics[width=0.20\linewidth]{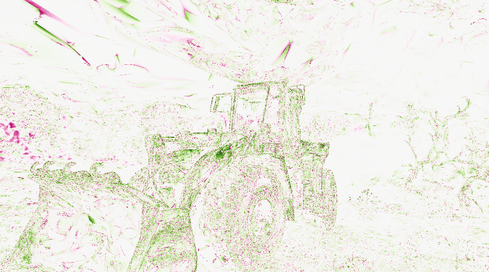}} 
&
\makecell{\includegraphics[width=0.20\linewidth]{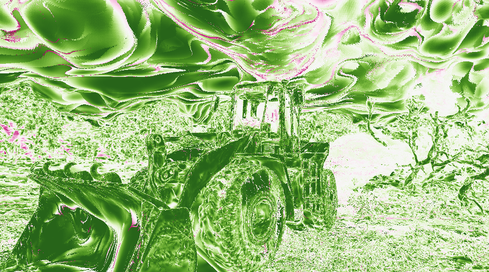}} \\[-0.25mm]
&\textbf{0.512} & {0.747}\\[-0.125mm]
\makecell{\includegraphics[width=0.20\linewidth]{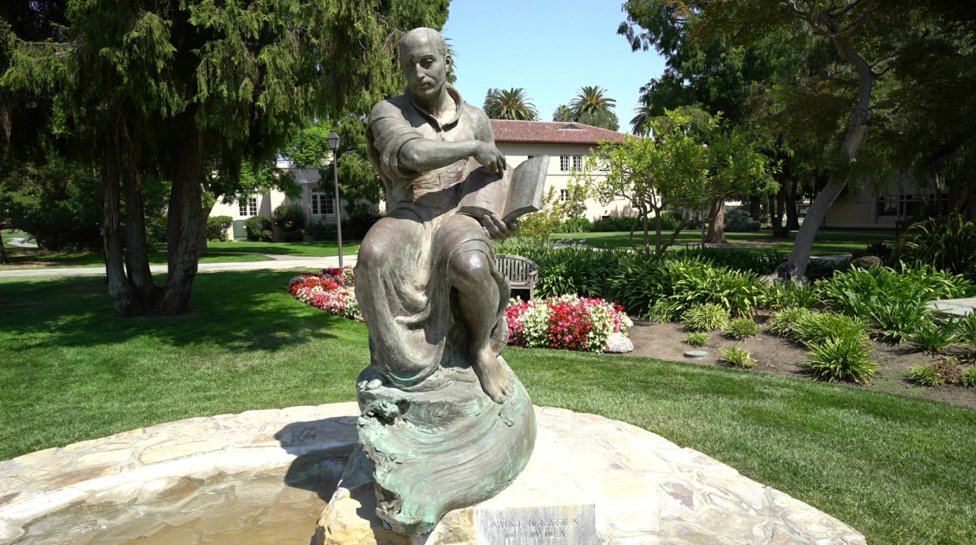}} 
&
\makecell{\includegraphics[width=0.20\linewidth]{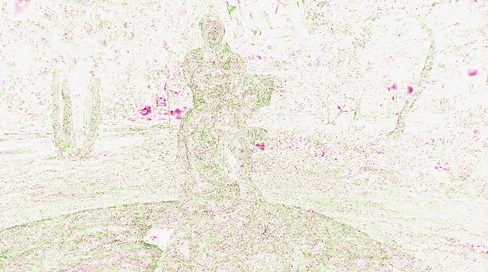}} 
&
\makecell{\includegraphics[width=0.20\linewidth]{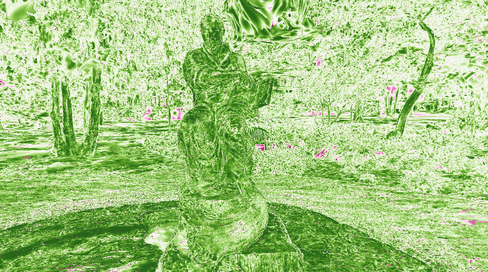}} \\[-0.25mm]
&\textbf{0.510} & {0.736}\\[-0.125mm]
\makecell{\includegraphics[width=0.20\linewidth]{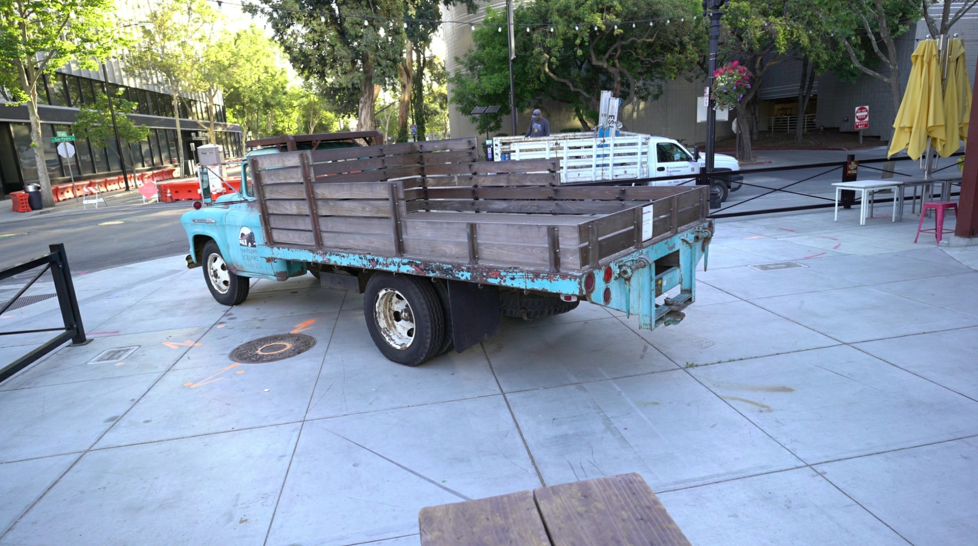}} 
&
\makecell{\includegraphics[width=0.20\linewidth]{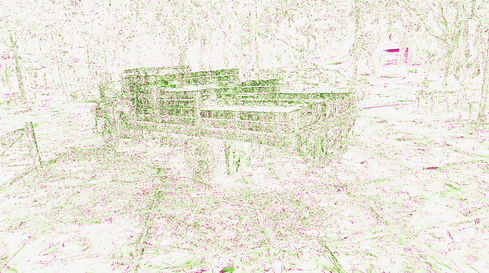}} 
&
\makecell{\includegraphics[width=0.20\linewidth]{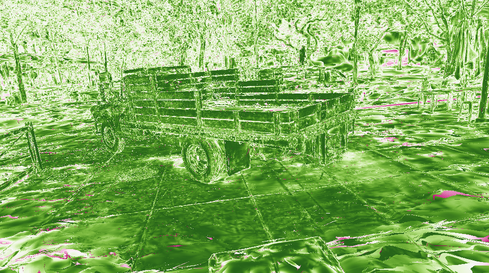}} \\[-0.25mm]
&\textbf{0.517} & {0.809}\\[-0.125mm]
\multicolumn{3}{c}{\includegraphics[width=0.54\linewidth]{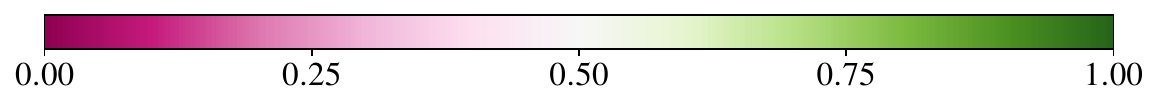}}
\end{tabular}
}
  \caption{\label{fig:small_opacity_comparison}%
    \textbf{Level-Set Alignment Comparison}:
\revised{
We render the opacity at the depth for both \gof and our method. As can be seen, GOF vastly overestimates depth values, which leads to high opacity values.
Benefiting from sorting, a revised depth estimate and direct opacity field supervision, our depth values closely align with the 0.5 level set.
}{
We render $O_N(\rayd)$ for our method and \gof, as well as our method with exact depth (white is $0.5$, see the inset colorbar).
Overall, as we can see by the images as well as the reported statistics, the depth and the $0.5$ level set are closely aligned; \gof vastly overestimates the depth values, leading to high opacity values.
The reported statistics were averaged over the entire test set.}
  }
\end{figure}

Accurately identifying surfaces for large, unbounded environments remains a challenging task due to the much higher computational complexity and lack of supervision.
Current methods such as Binary Opacity Grids \cite{reiser2024bog} apply marching cubes \cite{lorensen1987marching}, requiring dense evaluation and costly mesh simplification strategies.
Recently, Gaussian Opacity Fields (GOF) \cite{yu2024gof} has greatly accelerated unbounded surface reconstruction.
\revised{\gof first establishes an opacity field from 3D Gaussians, and extracts a mesh by finding its level set.
Ideally, this level set and the depth would coincide.
However, as can be seen in Fig. \ref{fig:small_opacity_comparison}, both properties diverge drastically; we will demonstrate that this is caused by imprecise depth estimates, as well as approximate global sorting of Gaussians.}{
However, the resulting quality is still limited by imprecise depth estimates and approximate sorting.}
Additionally, the required processing time still exceeds an hour for large outdoor environments.

To this end, we present Sorted Opacity Fields (SOF), a method for rapid, high-fidelity unbounded mesh extraction from 3D Gaussians.
Building on GOF, we first introduce hierarchical resorting \cite{radl2024stopthepop} to ensure robust depth estimates.
Additionally, we propose a more precise depth estimate, which ensures close alignment with the $0.5$ level set.
We further enhance mesh quality by 
(1) encouraging foreground Gaussians towards surface-aligned disks whilst allowing background Gaussians to be more isotropic, and,
(2) encouraging depth/level-set alignment.
Finally, we propose a new parallelization scheme and several optimization for the Marching Tetrahedra algorithm from \citet{yu2024gof}, cutting meshing time by up to an order of magnitude.

Summarized, our work makes the following contributions:
\begin{itemize}
    \item We provide a comprehensive analysis of current 3DGS-based surface reconstruction methods and their shortcomings.
    \item We identify sorting as a culprit for incorrect depth computations and derive a more appropriate depth estimate.
    \item We propose an adaptive extent loss, combined with direct opacity field supervision, leading to high-fidelity unbounded meshes.
    \item We propose a new parallelization scheme and optimizations for Marching Tetrahedra, thereby improving meshing speed by up to $10\times$.
\end{itemize}

Overall, our method improves quality for unbounded mesh extraction whilst reducing overall processing time (optimization \& meshing) by over $3\times$ (\cf Fig. \ref{fig:teaser}). 
{Our code is publicly available at {\color{blue}\url{https://github.com/r4dl/SOF}}.}


\section{Related Work}
\label{sec:related}
In this section, we revisit prior art in radiance fields and surface reconstruction.

\paragraph{Radiance Fields.}
The advent of Neural Radiance Fields (NeRF) \cite{Mildenhall2020NeRF} marked a significant advancement in the field of novel view synthesis; leveraging large multi-layer perceptrons and differentiable volume rendering, novel views could be rendered at previously unattainable quality.
Subsequent follow-up works have addressed reconstruction of unbounded scenes \cite{barron2022mipnerf360, barron2023zip}, faster inference \cite{mueller2022instant, fridovich2022plenoxels, Chen2022ECCV}, common rendering artifacts \cite{barron2021mipnerf, barron2023zip},
{and also applied NeRF in various other domains, such as 3D scene editing \cite{nguyen2022snerf, radl2024laenerf, jambon2023nerfshop}.}

However, the field was again revolutionized by 3D Gaussian Splatting (3DGS) \cite{kerbl20233dgs}, which instead models a scene as a set of emissive 3D Gaussian distributions.
This work has resulted in a plethora of follow-up research, focusing on various rendering artifacts \cite{radl2024stopthepop, Yu2024MipSplatting, huang2024optimal, Tu2025VRSplat}, faster optimization \cite{mallick2024taming}, and more lightweight representations \cite{fan2023lightgaussian, papantonakis2024reducing}.
Many works \cite{kheradmand2024mcmc, bulo2024revising, yu2024gof} investigated and improved adaptive density control for Gaussians.
While most works still adhere to the EWA splatting formulation \cite{zwicker2001EWA}, recent methods have explored 3D evaluation due to its increased robustness \cite{hahlbohm2025perspective, steiner2025aaagaussians, yu2024gof, loccoz20243dgrt}. 

\paragraph{Surface Reconstruction.}
Many works have investigated image-based surface reconstruction.
Building on NeRF \cite{Mildenhall2020NeRF}, approaches have investigated occupancy networks \cite{Oechsle2021unisurf} or signed distance fields (SDF) \cite{yariv2021volsdf, Yu2022MonoSDF}.
More recent methods have improved the aforementioned in various ways \cite{yariv2023baked, li2023neuralangelo}; however, the state-of-the-art Neuralangelo (NA) \cite{li2023neuralangelo} requires 100+ GPU hours for optimization.
Particularly noteworthy is Binary Opacity Grids \cite{reiser2024bog}, which extracts a mesh from a proposal network by forcing opacity along a ray towards a Dirac impulse for high-quality mesh rendering.

Benefiting from the explicit nature of 3D Gaussians, many works have investigated extraction of surfaces from the resulting point clouds.
Most works either try to align 3D primitives with the reconstructed surface \cite{guedon2023sugar} or use 2D Gaussians directly \cite{Dai2024GaussianSurfels, huang20242dgs}.
Recent methods have explored the use of multi-view stereo constraints to improve surface reconstruction \cite{wang2024GausSurf, chen2024pgsr}.
All of the aforementioned method leverage TSDF fusion \cite{curless1996volumetric}, which only extracts faithful meshes for foreground objects.
\new{Another line of work investigates hybrid methods, combining 3D Gaussians with distance fields \cite{li2025udf, lyu20243dgsr, yu20243dgssdf}.
Additionally, other point-based representations \cite{chen2024dipole, jiang2025geometryfieldsplatting} and stochastic-based methods \cite{miller2024objectsvolumesstochasticgeometry} have also been explored.
} 
\revised{Despite these efforts, extraction}{Extraction} of high-fidelity unbounded meshes from 3D Gaussians \revised{remains}{is} a relatively underexplored topic.
Gaussian Opacity Fields (\gof) \cite{yu2024gof} directly extracts the level set leveraging a Delaunay Triangulation of 3D primitives \cite{kulhanek2023tetranerf}; as we will demonstrate, the extraction step is costly, and the resulting meshes often lack intricate details.
Our Sorted Opacity Fields extract more detailed meshes while drastically improving performance.

\section{Method}
We present Sorted Opacity Fields, a novel method for extracting high-quality, unbounded meshes from 3D Gaussians.
Our work extends Gaussian Opacity Fields \cite{yu2024gof} by incorporating hierarchical resorting \cite{radl2024stopthepop} and more precise depth estimates (\cf Sec. \ref{sec:sof}).
We additionally incorporate novel losses to aid optimization (Sec. \ref{sec:optimization}), and finally propose a fast Marching Tetrahedra implementation (Sec. \ref{sec:fast_march_tets}).
\lukas{Should we have a pipeline figure?}
\markus{no}

\subsection{Preliminaries}
\label{sec:prelims}
In this section, we revisit 3D Gaussian Splatting and its variants.

\paragraph{3D Gaussian Splatting.}
\gs \cite{kerbl20233dgs} represents a scene as a collection of $N$ 3D Gaussians, each characterized by a position $\bm{\mu}_i \in \R^3$, scale $\mathbf{s}_i \in \R^3_+$ and orientation $\mb{R}_i \in \R^{3 \times 3}$.
The covariance matrix $\bm{\Sigma}_i$ is computed as $\mb{R}_i\mb{S}_i\mb{S}_i^T\mb{R}_i^T$, with $\mb{S}_i = \text{diag}(\mb{s}_i)$.
Each primitive is also equipped with a scalar opacity value $\opa_i \in [0,1]$, and a set of Spherical Harmonics coefficients $\mb{f}_i \in \R^{48}$. 
For rendering, the pixel color is accumulated via alpha blending:
\begin{equation}
    \mathbf{C}(\mb{p}) = \sum_{i=0}^{N-1} \mb{c}_i \opa_i \mathcal{G}_i(\mb{p}) T_i  \label{eq:alphablend}, 
\end{equation}
with $\mathcal{G}_i$ the value of the $i$-th Gaussian at pixel $\mb{p}$ and $T_i = \prod_{j=0}^{i-1} (1 - \opa_j \mathcal{G}_j(\mb{p}))$ the transmittance.
Further, Gaussians are assigned to \emph{tiles} based on their image plane extent, significantly reducing the number of primitives per-pixel and thereby improving rendering speed.
\paragraph{StopThePop.}
Recently, \citet{radl2024stopthepop} demonstrated significantly improved view-consistency by hierarchically resorting Gaussians for each pixel, avoiding so-called popping artifacts.
Instead of relying on a global sort based on the view-space depth of $\mue_i$, StopThePop performs per-pixel resorting based on each primitives point of maximum contribution along view rays $\mb{r}(t) = \mb{o} + t\mb{d}$, with
\begin{equation}
    t_i = \frac{\mb{d}^T \bm{\Sigma}_i^{-1}(\mue_i - \mb{o})}{\mb{d}^T \bm{\Sigma}_i^{-1}\mb{d}} \label{eq:stp_depth},
\end{equation}
thereby avoiding rendering artifacts.
As 3DGS does, StopThePop also evaluates projected 2D Gaussians on the image plane; 
however, StopThePop exhibits smoother depth maps due to its improved view-consistency.
To compensate for the computational overhead of resorting, StopThePop aggressively culls Gaussians from tiles;
for instance, instead of using a $3\sigma$ bounding-box for each Gaussian, a tighter bound $\mathcal{E}_i$ can be computed by considering the opacity:
\begin{equation}
    \mathcal{E}_i = \sqrt{2 \ln \left( 255\ \opa_i \right)}\label{eq:stp_bounding} .
\end{equation}

\paragraph{Gaussian Opacity Fields.}
\citet{yu2024gof} adhere to the same scene representation as \cite{kerbl20233dgs, radl2024stopthepop}, but evaluate the Gaussians 3D, at the point of maximum contribution along each view ray $\mb{r}(t)$.
Instead of performing evaluation in world space, \gof first maps $\mb{o}$ and $\mb{d}$ to Gaussian space with
\begin{align}
    \mb{o}_g &= \mb{S}_i^{-1} \mb{R}_i (\mb{o}- \mue_i), \\
    \mb{d}_g &= \mb{S}_i^{-1} \mb{R}_i \mb{d}, \\
    \mb{r}_g(t) &= \mb{o}_g + t \mb{d}_g.
\end{align}
The contribution along the ray is now a 1D Gaussian, with
\begin{align}
    \mathcal{G}_i^{\text{1D}} (t) &= \exp\left(-\frac{1}{2} \mb{r}_g(t)^T \mb{r}_g(t)\right) \label{eq:gaussspace_quadratic} \\
    &= \exp\left(-\frac{1}{2} \left({\mb{d}_g^T \mb{d}_g} t^2 + {2\mb{d}_g^T \mb{o}_g} + {\mb{o}_g^T \mb{o}_g}\right)\right) \label{eq:att_bt_c}.
\end{align}
Now, defining $A_i = \mb{d}_g^T \mb{d}_g$, $B_i=2\mb{d}_g^T \mb{o}_g$ and $C_i=\mb{o}_g^T \mb{o}_g$, the maximum of Eqn. \eqref{eq:gaussspace_quadratic} is attained at
\begin{equation}
    t^*_i = - \frac{B_i}{2A_i} \label{eq:max_t},
\end{equation}
which is equivalent to Eqn. \eqref{eq:stp_depth}.
%
Combining Eqns. \eqref{eq:gaussspace_quadratic} and \eqref{eq:max_t}, we arrive at
\begin{equation}
    \mathcal{G}_i^{\text{1D}} (t^*_i) = \exp\left( -\frac{1}{2}\left(-\frac{B_i^2}{4A_i} + C_i\right)\right), \label{eq:min_value}
\end{equation}
where $t^*_i$ is the depth of the Gaussian.
The color for a pixel $\mb{p}$ is now
\begin{equation}
    \mathbf{C}(\mathbf{p}) = \sum_{i=0}^{N-1}\mb{c}_i \alpha_i T_i,
\end{equation}
with $\alpha_i = \opa_i \mathcal{G}_i^{\text{1D}} (t^*_i)$ and $\mathbf{c}_i$ the Gaussians color, derived from $\mathbf{f}_i$.

Equipped with the ability to evaluate 3D Gaussians at arbitrary positions by inserting $t$ in Eqn. \eqref{eq:att_bt_c}, \gof defines the opacity of a point $\mb{x} \in \R^3$ observed from a camera $\mb{o}$ as
\begin{equation}
    {O}_N(\mathbf{x}) = \sum_{i=0}^{N-1} \opa_i \mathcal{G}_i^{\text{1D}}({\teval}) \prod_{j=1}^{i-1} \left(1 - \opa_j \mathcal{G}^{\text{1D}}_i({\teval})\right), \label{eq:opacityfield}
\end{equation}
with $\teval = \min\left(t^*_i, t\right)$ for $\mb{x} = \mb{o} + t\mb{d}$.
However, $\mb{x}$ may not be observed at all, \eg due to occlusion or being outside the view frustum.
Thus, we define the opacity of a point as the minimum opacity observed over all training views:
\begin{equation}
    \mb{O}(\mb{x}) = \underset{\{\mb{o}, \mb{d}\}}{\min}\ O_N(\mb{o} + t\mb{d}) \label{eq:opafield_min}.
\end{equation}
See \ifACM{App. A.1}{\appref{app:gof:gaussian_space}} for details on an efficient CUDA implementation.
\begin{figure}[t!]
    \centering
    \includegraphics[width=0.95\linewidth]{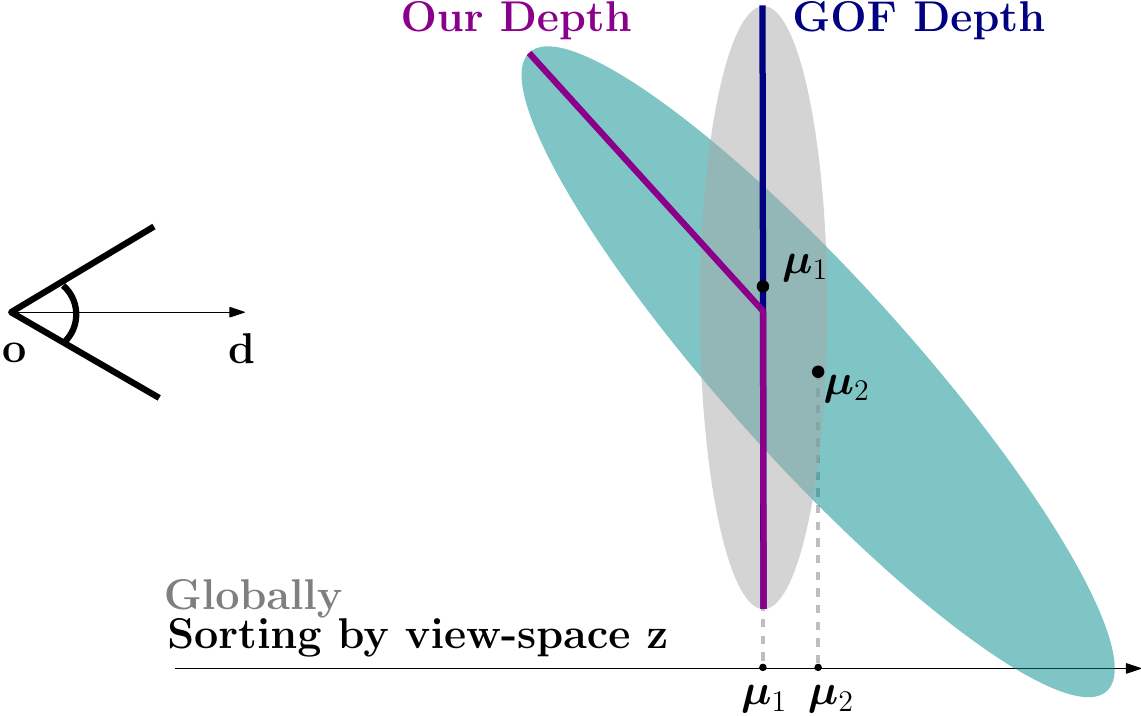}
    \caption{
\textbf{Sorting Motivation}: 
\revised{3DGS sorts Gaussian globally by the $z$-depth of their mean in view space, thus assuming all Gaussians have identical orientation. Instead, we perform per-ray resorting \cite{radl2024stopthepop}, which improves sorting accuracy and thus the robustness of depth estimates.}{Sorting 3D Gaussians along view rays according to the $z$-depth in view space (as done by 3DGS) results in inaccurate depth values.  We leverage hierarchical resorting \cite{radl2024stopthepop} for improved depth accuracy.}
}
\label{fig:sortingrequired}
\end{figure}

\subsection{\revised{Revisiting Depth Computations}{Sorted Opacity Fields}}
\label{sec:sof}

While \gof \cite{yu2024gof} manages to deliver convincing results for mesh extraction in bounded and unbounded scenes, we identify that incorrectly sorted Gaussians result in inconsistent depth estimates, causing a misalignment between depth and the underlying opacity field.
Therefore, we incorporate hierarchical resorting \cite{radl2024stopthepop}, which significantly improves depth accuracy (see Fig. \ref{fig:sortingrequired}).
By considering the opacity field defined in Eqn. \eqref{eq:opacityfield}, we derive a more precise depth estimate, which provides closer alignment with the $0.5$ level set.

\paragraph{Exact Depth.}
\gof \cite{yu2024gof} defines the depth of a ray as 
\begin{equation}
    \rayd = \sum_{i=0}^{N-1} \mathcal{I}_{\{T_i > 0.5, T_{i+1} < 0.5\}} t^*_i,
\end{equation}
with $\mathcal{I}_{\{\cdot\}}$ denoting the indicator function.
The opacity at this point is exactly  $1 - T_{i+1}$, assuming primitives do not overlap and sorting is correct.
While this already exceeds $0.5$ in virtually all cases, Gaussians often overlap in regions with high primitives counts, for which $1 - T_{i+1}$ is now a lower bound\revised{}{; once again}, assuming that sorting is correct. 

To this end, we derive a more precise depth estimate.
We want to find the exact depth where the transmittance $T_i$ reaches $0.5$; as above, this is only possible if $T_i > 0.5$ and $T_{i+1} < 0.5$.
For the Gaussian which satisfies these conditions, we want to find a $\rayd^*$ such that
\begin{equation}
        T_{i}\left(1-\opa_i \mathcal{G}_i^{1D}\left(\rayd^*\right)\right) = 0.5.
\end{equation}
Inserting Eqn. \eqref{eq:gaussspace_quadratic}, it can be shown (\cf \ifACM{App. A.3}{\appref{app:exact_depth}} for the exact derivation) that the exact depth $\rayd^*$ is given 
\begin{equation}
         \rayd^* = \rayd - \frac{\sqrt{B_i^2 - 4A_i \left( C_i + 2\ln\left(\frac{T_{i} - 0.5}{\alpha_{i}}\right)\right)}}{2A}.
\end{equation}
Under the assumption of non-overlapping and correctly sorted Gaussians, this exactly identifies the $0.5$ level set of our opacity field. 
See Fig. \ref{fig:exactdepth} for a visual explanation.

While this assumption is violated even with our hierarchical resorting, we experimentally find that this depth formulation is much more appropriate and leads to improved results.
\begin{figure}[ht!]
    \centering
    \includegraphics[width=0.95\linewidth]{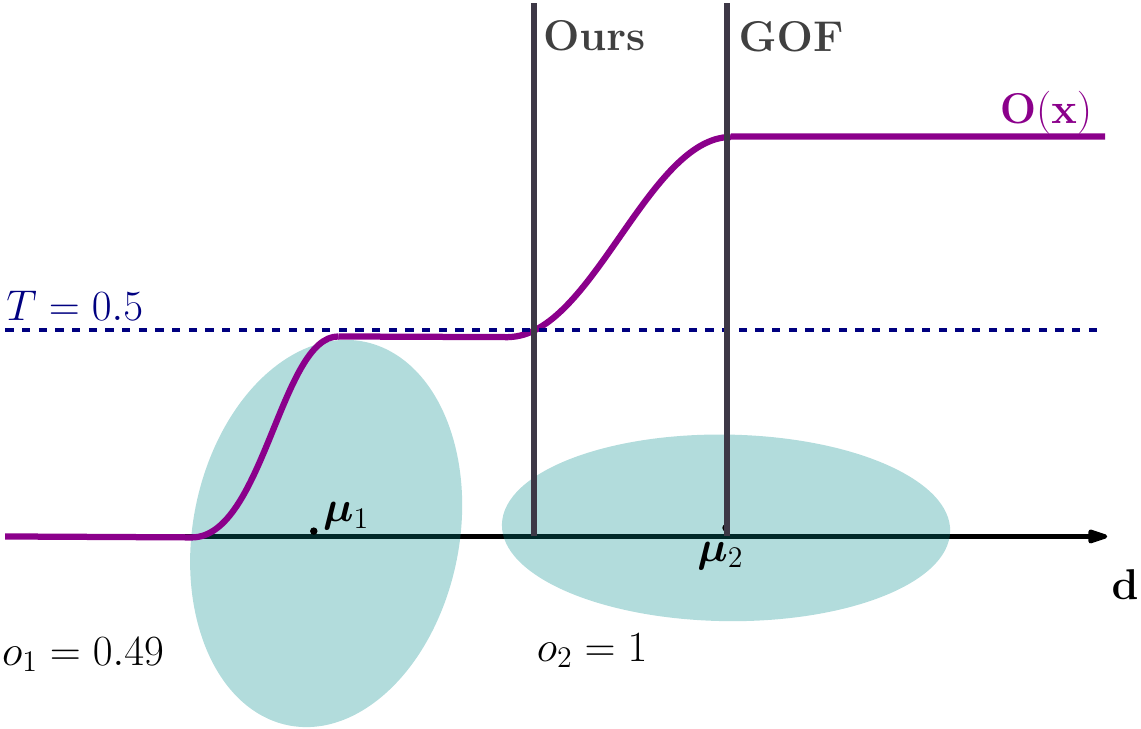}
    \caption{\textbf{Illustration of our exact depth computation.} 
    Whereas GOF consistently overestimates their depth to always be located at the maximum contribution of a Gaussian, we determine the depth by considering the $0.5$ level set of our opacity field $\mb{O}$.}
    \label{fig:exactdepth}
\end{figure}
%



\subsection{Optimization}
\label{sec:optimization}
In this section, we analyze the current losses included in \gof \cite{yu2024gof} and their issues;
we propose a novel extent loss which encourages flat, surface-aligned Gaussians for close objects while allowing for larger, less anistropic Gaussians required for unbounded meshes in the background.
We also propose a method for direct opacity field supervision during optimization, and include a normal smoothness loss for better geometry reconstruction.


\paragraph{Distortion Loss and Depth-Normal Consistency Loss.}
Following both \cite{huang20242dgs, yu2024gof}, we utilize the distortion loss and the depth-normal consistency loss.
The distortion loss $\mathcal{L}_{\text{dist}}$ is defined as
\begin{align}
    \mathcal{L}_{\text{dist}} = \sum_{i=0}^{N-1} \sum_{j=0}^{N-1} w_i w_j \left\|\text{NDC}\left(t_i\right) - \text{NDC}\left(t_j\right)\right\|^2,
\end{align}
with $w_i = \alpha_i T_i$ and $\text{NDC}(\cdot)$ the mapping to NDC space, where the near and far plane set to $\near=0.2,\ \far=100$, respectively.

However, in contrast to the standard 3DGS rasterizer \cite{kerbl20233dgs}, the StopThePop rasterizer performs its backward pass in front-to-back ordering to accomodate sorting.
It can be shown (see \ifACM{App. C.1}{\appref{app:deriv:distortion_loss}} for the detailed derivation) that the required gradients 
can be computed in this order as well.
Finally, we find that detaching $w_i, w_j$, as done by GOF, results in worse surface reconstruction results given our improved sorting accuracy.

The depth-normal consistency loss $\mathcal{L}_{\text{normal}}$ is defined as
\begin{equation}
        \mathcal{L}_{\text{normal}} = \sum_{i=0}^{N-1} w_i \left( 1 - \mb{n}_i^T \mb{N}\right),
\end{equation}
with $\mb{N}$ being the normal of the pixel, derived using finite differences from the rendered depth, and $\mb{n}_i$ the normal of the $i$-th Gaussian (as defined in \citet{yu2024gof}).

\paragraph{Extent Loss.}
Although the distortion loss drastically improves reconstruction quality, we observe that the scale of each Gaussian is not taken into consideration; 
for a Gaussian with $\mb{n}_i$ perpendicular to the view direction, the scale itself has no impact on $\mathcal{L}_{\text{dist}}$.
Therefore, a distortion loss would be minimized by a setup of two closely aligned Gaussians with vastly different variances along view rays.
One solution would be to minimize the scales of Gaussians along a single axis \cite{jiang2023gaussianshader} or the direct use of 2D Gaussian disks \cite{huang20242dgs, Dai2024GaussianSurfels};
however, this directly results in worse background reconstruction, as the number of primitives allocated in the background is significantly reduced.

%
\begin{figure}
    \centering
    \includegraphics[width=0.95\linewidth]{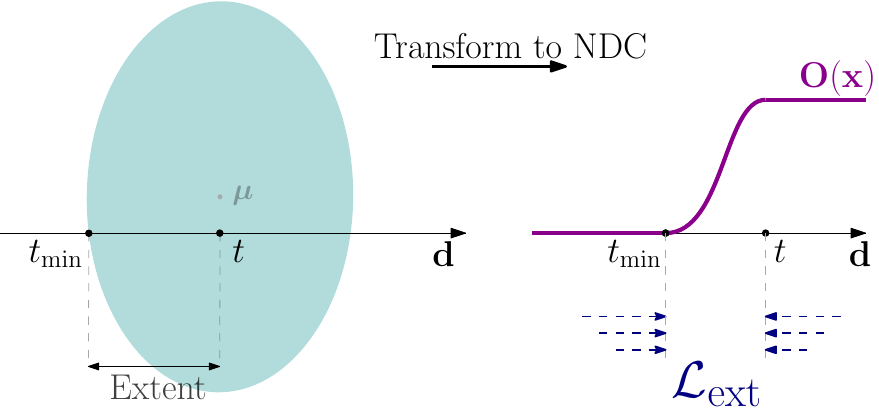}
    \caption{
\textbf{Illustration of our extent loss} $\mathcal{L}_{\text{ext}}$.
We map both the mean and the minimum contribution $t_{\text{min}}$ to NDC-space, where we penalize their extent;
this results in a loss primarily focused on foreground Gaussians.
}
    \label{fig:extent}
\end{figure}
To this end, we propose a novel \emph{Extent Loss}, which penalizes the extent of Gaussians in NDC space (see Fig. \ref{fig:extent} for an illustration).
This encourages flat Gaussians in the foreground to properly represent surfaces, while allowing background Gaussians to be larger, enabling high-fidelity unbounded meshes.

Our loss can be defined as
\begin{equation}
    \mathcal{L}_{\text{ext}} = \frac{\near \far}{\far-\near}\sum_{i=0}^{N-1} w_i\ \frac{2A_i\sqrt{B_i^2 - 4A_i\left(C_i - \mathcal{E}_i^2\right)}}{B_i^2}.
\end{equation}
Note that this version corresponds to the linearized extent loss in NDC (\cf \ifACM{App. C.2}{\appref{app:deriv:extent_loss}} for the full derivation).

\paragraph{Direct Opacity Field Supervision.}
During meshing, the $0.5$ level-set of the opacity field is extracted as the resulting surface;
however, during optimization, no direct supervision to the opacity field is applied.
Our key insight is that we can enforce alignment of the depth $\rayd^*$ and the $0.5$ level set by minimizing
%
\begin{equation}
    \mathcal{L}_{\text{opa}} = \left\|O_N\left({\mb{o} + \rayd^* \mb{d}}\right) - 0.5\right\|^2.
\end{equation}
%
However, the depth is unknown until $T_{i+1} < 0.5$, and, by the definition of our exact depth, $t_j$ may be larger than $\rayd^*$, even if $T_j > 0.5$.

Instead of using a straightforward, but inefficient two-pass solution, we compute $O_N$ as
%
\begin{equation}
    O_N = O_k + T_k \sum_{i=k}^{N-1} \alpha_i \prod_{j=k}^{i-1} (1 - \alpha_j),
\end{equation}
where $k$ is the index of the Gaussian for which $T_k > 0.5$, $T_{k+1} < 0.5$ holds.
This can be implemented efficiently by accumulating $\sum_{i=k}^{N-1} \alpha_i \prod_{j=k}^{i-1} (1 - \alpha_j)$ after the depth is obtained, then computing the remaining $O_k, T_k$ in a second pass, where we can re-use all results for Gaussian preprocessing.
Note that the second pass is very efficient, as we do not need to perform alpha blending for pixel colors, and we only need to consider Gaussian up to index $k$.

\paragraph{Normal Smoothness.}

Finally, we incorporate a normal smoothness loss following \citet{gao2024relightable3d}, formally defined as 
\begin{equation}
    \mathcal{L}_{\text{smooth}} = \|\nabla \mathbf{N}\|\exp\left(-\|\nabla \mathbf{I}\|\right),
\end{equation}
with $\nabla$ denoting finite-differences and $\mathbf{I}$ the ground-truth image.

Our final loss $\mathcal{L}$ is thus composed as a combinations of losses from GOF ($\mathcal{L}_{\text{GOF}}$) complemented with our novel losses:
\begin{align}
    \mathcal{L}_{\text{\gof}} &= \mathcal{L}_{\text{rgb}} 
    + \lambda_{\text{dist}} \mathcal{L}_{\text{dist}} 
    + \lambda_{\text{normal}} \mathcal{L}_{\text{normal}}, \\
    \mathcal{L} &= \mathcal{L}_{\text{\gof}}
    + \lambda_{\text{ext}} \mathcal{L}_{\text{ext}}
    + \lambda_{\text{opa}} \mathcal{L}_{\text{opa}} + \lambda_{\text{smooth}} \mathcal{L}_{\text{smooth}}.
\end{align}

\subsection{Fast Marching Tetrahedra}
\label{sec:fast_march_tets}

For mesh extraction in unbounded scenes, we follow \gof \cite{yu2024gof} and use Deep Marching Tetrahedra \cite{shen2021dmtet} with 8 binary search iterations to identify the $0.5$ level set.

The current Marching Tetrahedra algorithm first processes all Gaussians, as done in the standard 3DGS rendering pipeline, then processes all 3D points $\mb{x}_i \in \R^3$, assigning them to individual pixels based on their projected position $\mb{u}_i \in \R^2$.
Next, for each pixel, all points assigned to this pixel are evaluated according to Eqn. \eqref{eq:opacityfield}, and the results are stored.
This setup results in (1) an uncontrollable number of Gaussians assigned to individual pixels, as well as (2) an unbalanced per-pixel load.


\paragraph{Ray/Tile Scheduling.}
To alleviate the aforementioned issues, we propose a different parallelization strategy:
Instead of assigning points to pixels and finding the contributing Gaussians via the pixel's tile, we opt for directly assigning points to individual tiles.

First, depending on $\mb{u}_i$, we assign points to individual tiles; 
clearly, each point may only belong to a single tile, but each tile may have an arbitrary number of points.
Next, we create key combinations of [\texttt{tile\_id}, \texttt{depth}] for each point, which we sort, ensuring a balanced per-block workload.
The number of blocks to launch for a tile is
\begin{equation}
    N_{\text{blocks}} = \lceil N_{P} / 256 \rceil,
\end{equation}
with $N_{P}$ denoting the number of points per tile.
An inclusive sum over all $N_{\text{blocks}}$ gives the total number of blocks to launch.
Finally, we create a look-up-table, which yields the correct tile id for each block, used for accessing the contributing Gaussians.

The per-thread workload is now evenly distributed across blocks due to evaluating a single point $\mb{x}_i$ per thread and previous depth sorting, and the algorithm works efficiently for any set of points.
For an overview of our parallelization scheme, \cf Fig. \ref{fig:fast_marching_tets}.

\begin{figure}
    \centering
    \includegraphics[width=\linewidth]{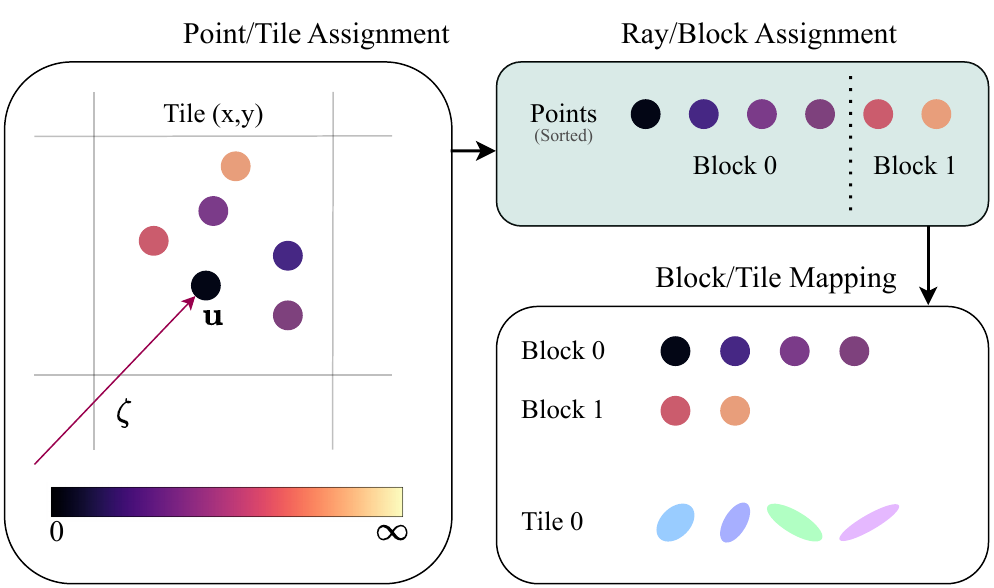}
    \caption{
\textbf{Pipeline Overview} for our Fast Marching Tetrahedra Ray/Tile mapping setup, assuming a block size of $4$ rays. 
Instead of mapping points to pixels, we directly map points to tiles, thereby ensuring a balanced load.
}
    \label{fig:fast_marching_tets}
\end{figure}

\paragraph{Tighter Gaussian Bounding.}
In contrast to the fixed $3\sigma$ cutoff, we use an adaptive, opacity-based Gaussian cutoff $\mathcal{E}_i$ (\cf Eqn. \eqref{eq:stp_bounding}).
%
%
We also observe that Gaussians with $\opa_i < \tfrac{1}{255}$ are never rendered, and thus no longer receive gradients; 
the number of unrendered primitives is $\approx 9\%$ after optimization due to the 3D filter proposed by \citet{Yu2024MipSplatting}.
We do not consider these points for the tetrahedral grid generation, which boosts performance while retaining reconstruction quality; see \ifACM{App. B.2}{\appref{app:tets_bounding}} for more details.

\paragraph{Early Stopping.}
We also propose additional early stopping strategies to further improve performance.
Our primary insight is that the sort order is irrelevant when evaluating Eqn. \eqref{eq:opacityfield} (see \ifACM{A.2}{\appref{app:opacityfield:nosorting}} for a proof).
Therefore, we perform sorting based on the minimal $z$-depth at which the Gaussian contributes, which is given as
\begin{equation}
    z_{\text{min}} = \left(\mathbf{W}\mue_i \right)_z - \mathcal{E}_i \lambda_3,
\end{equation}
with $\lambda_3$ the eigenvalue of the $z$-axis of $\bm{\Sigma}_i$ in view space and $\mathbf{W}$ the camera matrix.
We then perform early stopping if $z_{\text{min}} > t$.

Finally, during the binary search, we iteratively refine the mesh to better align with the $0.5$ level set.
However, for each point, we simply need to know whether $\mb{O}(\mb{x}_i)$ exceeds $0.5$, or not.
Thus, we perform early stopping once $O_N(\mb{x}_i) > 0.5$, as the exact value is not required.
We can also prune points for which we have observed that $O_N(\mb{x}_i) < 0.5$, as $\mb{O}(\mb{x}_i)$ will never exceed $0.5$ thereafter.

\section{Results and Evaluation}
In this section, we evaluate our Sorted Opacity Fields for Surface Reconstruction and Novel View Synthesis.

\paragraph{Implementation Details.}
We base our implementation on StopThePop \cite{radl2024stopthepop}, and integrate 3D evaluation, densification, normal and distortion losses following \gof \cite{yu2024gof}.
We also adopt the appearance model from VastGaussian \cite{lin2024vastgaussian} for all geometry reconstruction results.
For faster optimization, we use the \emph{fused-ssim} implementation from \citet{mallick2024taming}, and additionally compute the 3D filter proposed by \citet{Yu2024MipSplatting} in custom CUDA kernels.
We also use the same densification strategy as \gof; however, we find that the choice of densification strategy significantly impacts the resulting mesh/reconstruction quality (\cf \ifACM{App. A.4}{\appref{app:densification}} for results with state-of-the-art MCMC-densification \cite{kheradmand2024mcmc}).

We use $\lambda_{\text{dist}} = 100$ for unbounded scenes and $\lambda_{\text{dist}} = 1000$ for bounded scenes.
We set $\lambda_{\text{normal}} = 0.05$ for all scenes, 
and let $\lambda_{\text{smooth}} = 0.01$, $\lambda_{\text{ext}} = 0.1$ and $\lambda_{\text{opa}} = 0.04$.
Note that all losses, except for $\mathcal{L}_{\text{rgb}}$, are inactive for the first 15K iterations.

\subsection{Surface Reconstruction}
For geometry reconstruction, we used both the DTU dataset \cite{jensen2014large} and the Tanks \& Temples dataset \cite{Knapitsch2017tanks} to test our method for both bounded and unbounded scenes.
We also considered the Mip-NeRF 360 dataset \cite{barron2022mipnerf360} for qualitative comparisons, as there is no ground-truth data available.
\new{Note that our primary comparison method is \gof, as all other methods do not extract meshes for background regions.}

\paragraph{Bounded Meshes.}
\begin{table*}[t!]
    \centering
    \caption{
    \textbf{Full geometry reconstruction results for the DTU dataset} \cite{jensen2014large}. 
Results with~\textsuperscript{$\bm\dagger$} were taken from GOF \cite{yu2024gof}, results with~\textsuperscript{$\bm\ddagger$} were reproduced with the current, publicly available codebase. 
We report Chamfer Distance (lower scores are better).
    }
\resizebox{.98\linewidth}{!}{
\begin{tabular}{lrrrrrrrrrrrrrrrrr}
\toprule
Method & 24 & 37 & 40 & 55 & 63 & 65 & 69 & 83 & 97 & 105 & 106 & 110 & 114 & 118 & 122 & Avg & Time\\
\midrule
VolSDF\textsuperscript{$\bm\dagger$} & 1.14 & 1.26 & 0.81 & 0.49 & 1.25 & \cellcolor{tab_color!15} 0.70 & 0.72 & 1.29 & \cellcolor{tab_color!15} 1.18 & 0.70 & 0.66 & 1.08 & 0.42 & \cellcolor{tab_color!15} 0.61 & 0.55 & 0.86 & >12h\\
NeuS\textsuperscript{$\bm\dagger$} & 1.00 & 1.37 & 0.93 & 0.43 & \cellcolor{tab_color!15} 1.10 & \cellcolor{tab_color!32} 0.65 & \cellcolor{tab_color!32} 0.57 & 1.48 & \cellcolor{tab_color!32} 1.09 & 0.83 & \cellcolor{tab_color!32} 0.52 & 1.20 & \cellcolor{tab_color!32} 0.35 & \cellcolor{tab_color!32} 0.49 & 0.54 & 0.84 & >12h\\
NA\textsuperscript{$\bm\dagger$} & \cellcolor{tab_color!49} 0.37 & \cellcolor{tab_color!32} 0.72 & \cellcolor{tab_color!32} 0.35 & \cellcolor{tab_color!49} 0.35 & \cellcolor{tab_color!49} 0.87 & \cellcolor{tab_color!49} 0.54 & \cellcolor{tab_color!49} 0.53 & 1.29 & \cellcolor{tab_color!49} 0.97 & 0.73 & \cellcolor{tab_color!49} 0.47 & \cellcolor{tab_color!49} 0.74 & \cellcolor{tab_color!49} 0.32 & \cellcolor{tab_color!49} 0.41 & \cellcolor{tab_color!49} 0.43 & \cellcolor{tab_color!49} 0.60 & >12h\\
3DGS\textsuperscript{$\bm\dagger$} & 2.14 & 1.53 & 2.08 & 1.68 & 3.49 & 2.21 & 1.43 & 2.07 & 2.22 & 1.75 & 1.79 & 2.55 & 1.53 & 1.52 & 1.50 & 1.96 & 11.2m\\
SuGaR\textsuperscript{$\bm\dagger$} & 1.47 & 1.33 & 1.13 & 0.61 & 2.25 & 1.71 & 1.15 & 1.63 & 1.62 & 1.07 & 0.79 & 2.45 & 0.98 & 0.88 & 0.79 & 1.33 & $\approx$1h\\
2DGS\textsuperscript{$\bm\ddagger$} & \cellcolor{tab_color!15} 0.52 & \cellcolor{tab_color!15} 0.80 & \cellcolor{tab_color!49} 0.34 & 0.42 & \cellcolor{tab_color!32} 0.97 & 0.89 & 0.82 & \cellcolor{tab_color!15} 1.24 & 1.25 & \cellcolor{tab_color!32} 0.63 & 0.65 & 1.97 & \cellcolor{tab_color!15} 0.42 & 0.69 & \cellcolor{tab_color!15} 0.49 & 0.81 & 10.9m\\
GOF\textsuperscript{$\bm\dagger$} & \cellcolor{tab_color!32} 0.50 & 0.82 & \cellcolor{tab_color!15} 0.37 & \cellcolor{tab_color!32} 0.37 & 1.12 & 0.74 & 0.73 & \cellcolor{tab_color!32} 1.18 & 1.29 & \cellcolor{tab_color!15} 0.68 & 0.77 & \cellcolor{tab_color!32} 0.90 & 0.42 & 0.66 & 0.49 & \cellcolor{tab_color!32} 0.74 & 18.4m\\
GOF\textsuperscript{$\bm\ddagger$} & 0.53 & 0.89 & 0.43 & \cellcolor{tab_color!15} 0.38 & 1.33 & 0.87 & 0.77 & 1.28 & 1.29 & 0.79 & 0.77 & 1.15 & 0.46 & 0.70 & 0.54 & 0.81 & $\approx$1h\\
Ours & 0.55 & \cellcolor{tab_color!49} 0.70 & 0.41 & 0.39 & 1.12 & 0.73 & \cellcolor{tab_color!15} 0.66 & \cellcolor{tab_color!49} 1.11 & 1.47 & \cellcolor{tab_color!49} 0.60 & \cellcolor{tab_color!15} 0.63 & \cellcolor{tab_color!15} 1.05 & 0.58 & 0.62 & \cellcolor{tab_color!32} 0.48 & \cellcolor{tab_color!32} 0.74 & 22.8m \\
\bottomrule
\end{tabular}
}
    \label{tab:dtu_results}
\end{table*}

%
We present our results for the DTU dataset in Table \ref{tab:dtu_results}.
We re-evaluated both \gof and 2DGS using TSDF fusion \cite{curless1996volumetric} with a voxel size of $0.002$ (indicated with \textsuperscript{$\bm\ddagger$}).
Note that \gof also uses TSDF fusion for bounded scenes\footnote{\new{See this \href{https://github.com/autonomousvision/gaussian-opacity-fields/issues/60}{Github issue} for details.}}; we also find that quality decreases when using Marching Tetrahedra.
For fairness, we also include the reported numbers.
As we can see, our method performs comparably to other explicit based methods.
Note that we only used the losses in $\mathcal{L}_{\text{\gof}}$, and disabled all other losses; as demonstrated in \ifACM{App. B.1}{\appref{app:ablation}}, our other losses result in slightly decreased surface reconstruction quality.

\paragraph{Unbounded Meshes.}
\begin{table}[ht!]
    \centering
        \setlength{\tabcolsep}{3pt}
    \caption{
    \textbf{Full geometry reconstruction results for the Tanks \& Temples dataset} \cite{Knapitsch2017tanks}.
    We evaluate the F1-score (higher is better) and include results for other methods from \gof \cite{yu2024gof} (marked with\textsuperscript{$\bm\dagger$}).
    Overall, our method achieves the best reconstruction quality of all tested 3DGS-based methods, combined with very fast optimization.
    }
\resizebox{.98\linewidth}{!}{
\begin{tabular}{lrrrrrrrr}
\toprule
Method & {{Barn}} & {{Caterp}} & {{Courth}} & {{Ignatius}} & {{Meetingr}} & {{Truck}} & {{Avg}} & Time\\
\midrule
Geo-NeuS\textsuperscript{$\bm\dagger$} & 0.33 & 0.26 & 0.12 & 0.72 & 0.20 & 0.45 & 0.38 & >24h \\
NeuS\textsuperscript{$\bm\dagger$} & 0.29 & 0.29 & 0.17 & \cellcolor{tab_color!32} 0.83 & 0.24 & 0.45 & 0.38 & >24h\\
NA\textsuperscript{$\bm\dagger$} & \cellcolor{tab_color!49} 0.70 & 0.36 & \cellcolor{tab_color!15} 0.28 & \cellcolor{tab_color!49} 0.89 & \cellcolor{tab_color!49} 0.32 & 0.48 & \cellcolor{tab_color!49} 0.50 & >24h\\
3DGS\textsuperscript{$\bm\dagger$} & 0.13 & 0.08 & 0.09 & 0.04 & 0.01 & 0.19 & 0.09 & 14.3m\\
SuGaR\textsuperscript{$\bm\dagger$} & 0.14 & 0.16 & 0.08 & 0.33 & 0.15 & 0.26 & 0.19 & >1h\\
2DGS\textsuperscript{$\bm\dagger$} & 0.36 & 0.23 & 0.13 & 0.44 & 0.16 & 0.26 & 0.30 & 15.5m\\
GOF\textsuperscript{$\bm\dagger$} & \cellcolor{tab_color!15} 0.51 & \cellcolor{tab_color!49} 0.41 & \cellcolor{tab_color!15} 0.28 & 0.68 & \cellcolor{tab_color!15} 0.28 & \cellcolor{tab_color!32} 0.59 & \cellcolor{tab_color!15} 0.46 & 24.2m \\
GOF & 0.48 & \cellcolor{tab_color!15} 0.40 & \cellcolor{tab_color!32} 0.29 & 0.67 & 0.27 & \cellcolor{tab_color!49} 0.60 & 0.45 & >1h\\
Ours & \cellcolor{tab_color!32} 0.54 & \cellcolor{tab_color!49} 0.41 & \cellcolor{tab_color!49} 0.30 & \cellcolor{tab_color!15} 0.74 & \cellcolor{tab_color!32} 0.31 & \cellcolor{tab_color!15} 0.56 & \cellcolor{tab_color!32} 0.47 & 16.7m\\
\bottomrule
\end{tabular}
}
    \label{tab:tnt_results}
\end{table}

%
We show the results for unbounded meshes, using our Fast Marching Tetrahedra algorithm, in Table \ref{tab:tnt_results}.
Our method quantitatively outperforms \gof and other explicit, 3DGS-based methods.
Particularly for \emph{Barn} and \emph{Ignatius}, our method attains much higher reconstruction quality than \gof.
We additionally provide comparisons with \gof for unbounded meshes in Fig. \ref{fig:mesh_comparison}.
As we can see, our method exhibits fewer artifacts compared to \gof, while recovering intricate geometric details (\cf Fig. \ref{fig:quality}).

\subsection{Novel View Synthesis}
We also evaluate our method in Novel View Synthesis, where we utilize the Mip-NeRF 360 dataset \cite{barron2022mipnerf360}.
For our comparison, we choose state-of-the-art reconstruction methods \cite{kheradmand2024mcmc, kerbl20233dgs, Yu2024MipSplatting, mallick2024taming, radl2024stopthepop} as well as recent works in surface reconstruction \cite{yu2024gof, huang20242dgs}.
For Ours and \gof, we disabled decoupled appearance modeling.

We use PSNR, SSIM, LPIPS \cite{zhang2018unreasonable} and \FLIP \cite{Andersson2020Flip} for our evaluation and reproduce the numbers for all methods; see the results in Table \ref{tab:m360_results}.
Our method remains competitive with \gof, with slightly reduced image quality metrics.
This aligns with the findings for StopThePop \cite{radl2024stopthepop}, where the indoor image metrics are worse compared to 3DGS, as the hierarchical rasterizer prevents "cheating" view-dependent effects.

\new{
We additionally provide a variant of our method which uses state-of-the-art MCMC densification \cite{kheradmand2024mcmc}.
For a fair comparison, we choose the primitive count to be identical as \emph{Ours}.
This variant improves image quality, particularly for indoor scenes, highlighting that densification plays an important role for standard image quality metrics.
}
\new{For more novel view synthesis results, \cf  \ifACM{App. B.1}{\appref{app:ablation}}.}
\begin{table}[ht!]
    \centering
        \setlength{\tabcolsep}{3pt}
    \caption{
\textbf{
Full Novel View Synthesis results for the Mip-NeRF 360 dataset}~\cite{barron2022mipnerf360}. 
Our method performs similarly to \gof for outdoor scenes, while obtaining lower image quality metrics for indoor scenes.
This can be attributed to per-pixel sorting (\cf 3DGS vs. StopThePop), which removes the ability to "fake" view-dependent effects with geometry.
    }
\resizebox{.99\linewidth}{!}{
\begin{tabular}{lrrrrrrrr}
\toprule
& \multicolumn{4}{c}{Mip-NeRF 360 Outdoor} & \multicolumn{4}{c}{Mip-NeRF 360 Indoor} \\
\cmidrule(lr){2-5}\cmidrule(lr){6-9}
Method 
& PSNR\textsuperscript{$\uparrow$} 
& SSIM\textsuperscript{$\uparrow$} 
& LPIPS\textsuperscript{$\downarrow$} 
& \FLIP \textsuperscript{$\downarrow$} 
& PSNR\textsuperscript{$\uparrow$} 
& SSIM\textsuperscript{$\uparrow$} 
& LPIPS\textsuperscript{$\downarrow$}
& \FLIP \textsuperscript{$\downarrow$} \\
\midrule
3DGS & 24.59 & 0.727 & 0.240 & 0.167 & 30.98 & 0.922 & 0.189 & \cellcolor{tab_color!15} 0.094 \\
Mip-Splatting & 24.72 & 0.731 & 0.240 & 0.168 & \cellcolor{tab_color!15} 31.06 & \cellcolor{tab_color!15} 0.925 & 0.187 & \cellcolor{tab_color!15} 0.094 \\
StopThePop & 24.60 & 0.728 & 0.235 & 0.167 & 30.62 & 0.921 & 0.186 & 0.099 \\
Taming-3DGS & \cellcolor{tab_color!32} 25.01 & 0.740 & 0.227 & \cellcolor{tab_color!32} 0.163 & \cellcolor{tab_color!32} 31.34 & \cellcolor{tab_color!32} 0.927 & \cellcolor{tab_color!32} 0.183 & \cellcolor{tab_color!49} 0.090 \\
3DGS-MCMC & \cellcolor{tab_color!49} 25.17 & \cellcolor{tab_color!49} 0.759 & \cellcolor{tab_color!49} 0.198 & \cellcolor{tab_color!49} 0.160 & \cellcolor{tab_color!49} 31.59 & \cellcolor{tab_color!49} 0.932 & \cellcolor{tab_color!49} 0.174 & \cellcolor{tab_color!32} 0.091 \\
2DGS & 24.22 & 0.705 & 0.285 & 0.174 & 30.10 & 0.912 & 0.212 & 0.103 \\
GOF & \cellcolor{tab_color!15} 24.79 & \cellcolor{tab_color!32} 0.745 & \cellcolor{tab_color!32} 0.208 & \cellcolor{tab_color!15} 0.165 & 30.47 & 0.918 & 0.189 & 0.100 \\
Ours & 24.78 & \cellcolor{tab_color!32} 0.745 & \cellcolor{tab_color!32} 0.208 & \cellcolor{tab_color!15} 0.165 & 30.13 & 0.916 & 0.188 & 0.104 \\
Ours (MCMC) & 24.70 & 0.738 & 0.214 & 0.168 & 30.41 & 0.921 & \cellcolor{tab_color!15} 0.184 & 0.103 \\
\bottomrule
\end{tabular}
}
    \label{tab:m360_results}
\end{table}
%
\begin{figure}[h]
\sffamily
\setlength{\tabcolsep}{1pt}%
\setlength{\fboxsep}{0pt}%
\setlength{\fboxrule}{0.5pt}%
\renewcommand{\arraystretch}{1.1}%
\centering

\centering
\resizebox{\linewidth}{!}{
\begin{tabular}{cccc}
\parbox[c]{0.5cm}{\centering\rotatebox{90}{\tiny Ground Truth}} &
\makecell{\includegraphics[width=0.235\linewidth]{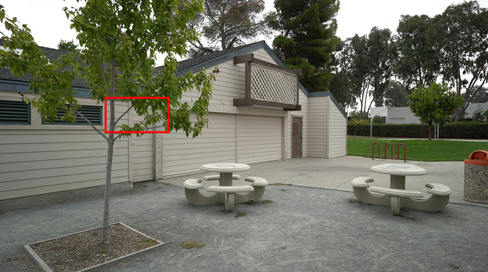}} &
\makecell{\includegraphics[width=0.235\linewidth]{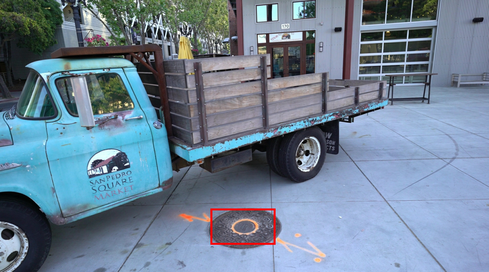}} & \makecell{\includegraphics[width=0.235\linewidth]{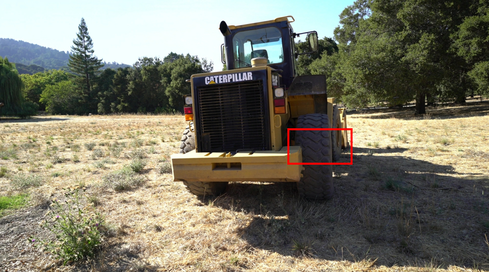}} \\
\\[-10pt]
\parbox[c]{0.5cm}{\centering\rotatebox{90}{\tiny GT Crop}} &
\makecell{\fcolorbox{red}{white}{\includegraphics[width=0.233\linewidth]{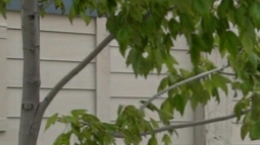}}} &
\makecell{\fcolorbox{red}{white}{\includegraphics[width=0.233\linewidth]{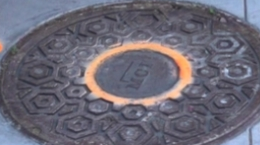}}} &
\makecell{\fcolorbox{red}{white}{\includegraphics[width=0.233\linewidth]{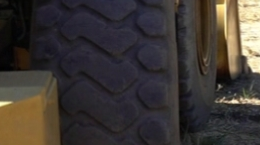}}} \\
\\[-10pt]
\parbox[c]{0.5cm}{\centering\rotatebox{90}{\tiny Ours Crop}} &
\makecell{\fcolorbox{red}{white}{\includegraphics[width=0.233\linewidth]{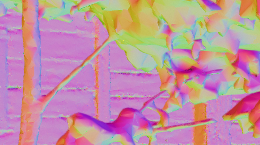}}} &
\makecell{\fcolorbox{red}{white}{\includegraphics[width=0.233\linewidth]{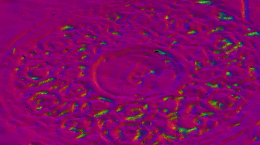}}} &
\makecell{\fcolorbox{red}{white}{\includegraphics[width=0.233\linewidth]{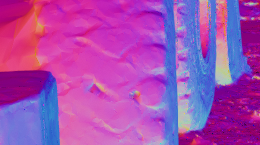}}} \\
\\[-10pt]
\parbox[c]{0.5cm}{\centering\rotatebox{90}{\tiny GOF Crop}} &
\makecell{\fcolorbox{red}{white}{\includegraphics[width=0.233\linewidth]{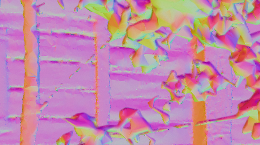}}} &
\makecell{\fcolorbox{red}{white}{\includegraphics[width=0.233\linewidth]{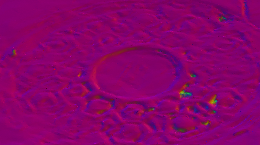}}} &
\makecell{\fcolorbox{red}{white}{\includegraphics[width=0.233\linewidth]{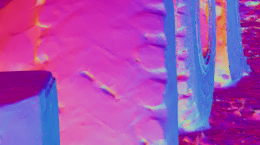}}} \\
\end{tabular}
}

\caption{\textbf{Qualitative Mesh Comparison:} 
Particularly for unbounded scenes, \gof tends to create overly smooth geometry, leading to a loss of intricate details (see the zoomed-in results).
Our method faithfully recovers fine details, while being much more efficient.
}
\label{fig:quality}
\end{figure}

\subsection{Performance Analysis}
We conduct a thorough performance evaluation of our method compared to \gof \cite{yu2024gof}, where we use the Tanks \& Temples dataset \cite{Knapitsch2017tanks} throughout.
\new{Note that we use the current \gof codebase for this comparison, which is $\approx 60\%$ faster compared to the original one (\cf \ifACM{App. A.1}{\appref{app:gof:gaussian_space}} for details).}
For all presented results, we used an NVIDIA RTX 4090 with CUDA 12.2.

%
\begin{table}[ht!]
    \centering
        \setlength{\tabcolsep}{3pt}
    \caption{
\new{
    \textbf{Detailed Training Statistics}.
We report per-iteration timings for the \emph{Barn} scene (trained for 30K iterations, with $\approx$ 800K Gaussians after optimization).
\textsuperscript{$\bm \ddagger$}The reported total training times also include other necessary overhead such as Adam Updates, I/O, etc. 
}}
\resizebox{.99\linewidth}{!}{
\begin{tabular}{lcccccc}
\toprule
Timings in ms & \multicolumn{3}{c}{Forward} & \multicolumn{2}{c}{Backward} \\
\cmidrule(lr){2-4}\cmidrule(lr){5-6}
Method & Preprocess & Render & Opacity & Render & Preprocess & Total\textsuperscript{$\bm \ddagger$} \\
\midrule
GOF & \cellcolor{tab_color!0}0.799 & \cellcolor{tab_color!0}28.367 & - & \cellcolor{tab_color!0}37.624 & \cellcolor{tab_color!0}0.501 & \cellcolor{tab_color!0}81.017 \\
\midrule
Ours & \cellcolor{tab_color!0}0.618 & \cellcolor{tab_color!0}\phantom{2}4.885 & \cellcolor{tab_color!0}1.294 & \cellcolor{tab_color!0}13.452 & \cellcolor{tab_color!0}0.528 & \cellcolor{tab_color!0}30.466 \\
w/o Opacity Cutoff & \cellcolor{tab_color!0}1.057 & \cellcolor{tab_color!0}\phantom{2}8.504 & \cellcolor{tab_color!0}2.545 & \cellcolor{tab_color!0}19.411 & \cellcolor{tab_color!0}0.520 & \cellcolor{tab_color!0}42.128 \\
w/o CUDA 3D Filter & \cellcolor{tab_color!0}0.616 & \cellcolor{tab_color!0}\phantom{2}4.860 & \cellcolor{tab_color!0}1.281 & \cellcolor{tab_color!0}13.381 & \cellcolor{tab_color!30}0.516 & \cellcolor{tab_color!0}31.705 \\
w/o \emph{fused-ssim} & \cellcolor{tab_color!0}0.615 & \cellcolor{tab_color!0}\phantom{2}4.858 & \cellcolor{tab_color!0}1.277 & \cellcolor{tab_color!0}13.480 & \cellcolor{tab_color!0}0.527 & \cellcolor{tab_color!0}31.339\\
w/o Exact Depth, $\mathcal{L}_{\text{opa}}$ &  \cellcolor{tab_color!30}0.601 & \cellcolor{tab_color!30}\phantom{2}4.766 & - &  \cellcolor{tab_color!30}13.352 & \cellcolor{tab_color!0}0.520 & \cellcolor{tab_color!30}28.735\\
\bottomrule
\end{tabular}
}
    \label{tab:train_timings}
\end{table}

\paragraph{\new{Training.}}
\label{app:training_time}
\new{
Despite the required computational overhead necessary for per-pixel resorting \cite{radl2024stopthepop} and our newly introduced losses, we are still $\approx3\times$ faster than \gof for optimization.
We conduct an additional training time ablation study to determine the culprit for our improved performance.
To this end, we train the \emph{Barn} scene from Tanks \& Temples \cite{Knapitsch2017tanks} for 30K iterations and report the obtained timings in Table \ref{tab:train_timings}.
}

As we can see, our improvement in rendering speed stems from the much faster \emph{Render} stage, which can be attributed to improved bounding and culling \cite{radl2024stopthepop}.
As the per-Gaussian overhead is increased due to additional depth/normal computations, this results in a significant performance improvement.

Additionally, the number of dead Gaussians is very high due to the 3D filter \cite{Yu2024MipSplatting}; therefore, disabling the opacity cutoff increases training time by $35\%$.
Our additional render pass for opacity is very efficient, as only few Gaussians need to be evaluated.
When we disable $\mathcal{L}_{\text{opa}}$ and our exact depth, our method is $10\%$ faster.
Finally, note that as we time CUDA kernels, pre-processing is faster when moving the 3D filter to PyTorch - however, the total timings still reflect the resulting speedup of our custom kernels.

\paragraph{Meshing.}
We conduct a thorough performance evaluation for our Fast Marching Tetrahedra algorithm.
We take trained models from \gof for all scenes in Tanks \& Temples \cite{Knapitsch2017tanks} to ensure a fair, point-cloud independent comparison.
We average per-iteration timings for point-preprocessing, Gaussian-processing and opacity field evaluation; all timings were averaged for the complete mesh extraction process.
The results are shown in Table \ref{tab:meshing_performance}.
\begin{table}[ht!]
    \centering
    \caption{
    \textbf{Marching Tetrahedra Performance Analysis}:
Our Fast Marching Tetrahedra algorithm outperforms \gof by a factor of $\approx 6.8\times$.
Timings were averaged over all scenes from the Tanks \& Temples dataset \cite{Knapitsch2017tanks}, using the same Gaussian point cloud (from \gof) for fairness; we report per-iteration measurements.
    }
        \setlength{\tabcolsep}{3pt}
\resizebox{.98\linewidth}{!}{
\begin{tabular}{lrrrr}
\toprule
Timings in ms & Process Points & Process Gaussians & Integrate & Total \\
\midrule
GOF & 1.17 & 1.07 & 501.12 & 503.38 \\
+ Tile/Ray Scheduling & 1.15 & \cellcolor{tab_color!30}0.99 & 415.99 & 418.13 \\
+ Min Z Bounding & 1.15 & \cellcolor{tab_color!30}0.99 & 257.29 & 259.44\\
+ Early Stopping & 1.15 & \cellcolor{tab_color!30}0.99 & 116.32 &  118.27 \\
+ Point Pruning & \cellcolor{tab_color!30} 0.91 & \cellcolor{tab_color!30}0.99 & \cellcolor{tab_color!30} 72.41 & \cellcolor{tab_color!30} 74.40 \\
\bottomrule
\end{tabular}
}
    \label{tab:meshing_performance}
\end{table}

As we can see, our full approach is $\approx 6.8\times$ faster than \gof, in terms of average iteration time.
Introducing our Tile/Ray Scheduling already improves performance, benefiting from a more evenly distributed workload across thread blocks.
With Min Z Bounding, performance once again improves drastically by removing unnecessary Gaussian evaluations.
%
%
Early stopping (as performed here for the $0.5$ level set) once again doubles performance, and point pruning leads to another large reduction in meshing times.

When adding up training and meshing time, our approach is more than $3\times$ faster compared to \gof (see \ifACM{App. B.3}{\appref{app:ablation:timing}} for a detailed breakdown), and results in better meshes, as seen previously.

\subsection{Ablation Studies}
In this section, we provide further experiments and analyze each introduced component; 
for more experiments, we refer to  \ifACM{App. B.1}{\appref{app:ablation}}.

\paragraph{Surface Reconstruction.}
We investigate the impact of our proposed components for unbounded scenes in Table \ref{tab:tnt_ablation}.
As we can see in \emph{(A) w/o Exact Depth}, disabling Exact Depth has the largest impact on reconstruction quality.
Besides, disabling $\mathcal{L}_{\text{opa}}, \mathcal{L}_{\text{ext}}, \mathcal{L}_{\text{smooth}}$ has a minor impact on final reconstruction quality.
As can be seen for \emph{(F)}, disabling both $\mathcal{L}_{\text{opa}}$, $\mathcal{L}_{\text{ext}}$ has a larger impact on reconstruction, highlighting that the losses benefit from each other.

\new{
Finally, we also provide reconstruction metrics using the scale minimization loss from \citet{jiang2023gaussianshader}, \ie
\begin{equation}
    \mathcal{L}_{\text{scale}} = \frac{1}{N}\sum_{i=0}^{N-1} \min\left(\mathbf{s}_i\right),
\end{equation}
with $\min\left(\mathbf{s}_i\right)$ denoting the minimum scalar element of the vector $\mathbf{s}_i \in \mathbb{R}_+^3$.
As can be seen in (\emph{(G) w/ $\mathcal{L}_{\text{scale}}$}), our full configuration, using our novel extent loss, achieves better results.
}
%
\begin{table}[ht!]
    \centering
    \caption{
    \textbf{Surface Reconstruction Ablation}: 
We remove individual components of our method and evaluate the F1-score for the Tanks \& Temples dataset \cite{Knapitsch2017tanks}.
Removing exact depth decreases quality drastically, highlighting the need for robust depth estimates.
As can also be seen, our extent loss produces better results compared to using the scale minimization loss.
    }
        \setlength{\tabcolsep}{3pt}
\resizebox{.99\linewidth}{!}{
\begin{tabular}{lrrrrrrrr}
\toprule
Method & {{Barn}} & {{Caterp}} & {{Courth}} & {{Ignatius}} & {{Meetingr}} & {{Truck}} & {{Avg}} \\
\midrule
Ours & \cellcolor{tab_color!49} 0.535 & \cellcolor{tab_color!49} 0.408 & \cellcolor{tab_color!32} 0.297 & \cellcolor{tab_color!32} 0.736 & \cellcolor{tab_color!49} 0.309 & \cellcolor{tab_color!49} 0.558 & \cellcolor{tab_color!49} 0.474 \\
(A) w/o Exact Depth & \cellcolor{tab_color!0} 0.468 & \cellcolor{tab_color!0} 0.396 & \cellcolor{tab_color!0} 0.272 & \cellcolor{tab_color!0} 0.706 & \cellcolor{tab_color!0} 0.286 & \cellcolor{tab_color!0} 0.545 & \cellcolor{tab_color!0} 0.445 \\
(B) w/o Attach Grad & \cellcolor{tab_color!0} 0.508 & \cellcolor{tab_color!15} 0.407 & \cellcolor{tab_color!15} 0.292 & \cellcolor{tab_color!0} 0.724 & \cellcolor{tab_color!0} 0.305 & \cellcolor{tab_color!0} 0.551 & \cellcolor{tab_color!0} 0.465 \\
(C) w/o $\mathcal{L}_{\text{opa}}$ & \cellcolor{tab_color!49} 0.535 & \cellcolor{tab_color!49} 0.408 & \cellcolor{tab_color!15} 0.292 & \cellcolor{tab_color!0} 0.734 & \cellcolor{tab_color!0} 0.305 & \cellcolor{tab_color!15} 0.552 & \cellcolor{tab_color!32} 0.471 \\
(D) w/o $\mathcal{L}_{\text{ext}}$ & \cellcolor{tab_color!0} 0.534 & \cellcolor{tab_color!15} 0.407 & \cellcolor{tab_color!0} 0.288 & \cellcolor{tab_color!32} 0.736 & \cellcolor{tab_color!32} 0.308 & \cellcolor{tab_color!0} 0.547 & \cellcolor{tab_color!0} 0.470 \\
(E) w/o $\mathcal{L}_{\text{smooth}}$ & \cellcolor{tab_color!0} 0.523 & \cellcolor{tab_color!0} 0.406 & \cellcolor{tab_color!49} 0.299 & \cellcolor{tab_color!49} 0.738 & \cellcolor{tab_color!15} 0.307 & \cellcolor{tab_color!32} 0.553 & \cellcolor{tab_color!32} 0.471 \\
(F) w/o $\mathcal{L}_{\text{opa}}, \mathcal{L}_{\text{ext}}$ & \cellcolor{tab_color!0} 0.527 & \cellcolor{tab_color!0} 0.405 & \cellcolor{tab_color!0} 0.283 & \cellcolor{tab_color!0} 0.733 & \cellcolor{tab_color!0} 0.304 & \cellcolor{tab_color!0} 0.544 & \cellcolor{tab_color!0} 0.466 \\
(G) w/ $\mathcal{L}_{\text{scale}}$ & \cellcolor{tab_color!49} 0.535 & \cellcolor{tab_color!0} 0.400 & \cellcolor{tab_color!0} 0.283 & \cellcolor{tab_color!0} 0.720 & \cellcolor{tab_color!0} 0.294 & \cellcolor{tab_color!0} 0.549 & \cellcolor{tab_color!0} 0.463 \\
\bottomrule
\end{tabular}
}
    \label{tab:tnt_ablation}
\end{table}

\begin{figure*}[ht!]
  \footnotesize\sffamily
\setlength{\tabcolsep}{1pt}%
\setlength{\fboxsep}{0pt}%
\setlength{\fboxrule}{0.5pt}%
\renewcommand{\arraystretch}{1.1}%
\resizebox{.99\linewidth}{!}{
\begin{tabular}{lccccc}
\multicolumn{1}{c}{Ground Truth} & \multicolumn{2}{c}{Ours} & \multicolumn{2}{c}{\gof} 
\\[-0.235mm]
\makecell{\includegraphics[width=0.235\linewidth]{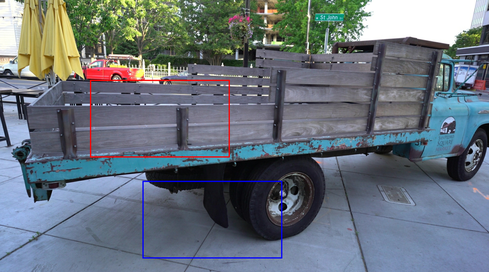}} 
&
\makecell{\includegraphics[width=0.235\linewidth]{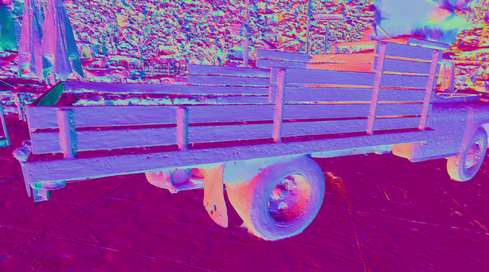}} 
&
\makecell{
\fcolorbox{red}{white}{\includegraphics[width=0.095\linewidth]{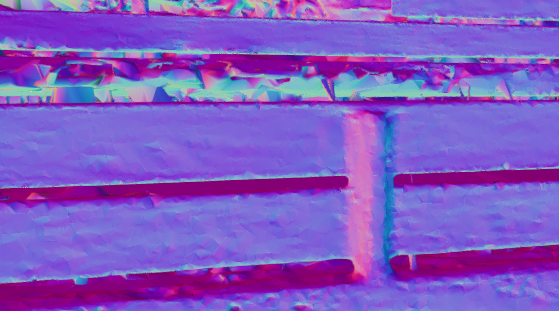}}\\[-0.235mm]
\fcolorbox{blue}{white}{\includegraphics[width=0.095\linewidth]{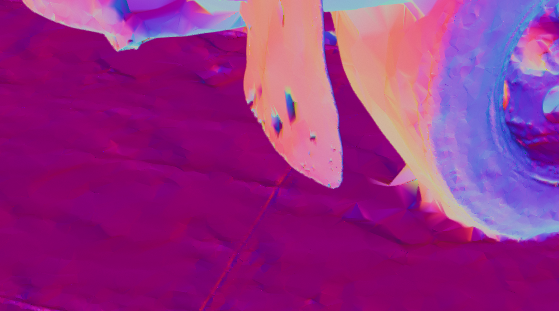}}} 
&
\makecell{\includegraphics[width=0.235\linewidth]{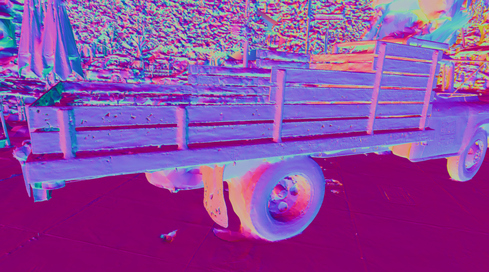}} 
&
\makecell{
\fcolorbox{red}{white}{\includegraphics[width=0.095\linewidth]{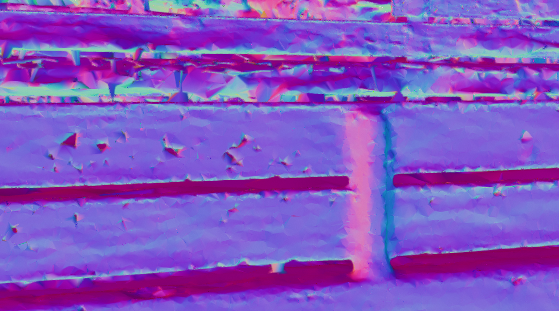}}\\[-0.235mm]
\fcolorbox{blue}{white}{\includegraphics[width=0.095\linewidth]{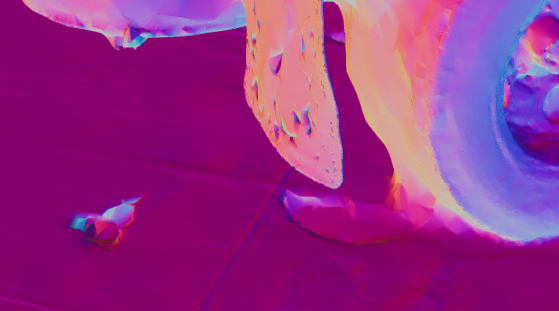}}} 
\\[-0.5mm]
Precision/Recall/F1-Score\textsuperscript{$\uparrow$} & 
\multicolumn{2}{c}{{0.552}/{0.564}/{0.558}} & 
\multicolumn{2}{c}{\textbf{0.584}/\textbf{0.609}/\textbf{0.596}}
\\[-0.5mm]
Optimization/Meshing (minutes)& 
\multicolumn{2}{c}{\textbf{20.7}/\textbf{5.2}} & 
\multicolumn{2}{c}{{52.9}/{22.9}} 
\\
\makecell{\includegraphics[width=0.235\linewidth]{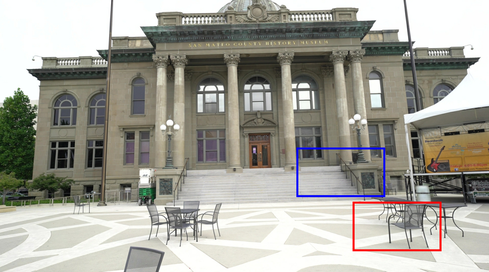}} 
&
\makecell{\includegraphics[width=0.235\linewidth]{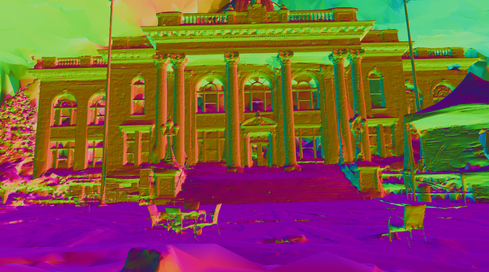}} 
&
\makecell{
\fcolorbox{red}{white}{\includegraphics[width=0.095\linewidth]{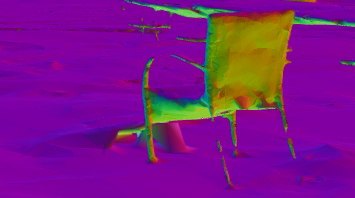}}\\[-0.235mm]
\fcolorbox{blue}{white}{\includegraphics[width=0.095\linewidth]{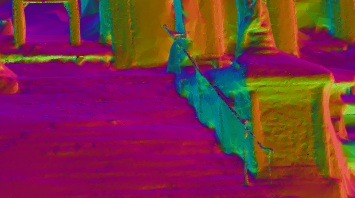}}} 
&
\makecell{\includegraphics[width=0.235\linewidth]{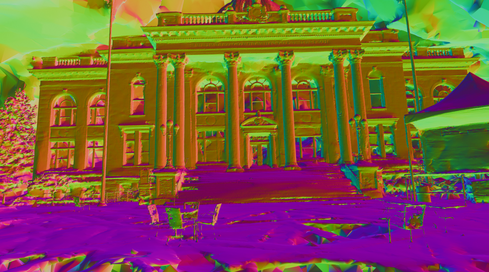}} 
&
\makecell{
\fcolorbox{red}{white}{\includegraphics[width=0.095\linewidth]{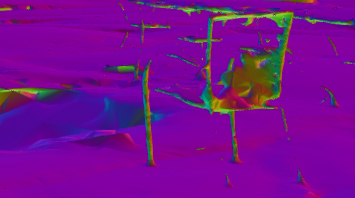}}\\[-0.235mm]
\fcolorbox{blue}{white}{\includegraphics[width=0.095\linewidth]{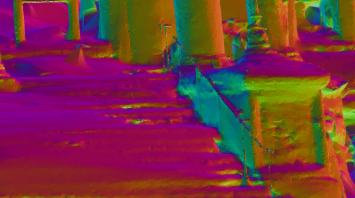}}} 
\\[-0.5mm]
Precision/Recall/F1-Score\textsuperscript{$\uparrow$} & 
\multicolumn{2}{c}{\textbf{0.509}/{0.210}/\textbf{0.297}} & 
\multicolumn{2}{c}{{0.441}/\textbf{0.214}/{0.288}}
\\[-0.5mm]
Optimization/Meshing (minutes)& 
\multicolumn{2}{c}{\textbf{11.9}/\textbf{7.3}} & 
\multicolumn{2}{c}{{34.0}/{34.9}} 
\\
\makecell{\includegraphics[width=0.235\linewidth]{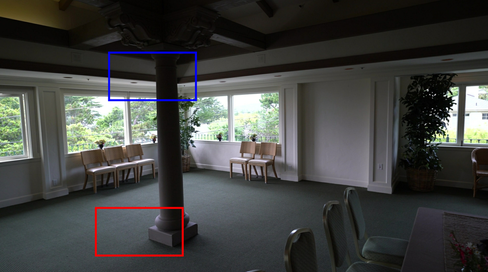}} 
&
\makecell{\includegraphics[width=0.235\linewidth]{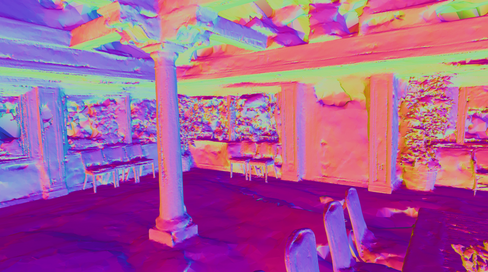}} 
&
\makecell{
\fcolorbox{red}{white}{\includegraphics[width=0.095\linewidth]{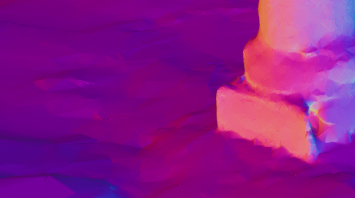}}\\[-0.235mm]
\fcolorbox{blue}{white}{\includegraphics[width=0.095\linewidth]{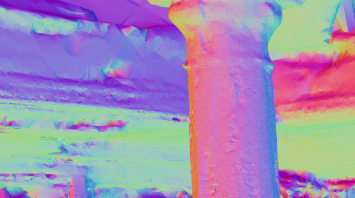}}} 
&
\makecell{\includegraphics[width=0.235\linewidth]{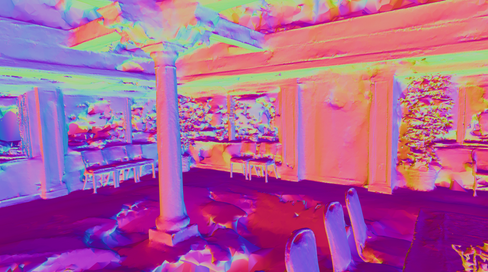}} 
&
\makecell{
\fcolorbox{red}{white}{\includegraphics[width=0.095\linewidth]{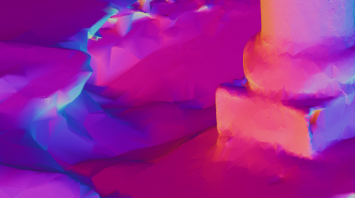}}\\[-0.235mm]
\fcolorbox{blue}{white}{\includegraphics[width=0.095\linewidth]{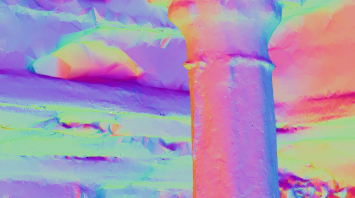}}} 
\\[-0.5mm]
Precision/Recall/F1-Score\textsuperscript{$\uparrow$} & 
\multicolumn{2}{c}{\textbf{0.433}/\textbf{0.240}/\textbf{0.309}} & 
\multicolumn{2}{c}{{0.403}/{0.2358}/{0.275}}
\\[-0.5mm]
Optimization/Meshing (minutes)& 
\multicolumn{2}{c}{\textbf{15.6}/\textbf{4.7}} & 
\multicolumn{2}{c}{{52.1}/{19.1}} 
\\
\makecell{\includegraphics[width=0.235\linewidth]{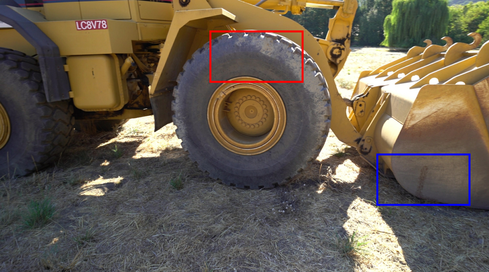}} 
&
\makecell{\includegraphics[width=0.235\linewidth]{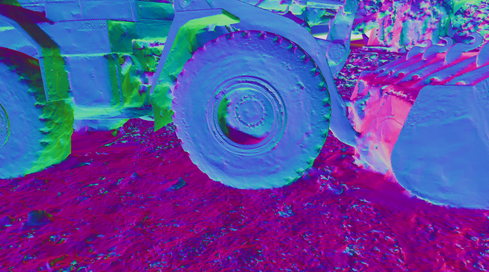}} 
&
\makecell{
\fcolorbox{red}{white}{\includegraphics[width=0.095\linewidth]{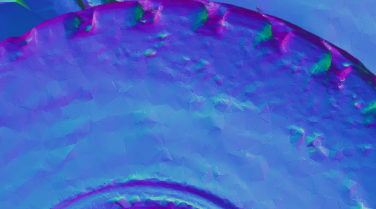}}\\[-0.235mm]
\fcolorbox{blue}{white}{\includegraphics[width=0.095\linewidth]{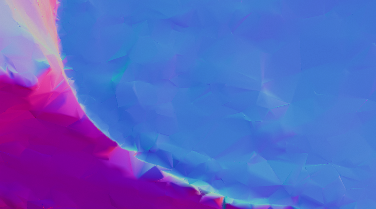}}} 
&
\makecell{\includegraphics[width=0.235\linewidth]{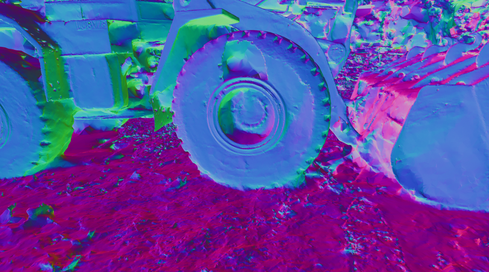}} 
&
\makecell{
\fcolorbox{red}{white}{\includegraphics[width=0.095\linewidth]{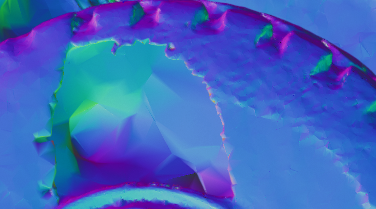}}\\[-0.235mm]
\fcolorbox{blue}{white}{\includegraphics[width=0.095\linewidth]{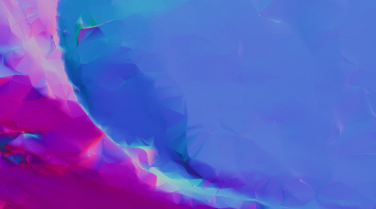}}} 
\\[-0.5mm]
Precision/Recall/F1-Score\textsuperscript{$\uparrow$} & 
\multicolumn{2}{c}{{0.399}/\textbf{0.417}/\textbf{0.408}} & 
\multicolumn{2}{c}{\textbf{0.404}/{0.400}/{0.402}}
\\[-0.5mm]
Optimization/Meshing (minutes)& 
\multicolumn{2}{c}{\textbf{14.4}/\textbf{2.5}} & 
\multicolumn{2}{c}{{39.2}/{20.5}} 
\\
\makecell{\includegraphics[width=0.235\linewidth]{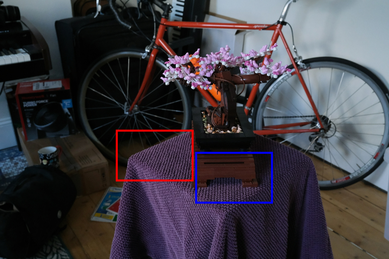}} 
&
\makecell{\includegraphics[width=0.235\linewidth]{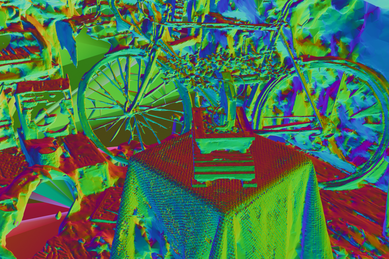}} 
&
\makecell{
\fcolorbox{red}{white}{\includegraphics[width=0.095\linewidth]{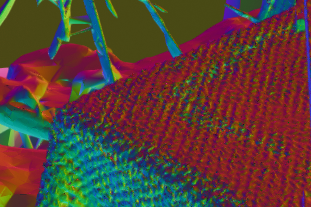}}\\[-0.235mm]
\fcolorbox{blue}{white}{\includegraphics[width=0.095\linewidth]{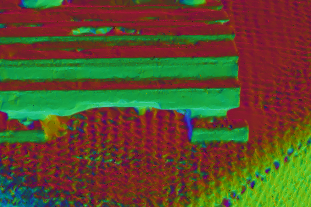}}} 
&
\makecell{\includegraphics[width=0.235\linewidth]{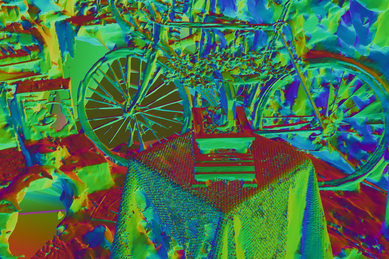}} 
&
\makecell{
\fcolorbox{red}{white}{\includegraphics[width=0.095\linewidth]{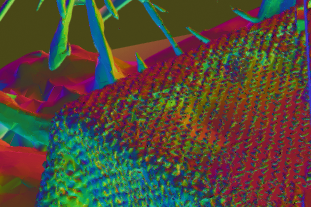}}\\[-0.235mm]
\fcolorbox{blue}{white}{\includegraphics[width=0.095\linewidth]{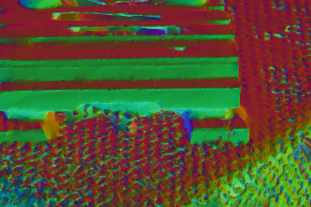}}} 
\\[-0.5mm]
Precision/Recall/F1-Score\textsuperscript{$\uparrow$} & 
\multicolumn{2}{c}{-} & 
\multicolumn{2}{c}{-}
\\[-0.5mm]
Optimization/Meshing (minutes)& 
\multicolumn{2}{c}{\textbf{27.1}/\textbf{0.6}} & 
\multicolumn{2}{c}{{60.3}/{7.1}} 
\\
\makecell{\includegraphics[width=0.235\linewidth]{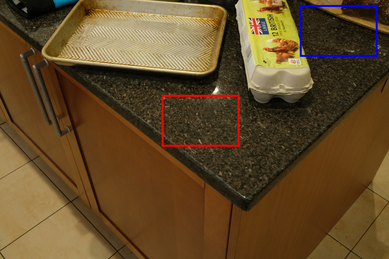}} 
&
\makecell{\includegraphics[width=0.235\linewidth]{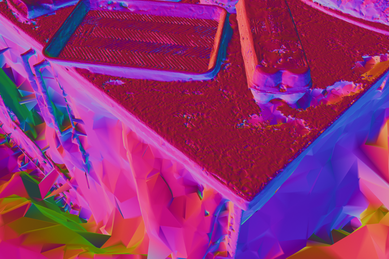}} 
&
\makecell{
\fcolorbox{red}{white}{\includegraphics[width=0.095\linewidth]{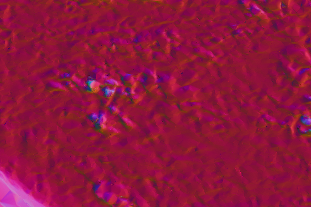}}\\[-0.235mm]
\fcolorbox{blue}{white}{\includegraphics[width=0.095\linewidth]{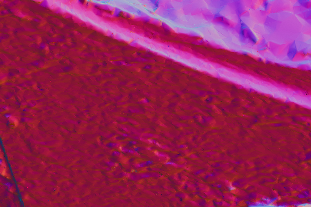}}} 
&
\makecell{\includegraphics[width=0.235\linewidth]{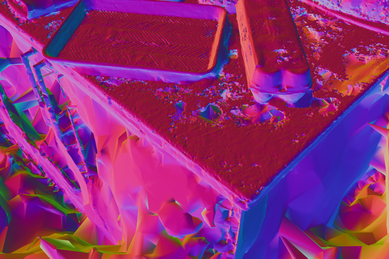}} 
&
\makecell{
\fcolorbox{red}{white}{\includegraphics[width=0.095\linewidth]{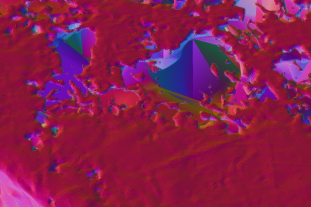}}\\[-0.235mm]
\fcolorbox{blue}{white}{\includegraphics[width=0.095\linewidth]{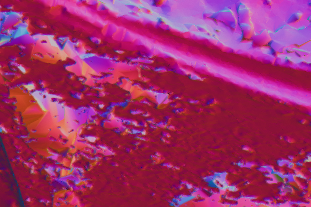}}} 
\\[-0.5mm]
Precision/Recall/F1-Score\textsuperscript{$\uparrow$} & 
\multicolumn{2}{c}{-} & 
\multicolumn{2}{c}{-}
\\[-0.5mm]
Optimization/Meshing (minutes)& 
\multicolumn{2}{c}{\textbf{30.1}/\textbf{0.7}} & 
\multicolumn{2}{c}{{70.2}/{7.7}} 
\\
\end{tabular}
}
  \caption{\label{fig:mesh_comparison}%
    \textbf{Qualitative Evaluation for Unbounded Mesh Extraction}:
We compare our unbounded meshes with those extracted by \gof using scenes from Tanks \& Temples \cite{Knapitsch2017tanks} as well as Mip-NeRF 360 \cite{barron2022mipnerf360}.
We report surface reconstruction metrics (if available) as well as optimization/meshing times (not including Tetrahedralization).
As we can see, our meshes are more detailed and exhibit fewer artifacts.
}
\end{figure*}

\begin{table}[ht!]
    \centering
        \setlength{\tabcolsep}{3pt}
    \caption{
\new{    \textbf{Detailed Opacity Statistics}:
\gof consistently overestimates the depth, leading to opacity values which vastly exceed $0.5$. 
Our method provides better estimates of the level set, benefiting from (1) improved sorting accuracy and (2) direct opacity field supervision.
Metrics were obtained for all test views, using scenes from the Tanks \& Temples dataset \cite{Knapitsch2017tanks}.
}}
\resizebox{.98\linewidth}{!}{
\begin{tabular}{lrrrrrrr}
\toprule
Method & Barn & Caterp & Courth & Ignatius & Meetingr & Truck & Average \\
\midrule
GOF & \cellcolor{tab_color!0} 0.83 & \cellcolor{tab_color!0} 0.75 & \cellcolor{tab_color!0} 0.83 & \cellcolor{tab_color!0} 0.74 & \cellcolor{tab_color!0} 0.83 & \cellcolor{tab_color!0} 0.81 & \cellcolor{tab_color!0} 0.80 \\
Ours (w/o Exact Depth) & \cellcolor{tab_color!0} 0.74 & \cellcolor{tab_color!0} 0.67 & \cellcolor{tab_color!0} 0.79 & \cellcolor{tab_color!0} 0.66 & \cellcolor{tab_color!0} 0.70 & \cellcolor{tab_color!0} 0.71 & \cellcolor{tab_color!0} 0.71 \\
Ours & \cellcolor{tab_color!32} 0.53 & \cellcolor{tab_color!32} 0.51 & \cellcolor{tab_color!32} 0.55 & \cellcolor{tab_color!32} 0.51 & \cellcolor{tab_color!32} 0.52 & \cellcolor{tab_color!32} 0.52 & \cellcolor{tab_color!32} 0.52 \\
\bottomrule
\end{tabular}
}
    \label{tab:tnt_opacity}
\end{table}

\paragraph{Exact Depth Computation.}
To compare our depth rendering strategy with the one from GOF, we evaluate the opacity according to Eqn. \eqref{eq:opacityfield}, with the median depth as used by \gof and our exact depth; we report these metrics averaged over all test views for the Tanks \& Temples dataset \cite{Knapitsch2017tanks}.

As can be seen in Table \ref{tab:tnt_opacity}, \gof exhibits high opacity values, which can be attributed to (1) approximate sorting and (2) their reliance on a non-robust depth estimate.
Our method provides close alignment with the $0.5$ level set, benefiting from hierarchical resorting, our exact depth estimate, as well as $\mathcal{L}_{\text{opa}}$.
Even when disabling exact depth, we still attain better alignment compared to \gof.
For a visualization, we refer to Fig. \ref{fig:small_opacity_comparison}.




\section{Discussion, Limitations and Future Work}

Generally, 3D Gaussians remain a poor target to extract a level set from, as identifying their exact level set requires expensive iterative evaluations (as performed in this work).
Another alternative is to use constant-density ellipsoids \cite{mai2024ever} or constant-density tetrahedral cells \cite{govindarajan2025radfoarm}, for which the exact level set can be computed efficiently;
however, these methods imply longer optimization times.
Therefore, moving towards fully correct volumetric rendering of 3D Gaussians \cite{talegaonkar2025volumetrically, condor2025dontsplat} may offer a practical compromise.

Another challenging current limitation is the use of low-order spherical harmonics, which do not allow for accurate modeling of complex view-dependent effects.
Thus, 3DGS resorts to incorrect Gaussian placement to "fake" correct view-dependent appearance with additional geometry, in turn posing a challenge to surface reconstruction methods (\cf Fig.
\ref{fig:failures}).
While recent work has already investigated improved approaches for view-dependent appearance \cite{liu2025beta, yang2024spec}, these methods still fall short in extreme scenarios, highlighting the need for a more robust solution.

\begin{figure}[h]
\sffamily
\setlength{\tabcolsep}{1pt}%
\setlength{\fboxsep}{0pt}%
\setlength{\fboxrule}{0.5pt}%
\renewcommand{\arraystretch}{1.1}%
\centering

\centering
\resizebox{\linewidth}{!}{
\begin{tabular}{lccc}
& Ground Truth & Ours & Crop \\
&
\makecell{\includegraphics[width=0.40\linewidth]{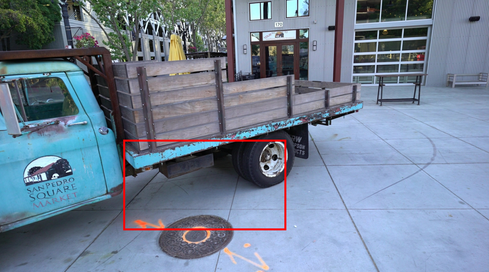}} 
&
\makecell{\includegraphics[width=0.40\linewidth]{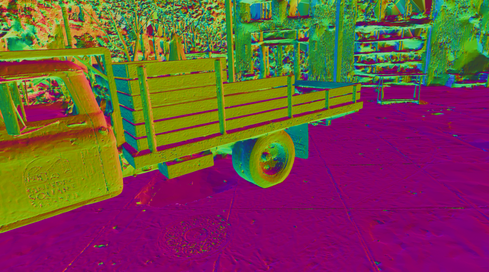}} 
&
\makecell{
\fcolorbox{red}{white}{\includegraphics[width=0.186\linewidth]{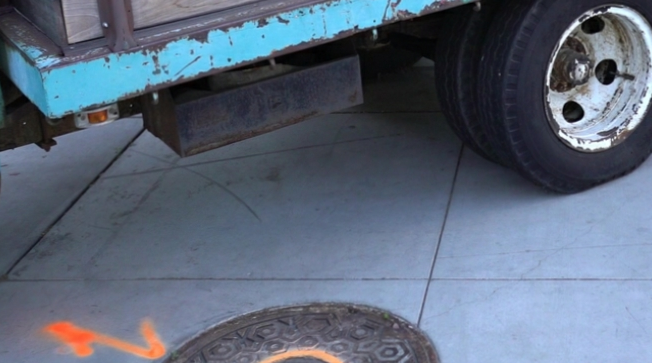}}\\[-0.235mm]
\fcolorbox{red}{white}{\includegraphics[width=0.186\linewidth]{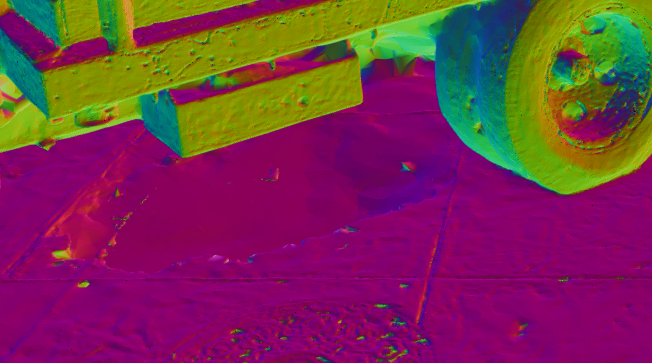}}} 
\\[-0.235mm]
&
\makecell{\includegraphics[width=0.40\linewidth]{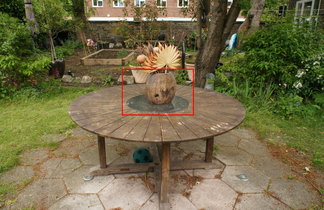}} &
\makecell{\includegraphics[width=0.40\linewidth]{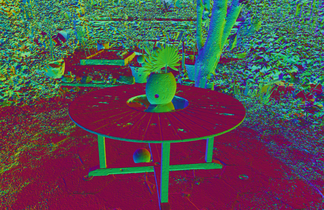}} &
\makecell{
\fcolorbox{red}{white}{\includegraphics[width=0.186\linewidth]{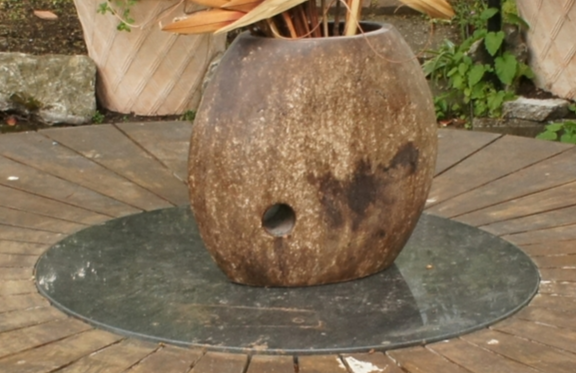}}\\[-0.5mm]
\fcolorbox{red}{white}{\includegraphics[width=0.186\linewidth]{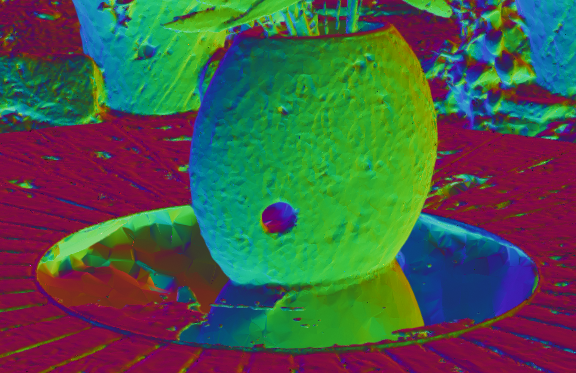}}}
\end{tabular}
}
\caption{\textbf{Failure Cases:} 3DGS-based methods struggle to model surfaces with view-dependent effects, resulting in holes due to a lack of primitive density, as well as inconsistent Gaussian orientation.
Note that these artifacts are specific to virtually all 3DGS-based methods, and stem from low-order spherical harmonics, as well as current image-based densification strategies.
}
\label{fig:failures}
\end{figure}

\new{
Another challenge is primitive density: 
Current densification methods for 3DGS \cite{kerbl20233dgs, yu2024gof, kheradmand2024mcmc} are image-based; thus, regions which are observed more often are generally reconstructed better, with more primitives.
However, this also implies that regions not observed as often have lower primitive counts, again posing a challenge to surface reconstruction methods.
This highlights the need for a more steerable densification procedure for 3DGS.
For more details, we refer to \ifACM{App. B.5}{\appref{app:failure_analysis}}.}

Finally, while we demonstrated improved quality for unbounded surface reconstruction, quality generally degrades compared to state-of-the-art implicit methods \cite{li2023neuralangelo} or recent \revised{works}{preprints} which explore bounded meshing with MVS-constraints \cite{wang2024GausSurf, chen2024pgsr}.
Future work could further improve the scene representation to bridge the current quality gap.


\section{Conclusion}
In this work, we rigorously analyzed Gaussian Opacity Fields and its shortcomings.
First, we show how \gof consistently overestimates the depth along a view ray, which leads to worse geometry reconstruction results; we present a more exact depth estimate, which remedies these issues and enables a better estimate of the $0.5$ level set during training.
In addition, we presented a novel extent loss along with direct opacity field evaluation to improve quality for meshing of unbounded scenes.
Although these losses incur a slight computational overhead, this is remedied by smart performance optimization, leading to reduced training times overall.
We also presented a novel \emph{Marching Tetrahedra} parallelization scheme, which is up to $10\times$ faster compared to \gof.
Our full method outperforms related work both qualitatively and quantitatively;
compared to current state-of-the art methods for unbounded surface reconstruction, our meshes exhibit more details, and our full approach is much more efficient.
{Our full implementation is publicly available at {\color{blue}\url{https://github.com/r4dl/SOF}}.}

\begin{acks}
This research was supported by the Austrian Science Fund \textit{FWF} [10.55776/I6663], the German Science Foundation \textit{DFG} [contract 528364066] and the \textit{Alexander von Humboldt Foundation} funded by the German Federal Ministry of Research, Technology and Space. For open access purposes, the author has applied a CC-BY public copyright license
to any author-accepted manuscript version arising from this submission.
We thank Thomas K\"ohler for helpful discussions and support during this project.
\end{acks}

\bibliographystyle{ACM-Reference-Format}
\bibliography{bibliography}

@String(CVPR= {IEEE Conf. Comput. Vis. Pattern Recog.})

@String(ICCV= {Int. Conf. Comput. Vis.})

@String(ECCV= {Eur. Conf. Comput. Vis.})

@String(NIPS= {Adv. Neural Inform. Process. Syst.})

@String(TOG= {ACM Trans. Graph.})

@String(SIG= {SIGGRAPH})

@String(SIGASIA= {SIGGRAPH Asia})

@String(TVCG  = {IEEE Trans. Vis. Comput. Graph.})

@String(CGF  = {Comput. Graph. Forum})

@String(CVPR  = {CVPR})

@String(ICCV  = {ICCV})

@String(ECCV  = {ECCV})

@String(NIPS  = {NeurIPS})

@String(TOG   = {ACM TOG})

@String(TVCG  = {IEEE TVCG})

@String{vis = {IEEE Visualization}}

@String{PACM = {Proc. ACM Comput. Graph. Interact. Tech}}

@inproceedings{yang2024spec,
  title={{Spec-Gaussian: Anisotropic View-Dependent Appearance for 3D Gaussian Splatting}},
  author={Yang, Ziyi and Gao, Xinyu and Sun, Yangtian and Huang, Yihua and Lyu, Xiaoyang and Zhou, Wen and Jiao, Shaohui and Qi, Xiaojuan and Jin, Xiaogang},
  booktitle=NIPS,
  year={2024}
}

@inproceedings{barron2023zip,
    title     = {{Zip-NeRF: Anti-Aliased Grid-Based Neural Radiance Fields}},
    author    = {Barron, Jonathan T. and Mildenhall, Ben and Verbin, Dor and Srinivasan, Pratul P. and Hedman, Peter},
    booktitle = ICCV,
    year      = {2023}
}

@inproceedings{barron2021mipnerf,
  title     = {{Mip-NeRF: A Multiscale Representation for Anti-Aliasing Neural Radiance Fields}},
  author    = {Jonathan T. Barron and Ben Mildenhall and Matthew Tancik and Peter Hedman and Ricardo Martin-Brualla and Pratul P. Srinivasan},
  booktitle = ICCV,
  year      = {2021}
}

@inproceedings{bulo2024revising,
author = {Rota Bul\`{o}, Samuel and Porzi, Lorenzo and Kontschieder, Peter},
title = {{Revising Densification in Gaussian Splatting}},
year = {2024},
booktitle = ECCV
}

@article{mueller2022instant,
	author    = {Thomas M\"uller and Alex Evans and Christoph Schied and Alexander Keller},
	title     = {{Instant Neural Graphics Primitives with a Multiresolution Hash Encoding}},
	journal   = TOG,
	volume    = {41},
	number    = {4},
	articleno = {102},
	numpages  = {15},
	year      = {2022},
}

@article{Tu2025VRSplat,
    author    = {Tu, Xuechang and Radl, Lukas and Steiner, Michael and Steinberger, Markus and Kerbl, Bernhard and de la Torre, Fornando},
    title     = {{VRsplat: Fast and Robust Gaussian Splatting for Virtual Reality}},
    journal   = PACM,
    volume    = {8},
    number    = {1},
    articleno = {1},
    year      = {2025},
}

@article{loccoz20243dgrt,
    author = {Nicolas Moenne-Loccoz and Ashkan Mirzaei and Or Perel and Riccardo de Lutio and Janick Martinez Esturo and Gavriel State and Sanja Fidler and Nicholas Sharp and Zan Gojcic},
    title = {{3D Gaussian Ray Tracing: Fast Tracing of Particle Scenes}},
    journal = TOG,
    year = {2024},
articleno = {232},
numpages = {19},
volume = {43},
number = {6},
}

@inproceedings{steiner2025aaagaussians,
      title={{AAA-Gaussians: Anti-Aliased and Artifact-Free 3D Gaussian Rendering}}, 
      author={Michael Steiner and Thomas Köhler and Lukas Radl and Felix Windisch and Dieter Schmalstieg and Markus Steinberger},
      year={2025},
      booktitle=ICCV,
}

@inproceedings{huang2024optimal,
	title       =   {{On the Error Analysis of 3D Gaussian Splatting and an Optimal Projection Strategy}},
	author      = {Letian Huang and Jiayang Bai and Jie Guo and Yuanqi Li and Yanwen Guo},
	booktitle   = ECCV,
	year        = {2024}
}

@article{hahlbohm2025perspective,
    author = {Hahlbohm, Florian and Friederichs, Fabian and Weyrich, Tim and Franke, Linus and Kappel, Moritz and Castillo, Susana and Stamminger, Marc and Eisemann, Martin and Magnor, Marcus},
    title = {{Efficient Perspective-Correct 3D Gaussian Splatting Using Hybrid Transparency}},
    journal = CGF,
  volume  = {44},
  number  = {2},
  year    = {2025},
}

@inproceedings{fridovich2022plenoxels,
	title     = {{Plenoxels: Radiance Fields without Neural Networks}},
	author    = {Fridovich-Keil, Sara and Yu, Alex and Tancik, Matthew and Chen, Qinhong and Recht, Benjamin and Kanazawa, Angjoo},
	booktitle = CVPR,
	year      = {2022}
}

@article{reiser2024bog,
author = {Reiser, Christian and Garbin, Stephan and Srinivasan, Pratul and Verbin, Dor and Szeliski, Richard and Mildenhall, Ben and Barron, Jonathan and Hedman, Peter and Geiger, Andreas},
title = {{Binary Opacity Grids: Capturing Fine Geometric Detail for Mesh-Based View Synthesis}},
year = {2024},
journal = TOG,
month = jul,
articleno = {149},
numpages = {14},
}

@inproceedings{Chen2022ECCV,
  author = {Anpei Chen and Zexiang Xu and Andreas Geiger and Jingyi Yu and Hao Su},
  title = {{TensoRF: Tensorial Radiance Fields}},
  booktitle = ECCV,
  year = {2022}
}

@inproceedings{talegaonkar2025volumetrically,
      title={{Volumetrically Consistent 3D Gaussian Rasterization}}, 
      author={Chinmay Talegaonkar and Yash Belhe and Ravi Ramamoorthi and Nicholas Antipa},
      year={2025},
      booktitle = CVPR,
}

@inproceedings{zwicker2001EWA,
	title      = {{EWA Volume Splatting}},
	booktitle  = vis,
	author     = {Zwicker, Mathias and Pfister, Hanspeter and van Baar, Jeroen and Gross, Markus},
	year       = {2001}
}

@article{yu2024gof,
author = {Yu, Zehao and Sattler, Torsten and Geiger, Andreas},
title = {{Gaussian Opacity Fields: Efficient Adaptive Surface Reconstruction in Unbounded Scenes}},
year = {2024},
volume = {43},
number = {6},
journal = TOG,
month = nov,
articleno = {271},
numpages = {13},
}

@inproceedings{yariv2023baked,
author = {Yariv, Lior and Hedman, Peter and Reiser, Christian and Verbin, Dor and Srinivasan, Pratul P. and Szeliski, Richard and Barron, Jonathan T. and Mildenhall, Ben},
title = {{BakedSDF: Meshing Neural SDFs for Real-Time View Synthesis}},
year = {2023},
booktitle = SIG,
}

@article{Andersson2020Flip,
	author = {Andersson, Pontus and Nilsson, Jim and Akenine-M\"{o}ller, Tomas and Oskarsson, Magnus and \r{A}str\"{o}m, Kalle and Fairchild, Mark D.},
	title = {{FLIP: A Difference Evaluator for Alternating Images}},
	year = {2020},
	volume = {3},
	number = {2},
	articleno = {15},
	numpages = {23},
	journal = PACM
}

@article{condor2025dontsplat,
    author = {Condor, Jorge and Speierer, Sebastien and Bode, Lukas and Bozic, Aljaz and Green, Simon and Didyk, Piotr and Jarabo, Adrian},
    title = {{Don't Splat your Gaussians: Volumetric Ray-Traced Primitives for Modeling and Rendering Scattering and Emissive Media}},
    year = {2025},
    volume = {44},
    number = {1},
    articleno = {10},
    journal = TOG,
}

@inproceedings{mai2024ever,
      title={{EVER: Exact Volumetric Ellipsoid Rendering for Real-time View Synthesis}}, 
      author={Alexander Mai and Peter Hedman and George Kopanas and Dor Verbin and David Futschik and Qiangeng Xu and Falko Kuester and Jonathan T. Barron and Yinda Zhang},
      year={2025},
      booktitle=ICCV,
}

@inproceedings{liu2025beta,
      title={{Deformable Beta Splatting}}, 
      author={Rong Liu and Dylan Sun and Meida Chen and Yue Wang and Andrew Feng},
      year={2025},
      booktitle = SIG,
}

@misc{govindarajan2025radfoarm,
      title={{Radiant Foam: Real-Time Differentiable Ray Tracing}}, 
      author={Shrisudhan Govindarajan and Daniel Rebain and Kwang Moo Yi and Andrea Tagliasacchi},
      year={2025},
      eprint={2502.01157},
      archivePrefix={arXiv},
      primaryClass={cs.CV},
      url={https://arxiv.org/abs/2502.01157}, 
}

@article{hedman2018deepblending,
  author    = {Hedman, Peter and Philip, Julien and Price, True and Frahm, Jan-Michael and Drettakis, George and Brostow, Gabriel},
  title     = {{Deep Blending for Free-viewpoint Image-based Rendering}},
  journal   = TOG,
  volume    = {37},
  number    = {6},
  articleno = {257},
  numpages  = {15},
  year      = {2018}
}

@inproceedings{lorensen1987marching,
author = {Lorensen, William E. and Cline, Harvey E.},
title = {{Marching Cubes: A high resolution 3D surface construction algorithm}},
year = {1987},
booktitle = SIG,
}

@inproceedings{shen2021dmtet,
    title = {{Deep Marching Tetrahedra: a Hybrid Representation for High-Resolution 3D Shape Synthesis}},
    author = {Tianchang Shen and Jun Gao and Kangxue Yin and Ming-Yu Liu and Sanja Fidler},
    year = {2021},
    booktitle = NIPS
}

@inproceedings{Oechsle2021unisurf,
    title = {{UNISURF: Unifying Neural Implicit Surfaces and Radiance Fields for Multi-View Reconstruction}},
    author = {Oechsle, Michael and Peng, Songyou and Geiger, Andreas},
    booktitle = ICCV,
    year = {2021},
  }

@inproceedings{Yu2022MonoSDF,
  author    = {Yu, Zehao and Peng, Songyou and Niemeyer, Michael and Sattler, Torsten and Geiger, Andreas},
  title     = {{MonoSDF: Exploring Monocular Geometric Cues for Neural Implicit Surface Reconstruction}},
  booktitle   = NIPS,
  year      = {2022},
}

@article{radl2024stopthepop,
	author    = {Radl, Lukas and Steiner, Michael and Parger, Mathias and Weinrauch, Alexander and Kerbl, Bernhard and Steinberger, Markus},
	title     = {{StopThePop: Sorted Gaussian Splatting for View-Consistent Real-time Rendering}},
	journal   = TOG,
	number    = {3},
	volume    = {43},
	articleno = {64},
	year      = {2024}
}

@Article{papantonakis2024reducing,
    author       = {Papantonakis, Panagiotis and Kopanas, Georgios and Kerbl, Bernhard and Lanvin, Alexandre and Drettakis, George},
    title        = {{Reducing the Memory Footprint of 3D Gaussian Splatting}},
    journal      = PACM,
    number       = {1},
    volume       = {7},
    year         = {2024},
}

@inproceedings{yariv2021volsdf,
  title={{Volume Rendering of Neural Implicit Surfaces}},
  author={Yariv, Lior and Gu, Jiatao and Kasten, Yoni and Lipman, Yaron},
  booktitle=NIPS,
  year={2021}
}

@inproceedings{fan2023lightgaussian,
	author    = {Zhiwen Fan and Kevin Wang and Kairun Wen and Zehao Zhu and Dejia Xu and Zhangyang Wang},
	title     = {{LightGaussian: Unbounded 3D Gaussian Compression with 15x Reduction and 200+ FPS}},
	year      = {2024}, 
	booktitle = NIPS
}

@inproceedings{jensen2014large,
  title={{Large Scale Multi-View Stereopsis Evaluation}},
  author={Jensen, Rasmus and Dahl, Anders and Vogiatzis, George and Tola, Engil and Aan{\ae}s, Henrik},
  booktitle=CVPR,
  year={2014}
}

@article{Knapitsch2017tanks,
	author    = {Arno Knapitsch and Jaesik Park and Qian-Yi Zhou and Vladlen Koltun},
	title     = {{Tanks and Temples: Benchmarking Large-Scale Scene Reconstruction}},
	journal   = TOG,
	volume    = {36},
	number    = {4},
	articleno = {78},
	numpages = {13},
	year      = {2017}
}

@inproceedings{kulhanek2023tetranerf,
  title={{Tetra-NeRF: Representing Neural Radiance Fields Using Tetrahedra}},
  author={Kulhanek, Jonas and Sattler, Torsten},
  booktitle=ICCV,
  year={2023}
}

@inproceedings{gao2024relightable3d,
author = {Gao, Jian and Gu, Chun and Lin, Youtian and Li, Zhihao and Zhu, Hao and Cao, Xun and Zhang, Li and Yao, Yao},
title = {{Relightable 3D Gaussians: Realistic Point Cloud Relighting with BRDF Decomposition and Ray Tracing}},
year = {2024},
booktitle = ECCV,
}

@article{chen2024pgsr,
author={Chen, Danpeng and Li, Hai and Ye, Weicai and Wang, Yifan and Xie, Weijian and Zhai, Shangjin and Wang, Nan and Liu, Haomin and Bao, Hujun and Zhang, Guofeng},
journal= TVCG,
title={{PGSR: Planar-Based Gaussian Splatting for Efficient and High-Fidelity Surface Reconstruction }},
year={2025},
volume={31},
number={09},
}

@article{wang2024GausSurf,
    title={{GausSurf: Geometry-Guided 3D Gaussian Splatting for Surface Reconstruction}}, 
    author={Wang, Jiepeng and Liu, Yuan and Wang, Peng and Lin, Cheng and Hou, Junhui and Li, Xin and Komura, Taku and Wang, Wenping},
    journal={arXiv preprint arXiv:2411.19454},
    year={2024}
}

@inproceedings{Dai2024GaussianSurfels,
  author = {Dai, Pinxuan and Xu, Jiamin and Xie, Wenxiang and Liu, Xinguo and Wang, Huamin and Xu, Weiwei},
  title = {{High-quality Surface Reconstruction using Gaussian Surfels}},
  booktitle = SIGASIA,
  year = {2024},
}

@inproceedings{jiang2023gaussianshader,
  title={{GaussianShader: 3D Gaussian Splatting with Shading Functions for Reflective Surfaces}},
  author={Jiang, Yingwenqi and Tu, Jiadong and Liu, Yuan and Gao, Xifeng and Long, Xiaoxiao and Wang, Wenping and Ma, Yuexin},
  booktitle=CVPR,
  year={2024}
}

@inproceedings{kheradmand2024mcmc,
 author = {Kheradmand, Shakiba and Rebain, Daniel and Sharma, Gopal and Sun, Weiwei and Tseng, Yang-Che and Isack, Hossam and Kar, Abhishek and Tagliasacchi, Andrea and Yi, Kwang Moo},
 booktitle = NIPS,
 title = {{3D Gaussian Splatting as Markov Chain Monte Carlo}},
 year = {2024}
}

@inproceedings{Yu2024MipSplatting,
  author    = {Yu, Zehao and Chen, Anpei and Huang, Binbin and Sattler, Torsten and Geiger, Andreas},
  title     = {{Mip-Splatting: Alias-free 3D Gaussian Splatting}},
  booktitle   = CVPR,
  year      = {2024},
}

@inproceedings{curless1996volumetric,
  title={{A Volumetric Method for Building Complex Models from Range Images}},
  author={Curless, Brian and Levoy, Marc},
  booktitle=SIG,
  year={1996}
}

@inproceedings{li2023neuralangelo,
  title={{Neuralangelo: High-Fidelity Neural Surface Reconstruction}},
  author={Li, Zhaoshuo and M\"uller, Thomas and Evans, Alex and Taylor, Russell H and Unberath, Mathias and Liu, Ming-Yu and Lin, Chen-Hsuan},
  booktitle=CVPR,
  year={2023}
}

@inproceedings{guedon2023sugar,
    title={{SuGaR: Surface-Aligned Gaussian Splatting for Efficient 3D Mesh Reconstruction and High-Quality Mesh Rendering}},
    author={Gu{\'e}don, Antoine and Lepetit, Vincent},
    booktitle=CVPR,
    year={2024}
  }

@article{kerbl20233dgs,
	author    = {Kerbl, Bernhard and Kopanas, Georgios and Leimk{\"u}hler, Thomas and Drettakis, George},
	title     = {{3D Gaussian Splatting for Real-Time Radiance Field Rendering}},
	journal   = TOG,
	number    = {4},
	volume    = {42},
	year      = {2023}
}

@inproceedings{huang20242dgs,
author = {Huang, Binbin and Yu, Zehao and Chen, Anpei and Geiger, Andreas and Gao, Shenghua},
title = {{2D Gaussian Splatting for Geometrically Accurate Radiance Fields}},
year = {2024},
booktitle = SIG,
}

@inproceedings{zhang2018unreasonable,
	title     = {{The Unreasonable Effectiveness of Deep Features as a Perceptual Metric}},
	author    = {Zhang, Richard and Isola, Phillip and Efros, Alexei A and Shechtman, Eli and Wang, Oliver},
	booktitle = CVPR,
	year      = {2018}
}

@inproceedings{barron2022mipnerf360,
  author={Barron, Jonathan T. and Mildenhall, Ben and Verbin, Dor and Srinivasan, Pratul P. and Hedman, Peter},
  booktitle=CVPR, 
  title={{Mip-NeRF 360: Unbounded Anti-Aliased Neural Radiance Fields}}, 
  year={2022},
}

@inproceedings{Mildenhall2020NeRF,
	title      = {{NeRF: Representing Scenes as Neural Radiance Fields for View Synthesis}},
	author     = {Mildenhall, Ben and Srinivasan, Pratul P. and Tancik, Matthew and Barron, Jonathan T. and Ramamoorthi, Ravi and Ng, Ren},
	year       = {2020},
	booktitle  = ECCV
}

@inproceedings{mallick2024taming,
	author={Saswat Mallick and Rahul Goel and Kerbl, Bernhard and Vicente Carrasco, Francisco and Steinberger, Markus and De La Torre, Fernando},
	title={T{aming 3DGS: High-Quality Radiance Fields with Limited Resources}},
	booktitle = SIGASIA,
	year={2024}
}

@inproceedings{lin2024vastgaussian,
  title     = {VastGaussian: Vast 3D Gaussians for Large Scene Reconstruction},
  author    = {Lin, Jiaqi and Li, Zhihao and Tang, Xiao and Liu, Jianzhuang and Liu, Shiyong and Liu, Jiayue and Lu, Yangdi and Wu, Xiaofei and Xu, Songcen and Yan, Youliang and Yang, Wenming},
  booktitle = CVPR,
  year      = {2024}
}

@article{nguyen2022snerf,
	title     = {{SNeRF: Stylized Neural Implicit Representations for 3D Scenes}},
	author    = {Nguyen-Phuoc, Thu and Liu, Feng and Xiao, Lei},
	journal   = TOG,
	volume    = {41},
	number    = {4},
	pages     = {142:1--142:11},
	year      = {2022}
}

@inproceedings{radl2024laenerf,
	author    = {Radl, Lukas and Steiner, Michael and Kurz, Andreas and Steinberger, Markus},
	title     = {{LAENeRF: Local Appearance Editing for Neural Radiance Fields}},
	booktitle = CVPR,
	year      = {2024},
}

@article{jambon2023nerfshop,
  title     = {{NeRFshop: Interactive Editing of Neural Radiance Fields}},
  author    = {Jambon, Cl{\'e}ment and Kerbl, Bernhard and Kopanas, Georgios and Diolatzis, Stavros and Drettakis, George and Leimk{\"u}hler, Thomas},
  journal   = PACM,
  volume    = {6},
  number    = {1},
  year      = {2023},
  pages     = {1:1--1:21}
}

@inproceedings{li2025udf,
    title={{GaussianUDF: Inferring Unsigned Distance Functions through 3D Gaussian Splatting}}, 
    author={Shujuan Li and Yu-Shen Liu and Zhizhong Han},
    year = {2025},
    booktitle = CVPR
}

@inproceedings{yu20243dgssdf,
    title={{GSDF: 3DGS Meets SDF for Improved Rendering and Reconstruction}}, 
    author={Mulin Yu and Tao Lu and Linning Xu and Lihan Jiang and Yuanbo Xiangli and Bo Dai},
    year = {2024},
    booktitle = NIPS
}

@article{lyu20243dgsr,
author = {Lyu, Xiaoyang and Sun, Yang-Tian and Huang, Yi-Hua and Wu, Xiuzhe and Yang, Ziyi and Chen, Yilun and Pang, Jiangmiao and Qi, Xiaojuan},
title = {{3DGSR: Implicit Surface Reconstruction with 3D Gaussian Splatting}},
year = {2024},
volume = {43},
number = {6},
journal = TOG,
month = nov,
articleno = {198},
numpages = {12},
}

@article{chen2024dipole,
author = {Chen, Hanyu and Miller, Bailey and Gkioulekas, Ioannis},
title = {{3D Reconstruction with Fast Dipole Sums}},
year = {2024},
volume = {43},
number = {6},
journal = TOG,
articleno = {192},
numpages = {19},
}

@inproceedings{jiang2025geometryfieldsplatting,
      title={{Geometry Field Splatting with Gaussian Surfels}}, 
      author={Kaiwen Jiang and Venkataram Sivaram and Cheng Peng and Ravi Ramamoorthi},
      year={2025},
      booktitle=CVPR,
}

@inproceedings{miller2024objectsvolumesstochasticgeometry,
      title={{Objects as volumes: A stochastic geometry view of opaque solids}}, 
      author={Bailey Miller and Hanyu Chen and Alice Lai and Ioannis Gkioulekas},
      year={2024},
      booktitle=CVPR
}


\appendix

\renewcommand{\thefigure}{A.\arabic{figure}}
\renewcommand{\thetable}{A.\arabic{table}}
\setcounter{figure}{0}
\setcounter{table}{0}

\section{Additional Method Details}
\label{app:details}

In this section, we provide additional implementation details and mathematical derivations for our method.

\subsection{Efficient 3D Gaussian Evaluation}
\label{app:gof:gaussian_space}
The following performance optimization is already present in the current GOF codebase\footnote{\url{https://github.com/autonomousvision/gaussian-opacity-fields}} \cite{yu2024gof}; for completeness, we describe it here.
Note that we re-evaluate \gof using the current implementation for all presented results (unless indicated otherwise). 
\new{Note that the current \gof codebase we used for comparison is $\approx 60\%$ faster than the original one.}

Recall the definition of $A_i,B_i,C_i$ from Sec. 3.1:
\begin{align}
    A_i &= \mb{d}_g^T \mb{d}_g \\
    B_i &= 2\mb{d}_g^T \mb{o}_g \\
    C_i &= \mb{o}_g^T \mb{o}_g.
\end{align}
Decomposing $A_i = \mb{d}^T \mb{R}_i^T \mb{S}_i^{-1} \mb{S}_i^{-1} \mb{R}_i \mb{d}$, we see that this can also be rewritten as $\mb{d}^T \bm{\Sigma}^{-1} \mb{d}$.
Thus, as $\bm{\Sigma}^{-1}$ is a $3\times 3$ symmetric matrix and independent of $\mathbf{d}$, we can precompute it and store it in 6 values per Gaussian.

Similarly, we can precompute $\mb{o}^T \mb{S}_i^{-1} \mb{S}_i^{-1} \mb{R}_i$ for $B_i$ (which can be stored in 3 values) and precompute $C_i$ directly, as a scalar.
As a result, we can compute these $10$ values during Gaussian preprocessing instead of computing them on-the-fly, leading to improved performance overall.


\subsection{Sorting for Opacity Field Evaluation}
\label{app:opacityfield:nosorting}

Here, we mathematically demonstrate why sorting is not required for correct opacity field evaluation.
\begin{proposition}
Let $\alpha_i = o_i \mathcal{G}_i(t_i)$. Now, 
\[
O_{N} = \sum_{i=0}^{N-1} \alpha_i \prod_{j=0}^{i-1} (1 - \alpha_j)
\]
is independent of the sorting order.
\end{proposition}

\begin{proof}
We want to show that $O_N = 1 - \prod_{i=0}^{N-1} (1 - \alpha_j)$, which is order-independent due to the product.
To this end, we have
\begin{equation*}
    O_N = \alpha_0 + (1-\alpha_0) \sum_{i=1}^{N-1} \alpha_i \prod_{j=1}^{i-1} (1 - \alpha_j).
\end{equation*}
We proceed by induction.

\textbf{Base case} (\( N = 1 \)): In both cases, we clearly have $O_1 = \alpha_0$.

\textbf{Inductive step}: Assume the identity holds for \( N - 1 \), \ie,
\[
\sum_{i=1}^{N-1} \alpha_i \prod_{j=1}^{i-1}(1 - \alpha_j) = 1 - \prod_{i=1}^{N-1}(1 - \alpha_i). \tag{a} \label{eq:hypothesis}
\]
We now consider \( N \):
\begin{align*}
\sum_{i=0}^{N-1} \alpha_i \prod_{j=0}^{i-1}(1 - \alpha_j)
&= \alpha_0 + (1 - \alpha_0) \sum_{i=1}^{N-1} \alpha_i \prod_{j=1}^{i-1}(1 - \alpha_j) \\
&\overset{\eqref{eq:hypothesis}}{=} \alpha_0 + (1 - \alpha_0) \left(1 - \prod_{i=1}^{N-1}(1 - \alpha_i)\right) \\
&= \alpha_0 + (1 - \alpha_0) - (1 - \alpha_0) \prod_{i=1}^{N-1}(1 - \alpha_i) \\
&= 1 - \prod_{i=0}^{N-1}(1 - \alpha_i).
\end{align*}
Thus, $O_N$ is a product of $(1-\alpha_i)$, and thus order-independent.
\end{proof}

\subsection{Exact Depth Details}
\label{app:exact_depth}

\paragraph{Exact Depth Derivation.}
We want to find a $t$ such that
\begin{equation}
        T_{i}\left(1-\opa_i \mathcal{G}_i^{1D}(t)\right) = 0.5 \label{eq:app:condition_exact}.
\end{equation}
First, we isolate $\mathcal{G}_i^{1D}(t)$:
\begin{align*}
    T_{i}-T_{i}\opa_i \mathcal{G}_i^{1D}(t) &= 0.5 \\
    \mathcal{G}_i^{1D}(t) &= \frac{T_{i} - 0.5}{\alpha_i}.
\end{align*}
Next, we insert Eqn. (8)
and continue reformulating
\begin{align*}
    \exp\left(-\frac{1}{2}\left(A_i t^2 + B_it + C_i\right)\right) &= \frac{T_{i} - 0.5}{\alpha_i} \\
    A_it^2 + B_it + C_i& + 2\ln\left(\frac{T_{i} - 0.5}{\alpha_i}\right) = 0,
\end{align*}
which is a quadratic equation in $t$.
The solution is thus
\begin{equation}
        t_{1,2} = \rayd\pm\frac{ \sqrt{B_i^2 - 4A_i\left(C_i + 2 \ln\left(\frac{T_{i}-0.5}{\alpha_{i}}\right)\right)}}{2A_i}.
\end{equation}
Because we defined the Gaussian contribution to be maximal at $\rayd$, our exact depth is
\begin{equation}
        t = \rayd-\frac{ \sqrt{B_i^2 - 4A_i\left(C_i + 2 \ln\left(\frac{T_{i}-0.5}{\alpha_{i}}\right)\right)}}{2A_i}.
\end{equation}
%

\paragraph{Backward Details.}
It can be shown that the additional gradients due to the exact depth computations are given as
\begin{align}
    \frac{\partial \delta_t}{\partial A_i} &= \frac{B_i^2 - 2A_i\left(C_i 
+ 2 \ln\left(\frac{T_{i}-0.5}{\alpha_i} \right)
    \right)}{2 A_i^2 \delta_t},\\
    \frac{\partial \delta_t}{\partial B_i} &= \frac{\rayd}{\delta_t},\\
    \frac{\partial \delta_t}{\partial C_i} &= \frac{1}{\delta_t},
\end{align}
where $\delta_t = -\frac{ \sqrt{B_i^2 - 4A_i\left(C_i + 2 \ln\left(\frac{T_{i}-0.5+}{\alpha_i}\right)\right)}}{2A_i}$.
In our implementation, we detached these gradients, as they resulted in worse metrics overall (\cf Table \ref{tab:app:tnt_ablation}, 
\emph{(A)~Exact Depth Grad}).

\begin{table}[ht!]
    \centering
    \caption{
    \textbf{Additional Ablation Studies}:
We introduce different component to our method and report the F1-score for the Tanks \& Temples dataset \cite{Knapitsch2017tanks}.
}
\resizebox{.98\linewidth}{!}{
\begin{tabular}{lrrrrrrrr}
\toprule
Method & {{Barn}} & {{Caterp}} & {{Courth}} & {{Ignatius}} & {{Meetingr}} & {{Truck}} & {{Avg}} \\
\midrule
GOF & \cellcolor{tab_color!0} 0.484 & \cellcolor{tab_color!32} 0.402 & \cellcolor{tab_color!32} 0.288 & \cellcolor{tab_color!0} 0.674 & \cellcolor{tab_color!0} 0.275 & \cellcolor{tab_color!49} 0.596 & \cellcolor{tab_color!0} 0.453 \\
Ours & \cellcolor{tab_color!32} 0.535 & \cellcolor{tab_color!49} 0.408 & \cellcolor{tab_color!49} 0.297 & \cellcolor{tab_color!49} 0.736 & \cellcolor{tab_color!49} 0.309 & \cellcolor{tab_color!15} 0.558 & \cellcolor{tab_color!49} 0.474 \\
(A) Exact Depth Grad & \cellcolor{tab_color!49} 0.542 & \cellcolor{tab_color!0} 0.396 & \cellcolor{tab_color!0} 0.280 & \cellcolor{tab_color!32} 0.725 & \cellcolor{tab_color!32} 0.307 & \cellcolor{tab_color!0} 0.550 & \cellcolor{tab_color!32} 0.467 \\
(B) MCMC & \cellcolor{tab_color!0} 0.487 & \cellcolor{tab_color!0} 0.375 & \cellcolor{tab_color!0} 0.284 & \cellcolor{tab_color!0} 0.674 & \cellcolor{tab_color!0} 0.286 & \cellcolor{tab_color!0} 0.531 & \cellcolor{tab_color!0} 0.439 \\
(C) Recycling & \cellcolor{tab_color!0} 0.531 & \cellcolor{tab_color!32} 0.402 & \cellcolor{tab_color!32} 0.288 & \cellcolor{tab_color!15} 0.721 & \cellcolor{tab_color!15} 0.297 & \cellcolor{tab_color!32} 0.560 & \cellcolor{tab_color!32} 0.467 \\
(D) Recloning & \cellcolor{tab_color!32} 0.535 & \cellcolor{tab_color!0} 0.400 & \cellcolor{tab_color!0} 0.283 & \cellcolor{tab_color!0} 0.720 & \cellcolor{tab_color!0} 0.294 & \cellcolor{tab_color!0} 0.549 & \cellcolor{tab_color!0} 0.463 \\
(E) PGSR Appearance & \cellcolor{tab_color!0} 0.532 & \cellcolor{tab_color!0} 0.398 & \cellcolor{tab_color!32} 0.288 & \cellcolor{tab_color!0} 0.702 & \cellcolor{tab_color!0} 0.284 & \cellcolor{tab_color!0} 0.552 & \cellcolor{tab_color!0} 0.459 \\
\bottomrule
\end{tabular}
}
    \label{tab:app:tnt_ablation}
\end{table}

\begin{table*}[t!]
    \centering
    \caption{
    \textbf{Additional Novel View Synthesis results} for Tanks \& Temples \cite{Knapitsch2017tanks} and Deep Blending \cite{hedman2018deepblending};
we use the same scenes as done in \citet{kerbl20233dgs}.
As can be seen, our method, as well as our MCMC-variant, perform very well for the Deep Blending dataset.
    }
\resizebox{.999\linewidth}{!}{
\begin{tabular}{lrrrrrrrrrrrrrrrr}
\toprule
Method & \multicolumn{4}{c}{DrJohnson} & \multicolumn{4}{c}{Playroom} & \multicolumn{4}{c}{Train} & \multicolumn{4}{c}{Truck} \\
\cmidrule(lr){2-5}\cmidrule(lr){6-9}\cmidrule(lr){10-13}\cmidrule(lr){14-17}
 & PSNR\textsuperscript{$\uparrow$} & SSIM\textsuperscript{$\uparrow$} & LPIPS\textsuperscript{$\downarrow$} & \FLIP\textsuperscript{$\downarrow$} & PSNR\textsuperscript{$\uparrow$} & SSIM\textsuperscript{$\uparrow$} & LPIPS\textsuperscript{$\downarrow$} & \FLIP\textsuperscript{$\downarrow$} & PSNR\textsuperscript{$\uparrow$} & SSIM\textsuperscript{$\uparrow$} & LPIPS\textsuperscript{$\downarrow$} & \FLIP\textsuperscript{$\downarrow$} & PSNR\textsuperscript{$\uparrow$} & SSIM\textsuperscript{$\uparrow$} & LPIPS\textsuperscript{$\downarrow$} & \FLIP\textsuperscript{$\downarrow$} \\
\midrule
3DGS & 29.060 & 0.898 & 0.247 & 0.119 & 29.860 & 0.901 & 0.246 & 0.143 & 22.040 & 0.813 & 0.208 & \cellcolor{tab_color!15} 0.250 & 25.390 & 0.878 & 0.148 & \cellcolor{tab_color!15} 0.148 \\
Mip-Splatting & 29.150 & 0.902 & 0.243 & 0.118 & 30.170 & 0.909 & 0.243 & 0.142 & \cellcolor{tab_color!15} 22.160 & \cellcolor{tab_color!15} 0.818 & 0.205 & \cellcolor{tab_color!32} 0.249 & 25.480 & 0.886 & 0.147 & 0.149 \\
StopThePop & 29.420 & 0.903 & \cellcolor{tab_color!32} 0.234 & \cellcolor{tab_color!15} 0.115 & \cellcolor{tab_color!15} 30.300 & 0.905 & \cellcolor{tab_color!15} 0.235 & \cellcolor{tab_color!15} 0.138 & 21.480 & 0.808 & 0.204 & 0.267 & 24.940 & 0.878 & 0.142 & 0.164 \\
Taming-3DGS & \cellcolor{tab_color!15} 29.510 & \cellcolor{tab_color!49} 0.909 & \cellcolor{tab_color!32} 0.234 & 0.117 & 30.240 & \cellcolor{tab_color!15} 0.911 & \cellcolor{tab_color!15} 0.235 & 0.151 & \cellcolor{tab_color!32} 22.250 & \cellcolor{tab_color!32} 0.819 & 0.208 & 0.253 & \cellcolor{tab_color!32} 25.880 & \cellcolor{tab_color!15} 0.892 & 0.128 & \cellcolor{tab_color!32} 0.143 \\
3DGS-MCMC & \cellcolor{tab_color!32} 29.520 & 0.904 & \cellcolor{tab_color!32} 0.234 & 0.118 & 29.930 & 0.908 & \cellcolor{tab_color!32} 0.233 & 0.151 & \cellcolor{tab_color!49} 22.830 & \cellcolor{tab_color!49} 0.843 & \cellcolor{tab_color!49} 0.183 & \cellcolor{tab_color!49} 0.241 & \cellcolor{tab_color!49} 26.450 & \cellcolor{tab_color!49} 0.900 & \cellcolor{tab_color!49} 0.112 & \cellcolor{tab_color!49} 0.135 \\
2DGS & 28.950 & 0.900 & 0.256 & 0.121 & 30.230 & 0.907 & 0.255 & 0.142 & 21.190 & 0.791 & 0.250 & 0.268 & 25.100 & 0.873 & 0.173 & 0.154 \\
GOF & 28.760 & 0.900 & 0.247 & 0.121 & 30.280 & \cellcolor{tab_color!15} 0.911 & 0.239 & 0.140 & 21.590 & 0.817 & \cellcolor{tab_color!15} 0.203 & 0.265 & \cellcolor{tab_color!15} 25.520 & 0.888 & 0.133 & 0.150 \\
Ours & 29.380 & \cellcolor{tab_color!15} 0.907 & \cellcolor{tab_color!49} 0.232 & \cellcolor{tab_color!32} 0.114 & \cellcolor{tab_color!32} 30.650 & \cellcolor{tab_color!49} 0.915 & \cellcolor{tab_color!49} 0.230 & \cellcolor{tab_color!49} 0.136 & 20.800 & 0.804 & 0.212 & 0.284 & 25.100 & 0.890 & \cellcolor{tab_color!15} 0.127 & 0.163 \\
Ours (MCMC) & \cellcolor{tab_color!49} 29.880 & \cellcolor{tab_color!32} 0.908 & \cellcolor{tab_color!32} 0.234 & \cellcolor{tab_color!49} 0.113 & \cellcolor{tab_color!49} 31.030 & \cellcolor{tab_color!49} 0.915 & 0.242 & \cellcolor{tab_color!49} 0.136 & 20.910 & 0.810 & \cellcolor{tab_color!32} 0.201 & 0.279 & 25.260 & \cellcolor{tab_color!32} 0.893 & \cellcolor{tab_color!32} 0.119 & 0.162 \\
\bottomrule
\end{tabular}
}
    \label{tab:tntdb_results}
\end{table*}

\subsection{Densification}
\label{app:densification}
We identify that densification is actually a significant factor for the resulting mesh quality.
To this end, we integrate 3DGS-MCMC \cite{kheradmand2024mcmc}, a recent state-of-the-art densification method for 3DGS, and present the results in Table \ref{tab:app:tnt_ablation}, variant \emph{(B)~MCMC~Densification}.

As we can see, the results are worse when considering the same number of primitives.
Intuitively, 3DGS-MCMC simply moves low-opacity Gaussians to the location of high-opacity Gaussians, and modifies their opacity such that the change in output is minimal.
Compared to the densification strategy in \gof, this results in a more non-uniform distribution of primitives, which poses issues for the subsequent Marching Tetrahedra algorithm.

\subsection{Dead Gaussians.}
Gaussians with $\opa_i < \frac{1}{255}$ are no longer visible from any view (as their corresponding $\alpha_i$ can no longer exceed this threshold), and can thus no longer receive gradients.
When using the 3D filter from \citet{Yu2024MipSplatting}, this applies to even more primitives, as the opacity with the 3D filter applied $\hat{\opa}_i$ is
\begin{equation}
    \hat{\opa}_i = \opa_i \sqrt{\frac{|\bm{\Sigma}_i|}{|\bm{\Sigma}_i + s \mb{I}|}},   
\end{equation}
with $s > 0$, thus the scaling factor is smaller than 1 (\cf \citet{Yu2024MipSplatting} for details).
In our experiments, we find that $\approx 9\%$ of Gaussians are no longer visible by the end of training (\cf Table \ref{tab:app:deads} for details).
To this end, we experimentally devise two strategies to make full use of our Gaussian budget, after densification.

\begin{table}[ht!]
    \centering
    \caption{
    \textbf{Point Cloud Analysis}:
We report the number of primitives, percentage of unrendered primitives as well as the final number of mesh vertices for Ours and \gof.
    }
    \setlength{\tabcolsep}{4pt}
\resizebox{.98\linewidth}{!}{
\begin{tabular}{lrrrrrrrrr}
\toprule
\multicolumn{2}{l}{Method} & {{Barn}} & {{Caterp}} & {{Courth}} & {{Ignatius}} & {{Meetingr}} & {{Truck}} & {{Avg}} \\
\midrule
\multirow{3}{*}{Ours} & \#Gaussians & 847K & 1.06M & 386K & 2.65M & 942K & 2.19M & 1.35M \\
 & \% dead & 5.02\% & 8.54\% & 7.86\% & 8.07\% & 11.91\% & 8.94\% & 8.39\% \\
& \#Vertices & 10.19M & 11.58M & 3.30M & 27.77M & 9.14M & 22.45M & 14.07M \\
 \midrule
\multirow{3}{*}{GOF} & \#Gaussians & 835K & 1.06M & 390K & 2.63M & 962K & 2.16M & 1.34M \\
 & \% dead & 5.48\% & 7.59\% & 11.26\% & 8.77\% & 12.33\% & 10.61\% & 9.34\% \\
& \#Vertices & 10.05M & 11.55M & 4.02M & 27.76M & 8.61M & 22.68M & 14.03M \\
\bottomrule
\end{tabular}
}
    \label{tab:app:deads}
\end{table}



\paragraph{Recycling.}
Inspired by 3DGS-MCMC \cite{kheradmand2024mcmc}, we repurpose their cloning strategy to move dead Gaussians towards Gaussians with high opacity.

\paragraph{Recloning.}
Due to the tendency of 3DGS-MCMC to create a more non-uniform distribution of Gaussians, we also design a recloning strategy, inspired by \gof densification \cite{yu2024gof}.
As done previously, we sample the alive Gaussians with their opacity as their probability.
However, instead of directly placing the Gaussians at the mean of the sampled Gaussians, we sample another position using the selected Gaussian as the PDF.

As we can see in Table \ref{tab:app:tnt_ablation} (\cf \emph{(C)~Recycling}, \emph{(D)~Recloning}), both variants perform worse compared to simply ignoring these primitives altogether.

\section{Ablation Studies}
\label{app:ablation}
In this section, we present additional ablation studies and more experimental results.

\subsection{Additional Experiments}
\label{app:ablation}

\paragraph{PGSR Appearance.}
We additionally test the appearance model proposed by PGSR \cite{chen2024pgsr}.
Here, each image $i$ is assigned two learnable scalars $a_i, b_i \in \mathbb{R}$, and $\mathcal{L}_{\text{rgb}}$ is computed with the adjusted rendered image $\hat{\mathbf{I}}_{\text{PGSR}}$
\begin{equation}
    \hat{\mathbf{I}}_{\text{PGSR}} = \exp\left(a_i\right) \hat{\mathbf{I}} + b_i,
\end{equation}
with $\hat{\mathbf{I}}$ the rendered image.
We provide results for both DTU and Tanks \& Temples (\cf Table \ref{tab:app_dtu}, \emph{(G)} and Table \ref{tab:app:tnt_ablation}, \emph{(E)}).
As can be seen, we find that the appearance model based on VastGaussian \cite{lin2024vastgaussian}, as used in our final model, performs better compared to the PGSR appearance embedding.


\paragraph{DTU Ablation.}
We present additional ablation studies for the DTU dataset \cite{jensen2014large} in Table \ref{tab:app_dtu}; we also included the results for our re-evaluation of \gof and 2DGS.
We also show a qualitative comparison for bounded meshes in Fig. \ref{fig:dtu_ablation}.

\paragraph{Novel View Synthesis.}
In addition to our results for the Mip-NeRF 360 dataset \cite{barron2022mipnerf360}, we additionally include image quality metrics for 4 scenes sourced from Tanks \& Temples \cite{Knapitsch2017tanks} and Deep Blending \cite{hedman2018deepblending}.
We present the results in Table \ref{tab:tntdb_results}.

We also provide a qualitative comparison in Fig. \ref{fig:renders}.
As we can see, our method produces similar rendering compared to \gof.

\subsection{Bounding Strategies for Marching Tetrahedra}
\label{app:tets_bounding}
As discussed in the main material, the $3\sigma$ cutoff is not correct, and results in bounding-boxes which are too small (if $\opa \approx 1$) or overly large bounding boxes otherwise.
As also discussed in App. \ref{app:densification}, we observe a large number of "dead" primitives, which are no longer rendered due to their low opacity; thus, they also no longer receive gradients.
These Gaussians might correspond to floaters which were eliminated by the opacity resets \cite{kerbl20233dgs} or optimization; hence, they are not necessarily a good indicator of surfaces.

To this end, we perform an additional ablation study with different cutoff values and Gaussian culling.
In terms of grid generation strategies, we evaluate $3\sigma$, as used by \gof \cite{yu2024gof}, $3.33\sigma$ (obtained by letting $\opa=1$ in Eqn. (3)
, which produces overly large bounding boxes)
as well as our proposed bounding strategy.
For culling, we evaluate the standard approach of considering all Gaussians for tetrahedral grid generation as well as removing all Gaussians with $\opa < \opa_{\text{min}}$.
We show the results in Table \ref{tab:app:tnt_mesh}.

As we can see, these strategies have a very minor impact on the final quality when averaged over many scenes.
However, note that removing primitives for the initial tetrahedral grid generation positively impacts meshing times, which justifies our use of the $0.0039$ cutoff.
Conversely, while this results in a reduced number of initial points for the tetrahedral grid generation, this does not impact the vertex count, as can be seen in Table \ref{tab:app:deads}.
%
\begin{table}[ht!]
    \centering
    \caption{
    \textbf{Marching Tetrahedra Bounding Strategies}:
We evaluate 3 types of bounding boxes (StopThePop-bounding, $3\sigma$, $3.33\sigma$) and 2 types of culling strategies (no culling, $\frac{1}{255}$) and report the results for precision, recall and F1-score.
    }
\resizebox{.98\linewidth}{!}{
\begin{tabular}{llrrrr}
\toprule
Bounding & Cutoff && Precision\textsuperscript{$\uparrow$} & Recall\textsuperscript{$\uparrow$} & F1-score\textsuperscript{$\uparrow$} \\
\midrule
\multirow[c]{2}{*}{StopThePop} & - && \cellcolor{tab_color!32} 0.5404 & \cellcolor{tab_color!0} 0.4376 & \cellcolor{tab_color!15} 0.4732 \\
 & 0.0039 && \cellcolor{tab_color!49} 0.5405 & \cellcolor{tab_color!0} 0.4376 & \cellcolor{tab_color!49} 0.4733 \\
   \cmidrule(lr){3-6}
\multirow[c]{2}{*}{3$\sigma$} & - && \cellcolor{tab_color!0} 0.5374 & \cellcolor{tab_color!49} 0.4392 & \cellcolor{tab_color!49} 0.4733 \\
 & 0.0039 && \cellcolor{tab_color!0} 0.5388 & \cellcolor{tab_color!15} 0.4384 & \cellcolor{tab_color!15} 0.4732 \\
   \cmidrule(lr){3-6}
\multirow[c]{2}{*}{3.33$\sigma$} & - && \cellcolor{tab_color!0} 0.5366 & \cellcolor{tab_color!32} 0.4388 & \cellcolor{tab_color!0} 0.4728 \\
 & 0.0039 && \cellcolor{tab_color!15} 0.5389 & \cellcolor{tab_color!15} 0.4384 & \cellcolor{tab_color!15} 0.4732 \\
\bottomrule
\end{tabular}
}
    \label{tab:app:tnt_mesh}
\end{table}
\begin{table}[ht!]
    \centering
    \setlength{\tabcolsep}{4pt}
    \caption{
    \textbf{Detailed per-scene results for the Tanks \& Temples dataset}~\cite{Knapitsch2017tanks}.
    }
\resizebox{.98\linewidth}{!}{
\begin{tabular}{llrrrrrrr}
\toprule
Metric & Method & Barn & Caterp & Courth & Ignatius & Meetingr & Truck & Avg \\
\midrule
\multirow[c]{2}{*}{Precision} & Ours 
& \cellcolor{tab_color!32} 0.589 
& \cellcolor{tab_color!0} 0.399 
& \cellcolor{tab_color!32} 0.509 
& \cellcolor{tab_color!32} 0.768 
& \cellcolor{tab_color!32} 0.433 
& \cellcolor{tab_color!0} 0.552 
& \cellcolor{tab_color!32} 0.542 \\
 & GOF & \cellcolor{tab_color!0} 0.522 
 & \cellcolor{tab_color!32} 0.404 
 & \cellcolor{tab_color!0} 0.441 
 & \cellcolor{tab_color!0} 0.725 
 & \cellcolor{tab_color!0} 0.403 
 & \cellcolor{tab_color!32} 0.584 
 & \cellcolor{tab_color!0} 0.513 \\
  \midrule
\multirow[c]{2}{*}{Recall} & Ours 
& \cellcolor{tab_color!32} 0.490 
& \cellcolor{tab_color!32} 0.417 
& \cellcolor{tab_color!0} 0.210 
& \cellcolor{tab_color!32} 0.707 
& \cellcolor{tab_color!32} 0.240 
& \cellcolor{tab_color!0} 0.564 
& \cellcolor{tab_color!32} 0.438 \\
 & GOF & \cellcolor{tab_color!0} 0.451 
 & \cellcolor{tab_color!0} 0.400 
 & \cellcolor{tab_color!32} 0.214 
 & \cellcolor{tab_color!0} 0.631 
 & \cellcolor{tab_color!0} 0.208 
 & \cellcolor{tab_color!32} 0.609 
 & \cellcolor{tab_color!0} 0.419 \\
 \midrule
 \multirow[c]{2}{*}{F1-score} & Ours 
& \cellcolor{tab_color!32} 0.535 
& \cellcolor{tab_color!32} 0.408 
& \cellcolor{tab_color!32} 0.297 
& \cellcolor{tab_color!32} 0.736 
& \cellcolor{tab_color!32} 0.309 
& \cellcolor{tab_color!0} 0.558 
& \cellcolor{tab_color!32} 0.474 \\
 & GOF & \cellcolor{tab_color!0} 0.484 
 & \cellcolor{tab_color!0} 0.402 
 & \cellcolor{tab_color!0} 0.288 
 & \cellcolor{tab_color!0} 0.674 
 & \cellcolor{tab_color!0} 0.275 
 & \cellcolor{tab_color!32} 0.596 
 & \cellcolor{tab_color!0} 0.453 \\
\bottomrule
\end{tabular}
}
    \label{tab:app:tnt_perscene}
\end{table}

\subsection{Full Timing Breakdown}
\label{app:ablation:timing}
We provide a detailed total timings comparison for our method and \gof in Table \ref{tab:app:performance_total}.
Note that both methods use the Delaunay Triangulation from Tetra-NeRF \cite{kulhanek2023tetranerf}, which uses the CGAL library.
We explicitly isolated this stage, as we did not optimize it and still use the same implementation; however, note this the performance of this stage is influenced by the number of primitives.

As we can see, our method is much faster compared to \gof, particularly for the binary search, where our method reduces the average time from 30 to just 7 minutes.
The tetrahedralization stage also benefits from our culling strategies for tetrahedral grid generation, particularly if the number of "dead" Gaussians is high; however, we observed a constant percentage of dead Gaussians for Tanks \& Temples.
\begin{table*}[ht!]
    \centering
    \caption{\textbf{Detailed Timing Comparisons} for Ours and \gof \cite{yu2024gof} on the Tanks \& Temples dataset \cite{Knapitsch2017tanks}, including Optimization, Tetrahedralization \cite{kulhanek2023tetranerf} and Binary Search (in minutes).
Overall, our method is about $2\times$ faster on average, with improved mesh quality.
    }
\resizebox{.98\linewidth}{!}{
\begin{tabular}{l*{7}{cc}}
\toprule
Timings in minutes         & \multicolumn{2}{c}{Barn (837K)} 
         & \multicolumn{2}{c}{Caterpillar (1.06M)} 
         & \multicolumn{2}{c}{Courthouse (384K)}
         & \multicolumn{2}{c}{Ignatius (2.74M)}
         & \multicolumn{2}{c}{Meetingroom (938K)}
         & \multicolumn{2}{c}{Truck (2.15M)}
         & \multicolumn{2}{c}{Average} \\
\cmidrule(lr){2-3}
\cmidrule(lr){4-5}
\cmidrule(lr){6-7}
\cmidrule(lr){8-9}
\cmidrule(lr){10-11}
\cmidrule(lr){12-13}
\cmidrule(lr){14-15}
         & GOF & Ours
         & GOF & Ours 
         & GOF & Ours 
         & GOF & Ours 
         & GOF & Ours  
         & GOF & Ours 
         & GOF & Ours \\
\midrule
        Optimization 
& 40.1m & \cellcolor{tab_color!30}14.3m 
& 39.2m & \cellcolor{tab_color!30}14.4m 
& 34.0m & \cellcolor{tab_color!30}11.5m
& 50.6m & \cellcolor{tab_color!30}23.7m 
& 52.1m & \cellcolor{tab_color!30}15.6m 
& 52.9m & \cellcolor{tab_color!30}20.7m 
& 44.8m & \cellcolor{tab_color!30}16.7m 
\\
        Tetrahedralization \shortcite{kulhanek2023tetranerf} 
& \cellcolor{tab_color!30}\phantom{0}1.2m & \cellcolor{tab_color!30}\phantom{0}1.2m
& \phantom{0}1.6m & \cellcolor{tab_color!30}\phantom{0}1.5m
& \phantom{0}0.6m & \cellcolor{tab_color!30}\phantom{0}0.5m
& \phantom{0}4.3m & \cellcolor{tab_color!30}\phantom{0}3.8m
& \phantom{0}1.4m & \cellcolor{tab_color!30}\phantom{0}1.3m
& \phantom{0}3.1m & \cellcolor{tab_color!30}\phantom{0}2.9m
& \phantom{0}2.2m & \cellcolor{tab_color!30}\phantom{0}1.9m
\\
        Binary Search  
& 30.3m & \cellcolor{tab_color!30} \phantom{0}3.1m 
& 20.5m & \cellcolor{tab_color!30} \phantom{0}2.5m 
& 34.9m & \cellcolor{tab_color!30} \phantom{0}7.3m 
& 15.6m & \cellcolor{tab_color!30} \phantom{0}5.2m 
& 19.1m & \cellcolor{tab_color!30} \phantom{0}4.7m 
& 22.9m & \cellcolor{tab_color!30} \phantom{0}5.2m 
& 23.9m & \cellcolor{tab_color!30} \phantom{0}4.6m 
\\\midrule
        Total 
& 71.6m & \cellcolor{tab_color!30} 18.6m
& 61.3m & \cellcolor{tab_color!30} 18.4m
& 69.5m & \cellcolor{tab_color!30} 19.3m
& 70.5m & \cellcolor{tab_color!30} 32.7m
& 72.6m & \cellcolor{tab_color!30} 21.6m
& 78.9m & \cellcolor{tab_color!30} 28.8m
& 70.7m & \cellcolor{tab_color!30} 23.2m\\
\bottomrule
    \end{tabular}
}
    \label{tab:app:performance_total}
\end{table*}

\subsection{Per Scene Results}
In this section, we present detailed per-scene results for surface reconstruction and novel view synthesis.

\paragraph{TNT Dataset.}
We present detailed per-scene results for the Tanks \& Temples dataset \cite{Knapitsch2017tanks}, including precision and recall, compared to \gof in Table \ref{tab:app:tnt_perscene}.
As we can see, our method achieves better reconstruction results compared to \gof for all scenes, except for \emph{Truck}.
Particularly for this scene, we find that our method struggles to reconstruct a smooth ground (\cf Fig. 9), resulting in a reduction for all surface reconstruction metrics.

\paragraph{Mip-NeRF 360 Dataset.}
We present detailed per-scene results for the Mip-NeRF 360 dataset \cite{barron2022mipnerf360} in Table \ref{tab:m360_perscene}.

\subsection{\new{Failure Analysis}}
\label{app:failure_analysis}
\new{
Here, we investigate why our method performs less favorably for the \emph{Truck} scene.
Our analysis can be seen in Fig. \ref{fig:app:failureanalysis}.}

\new{
In \emph{Error Comparison: Recall}, we compare the recall for each point in the ground-truth mesh; to highlight the differences, we color-code the relative performance.
As can be seen, the region where \gof outperforms our method is the floor region.
To understand the underlying causes, we render the number of contributing Gaussians for each pixel for both methods, which reveals that \gof exhibits increased primitive density for this region.
This further highlights the need for a robust, steerable densification scheme for 3DGS.
}

\new{
Furthermore, when inspecting the training images, we observe that this region has significant view-dependent illumination, where Spherical Harmonics commonly used in 3DGS methods commonly struggle.
In such cases, 3DGS often "fakes" correct view-dependent appearance with geometry placed below semi-transparent surfaces.
}

\begin{figure}[!h]
  \tiny\sffamily
\setlength{\tabcolsep}{1pt}%
\setlength{\fboxsep}{0pt}%
\setlength{\fboxrule}{0.25pt}%
\renewcommand{\arraystretch}{1.1}%
\resizebox{.99\linewidth}{!}{
\begin{tabular}{ccc}
Ours & \gof & Error Comparison: Recall
\\[-0.25mm]
\makecell{
\fcolorbox{blue}{white}{\includegraphics[width=0.2\linewidth]{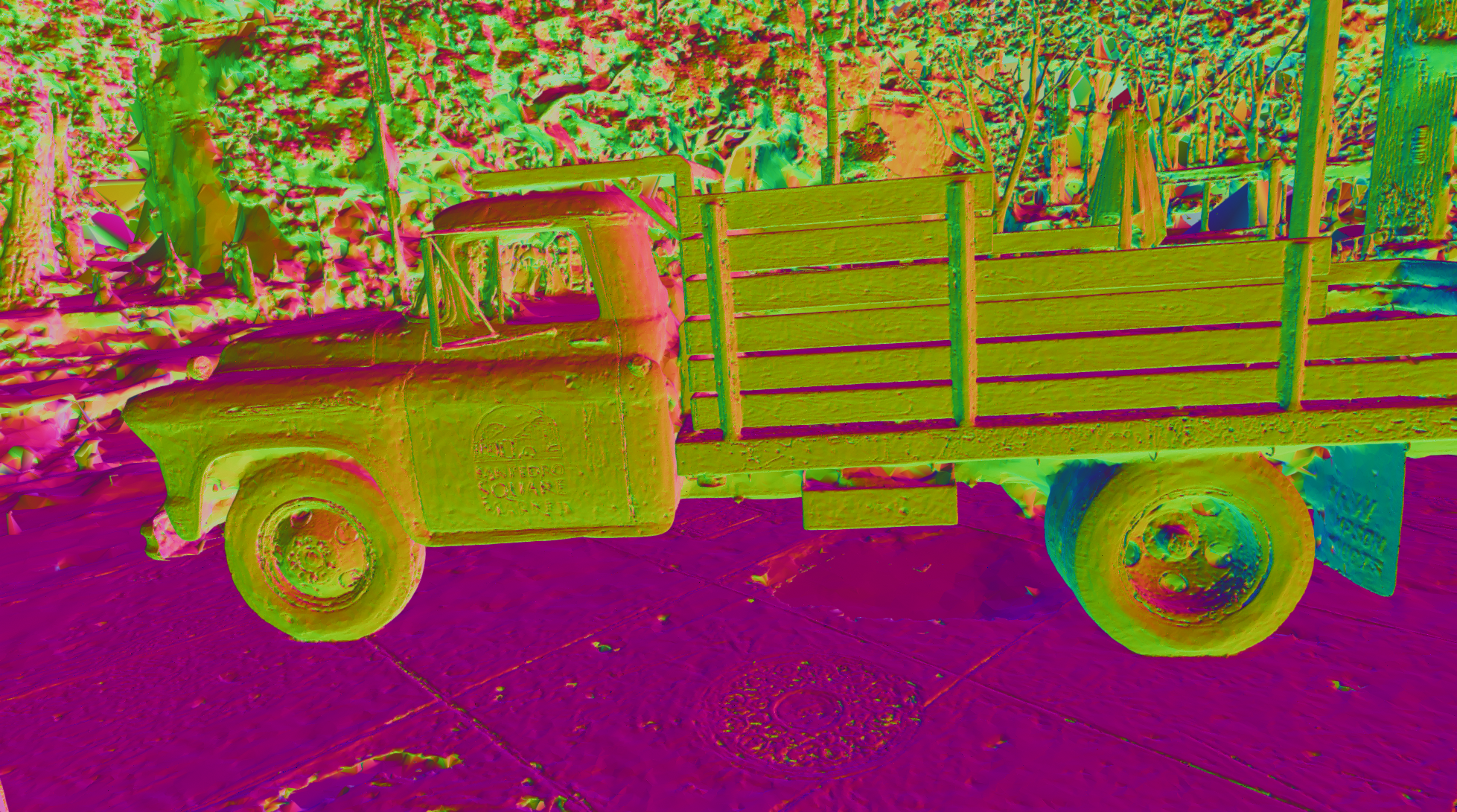}}\\[-0.25mm]
\fcolorbox{blue}{white}{\includegraphics[width=0.2\linewidth]{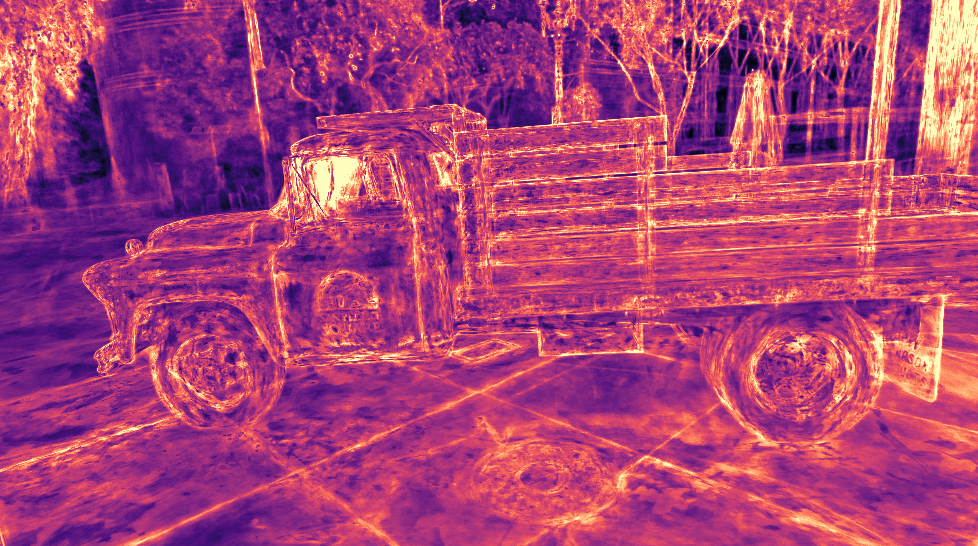}}} 
&
\makecell{
\fcolorbox{red}{white}{\includegraphics[width=0.2\linewidth]{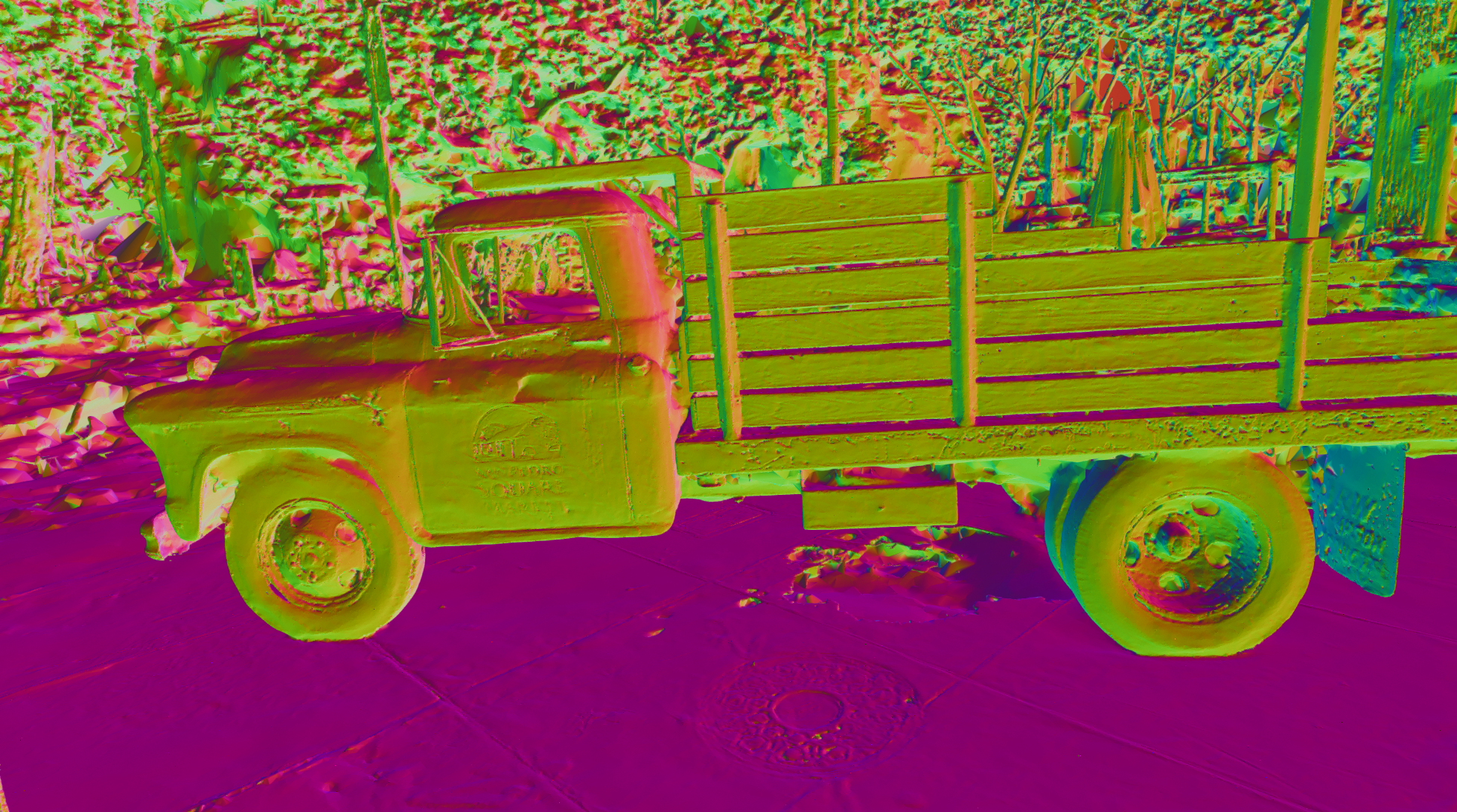}}\\[-0.25mm]
\fcolorbox{red}{white}{\includegraphics[width=0.2\linewidth]{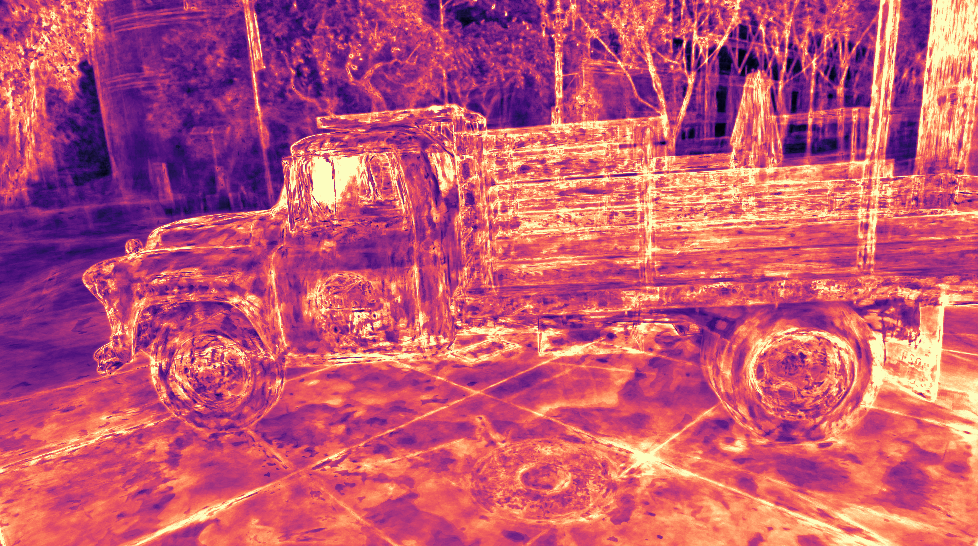}}
}  
&
\makecell{\includegraphics[width=0.355\linewidth]{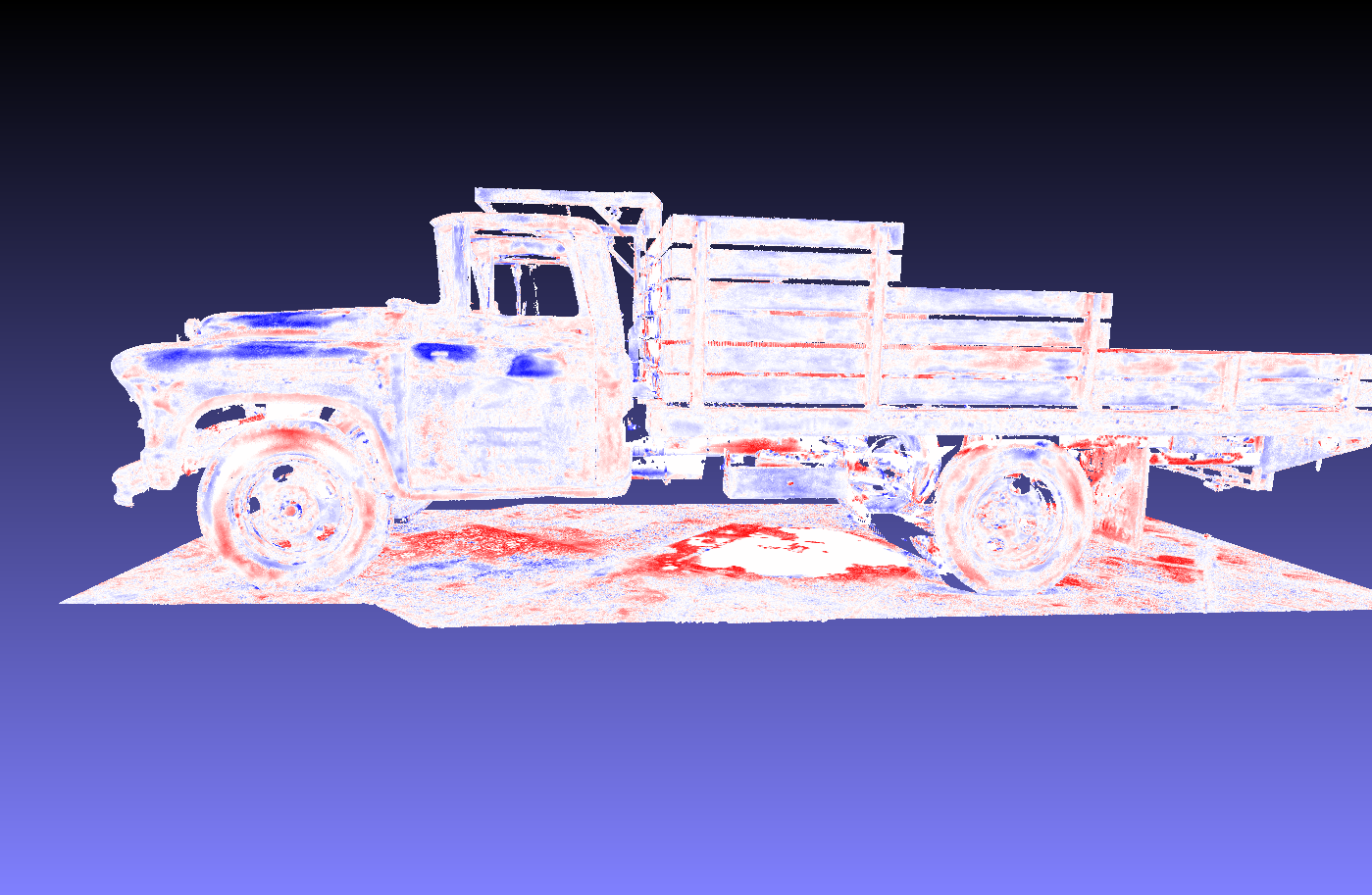}} \\[-0.25mm]
\multicolumn{3}{c}{\includegraphics[width=0.57\linewidth]{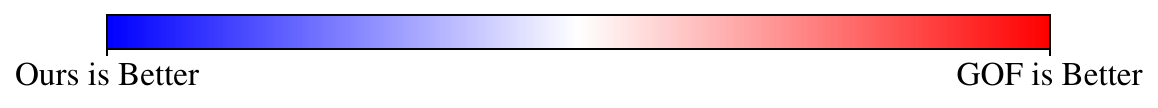}}
\end{tabular}
}
  \caption{\label{fig:app:failureanalysis}%
    \textbf{Detailed Failure Analysis}:
\new{
As can be seen in the Recall Comparison, the underperformance of our method relative to GOF for the \emph{Truck} scene can be attributed to the floor.
Rendering the number of contributing primitives reveals the underlying reason:
\gof benefits from more primitives for the floor, highlighting the need for a more steerable densification scheme.
}
  }
\end{figure}

\section{Derivations}
In this section, we provide additional mathematical derivations.

\subsection{Distortion Loss Derivation}
\label{app:deriv:distortion_loss}
The distortion loss \cite{huang20242dgs} is formally defined as 
\begin{equation}
    \mathcal{L}_{\text{dist}} = \sum_{i=0}^{N-1} \sum_{j=0}^{N-1} w_i w_j (d_i - d_j)^2\label{eq:dist1},
\end{equation}
with $d_i = \text{NDC}(t_i)$, and $t_i$ the depth of the $i$-th Gaussian.
As demonstrated by \citet{huang20242dgs}, the loss can be reformulated as
\begin{equation}
    \mathcal{L}_{\text{dist}} = \sum_{i=0}^{N-1} w_i (d_i^2 A_{i-1} + \mathcal{D}_{i-1} - 2 d_i D_{i-1}),
\end{equation}
with $A_i = \sum_{j=0}^{i} w_j$, $D_i = \sum_{j=0}^{i} w_j d_j$, and ${\mathcal{D}}_i = \sum_{j=0}^{i} w_j d_j^2$.

It can be shown that the required derivatives $\frac{\partial \mathcal{L}_{\text{dist}}}{\partial w_k}$, $\frac{\partial \mathcal{L}{\text{dist}}}{\partial d_k}$ are
\begin{align}
    \frac{\partial \mathcal{L}_{\text{dist}}}{\partial w_k} = d_i^2 A_{N-1} + \mathcal{D}_{N-1} - 2 d_i D_{N-1}, \\
    \frac{\partial \mathcal{L}_{\text{dist}}}{\partial d_k} = 2 \alpha_k \left( d_k A_{N-1} - D_{N-1} \right).
\end{align}
However, due to occlusion (\ie $\alpha_k$ being large reduces the loss for $\alpha_{k+1}$ due to $T_{k+1}$), the derivative w.r.t. $\alpha_k$ is actually (in back-to-front order) \cite{huang20242dgs}
\begin{equation}
    \frac{\partial \mathcal{L}_{\text{dist}}}{\partial \alpha_{k}} =
    \alpha_{k} \frac{\partial\mathcal{L}_{\text{dist}}}{\partial w_{k}} +
    (1 - \alpha_{k}) \frac{\partial \mathcal{L}_{\text{dist}}}{\partial w_{k+1}}.
\end{equation}

However, StopThePop \cite{radl2024stopthepop} performs the backward pass in front-to-back ordering, avoiding the otherwise implied memory overhead of storing per-ray sorted lists of Gaussians.

We can write $\frac{\partial \mathcal L}{\partial \alpha_{k}}$ as
\begin{align*}
    \frac{\partial \mathcal L}{\partial \alpha_{k}} &= 
\left( \frac{\partial \mathcal{L}_{\text{dist}}}{\partial w_{k}} - 
\alpha_{k+1}  \frac{\partial \mathcal{L}_{\text{dist}}}{\partial w_{k+1}} -
\sum_{i=k+2}^{N-1} \alpha_i \prod_{j=k+1}^{i-1} (1 - \alpha_{j})  \frac{\partial \mathcal{L}_{\text{dist}}}{\partial w_{i}} \right) T_k \\
&= 
\left( \frac{\partial \mathcal{L}_{\text{dist}}}{\partial w_{k}} - 
\frac{1}{T_{k+1}}\sum_{i=k+1}^{N-1} w_i \frac{\partial \mathcal{L}_{\text{dist}}}{\partial w_{i}}
\right) T_k.
\end{align*}

Now, we analyze the term $\sum_{i=k+1}^{N-1} w_i \frac{\partial \mathcal{L}_{\text{dist}}}{\partial w_{i}}$:
\begingroup
\small 
\begin{align*}
    {\sum_{i=k+1}^{N-1}\! w_i\! \frac{\partial \mathcal{L}_{\text{dist}}}{\partial w_{i}}} &= 
\sum_{i=k+1}^{N-1} w_i (d_i^2 A + \mathcal{D} - 2 d_i D), \\
    &= 
A\! \sum_{j=k+1}^{N-1}\! w_i d_i^2 + 
\mathcal{D} \sum_{j=1}^{N-1} w_i 
- 2D \sum_{j=k+1}^{N-1} w_i d_i, \\
    &= 
A \left(\mathcal{D} - \!\sum_{j=0}^{k}\! w_j d_j^2\right) 
+ \mathcal{D} \left(A - \!\sum_{j=0}^{k}\! w_j\right) 
- 2D \left(D - \!\sum_{j=0}^{k}\! w_j d_j\right) \\
\end{align*}
\endgroup
where we omitted the subscript $N-1$ for notational convenience.
%

Thus, we conclude that we can in fact compute $\frac{\partial \mathcal L}{\partial \alpha_{k}}$ on the fly in \emph{front-to-back ordering}, where we prepare for the next iteration by repeated subtraction starting from $A_{N-1}, D_{N-1}, \mathcal{D}_{N-1}$.
%

\subsection{Extent Loss Derivation}
\label{app:deriv:extent_loss}
The extent of a Gaussian along a view ray can be defined as
\begin{equation}
    \epsilon_i = \frac{ \sqrt{B_i^2 - 4A_i\left(C_i - 2 \ln\left(255\opa_i\right)\right)}}{2A_i},
\end{equation}
which is easily obtained by letting $\alpha_i = \frac{1}{255}$ and solving for $t$. 
However, instead of mapping both $t^*$ and $t^* - \epsilon_i$ to NDC and compute the distance there, we linearize the NDC-mapping:
\begin{equation}
    \text{NDC}(t^* - \epsilon_i) \approx \text{NDC}(t^*) - \text{NDC}'(t^*) (t^* - (t^* - \epsilon_i)).
\end{equation}
Inserting $\frac{\partial \text{NDC}}{\partial z}$ and reformulating, we obtain
\begingroup
\small 
\begin{align}
    \text{NDC}(t^*) - \text{NDC}(t^* - \epsilon_i) &\approx \frac{\far \near}{\left(\far - \near\right)}\frac{\epsilon_i}{(t^*)^2} \\
    &= \frac{\far \near}{\left(\far - \near\right)} \frac{2A_i \sqrt{B_i^2 - 4A_i\left(C_i - \mathcal{E}_i^2\right)}}{B_i^2}.
\end{align}
\endgroup
We now define our extent loss as the alpha-blended sum of individual NDC distances:
\begin{equation}
    \mathcal{L}_{\text{ext}} = \frac{\far \near}{\left(\far - \near\right)} \sum_{i=0}^{N-1} w_i \frac{2A_i \sqrt{B_i^2 - 4A_i\left(C_i - \mathcal{E}_i^2\right)}}{B_i^2}.
\end{equation}
Computing the gradients for this term is straightforward.

\clearpage
\newpage
\begin{figure*}
\footnotesize\sffamily
\setlength{\tabcolsep}{1pt}%
\setlength{\fboxsep}{0pt}%
\setlength{\fboxrule}{0.5pt}%
\renewcommand{\arraystretch}{1.1}%
\resizebox{.99\linewidth}{!}{
\begin{tabular}{cccccccc}
Ground Truth & Ground Truth & Mip-Splatting & Taming-3DGS & 3DGS-MCMC & GOF & Ours & Ours (MCMC) 
\\[-0.25mm]
\makecell{\includegraphics[width=0.302\linewidth]{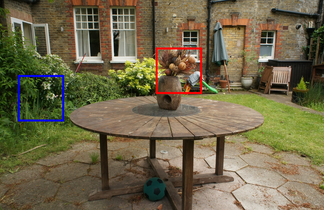}} 
&
\makecell{
\fcolorbox{red}{white}{\includegraphics[width=0.095\linewidth]{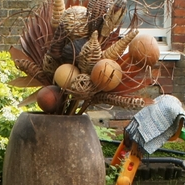}}\\[-0.25mm]
\fcolorbox{blue}{white}{\includegraphics[width=0.095\linewidth]{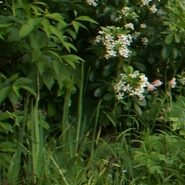}}} 
&
\makecell{
\fcolorbox{red}{white}{\includegraphics[width=0.095\linewidth]{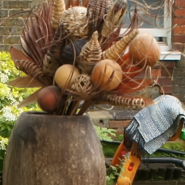}}\\[-0.25mm]
\fcolorbox{blue}{white}{\includegraphics[width=0.095\linewidth]{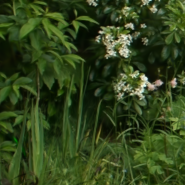}}} 
&
\makecell{
\fcolorbox{red}{white}{\includegraphics[width=0.095\linewidth]{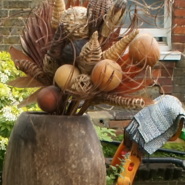}}\\[-0.25mm]
\fcolorbox{blue}{white}{\includegraphics[width=0.095\linewidth]{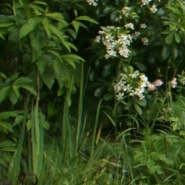}}} 
&
\makecell{
\fcolorbox{red}{white}{\includegraphics[width=0.095\linewidth]{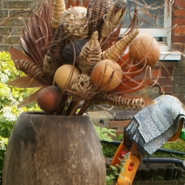}}\\[-0.25mm]
\fcolorbox{blue}{white}{\includegraphics[width=0.095\linewidth]{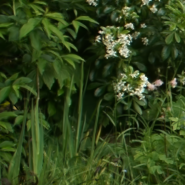}}} 
&
\makecell{
\fcolorbox{red}{white}{\includegraphics[width=0.095\linewidth]{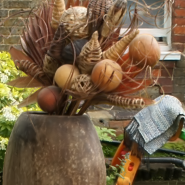}}\\[-0.25mm]
\fcolorbox{blue}{white}{\includegraphics[width=0.095\linewidth]{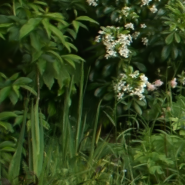}}} 
&
\makecell{
\fcolorbox{red}{white}{\includegraphics[width=0.095\linewidth]{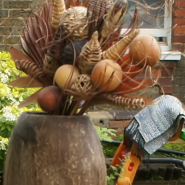}}\\[-0.25mm]
\fcolorbox{blue}{white}{\includegraphics[width=0.095\linewidth]{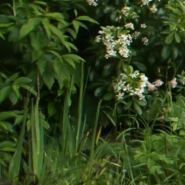}}} 
&
\makecell{
\fcolorbox{red}{white}{\includegraphics[width=0.095\linewidth]{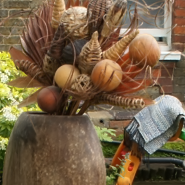}}\\[-0.25mm]
\fcolorbox{blue}{white}{\includegraphics[width=0.095\linewidth]{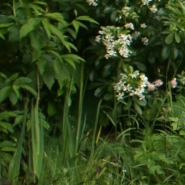}}} 
\\[-0.25mm]
PSNR\textsuperscript{$\uparrow$}/LPIPS\textsuperscript{$\downarrow$}
&& 
\textbf{28.15}/0.105 & 28.32/\textbf{0.099} & 27.89/0.105 & 28.01/0.102 & 27.04/0.119 & 27.25/0.136 \\
\makecell{\includegraphics[width=0.302\linewidth]{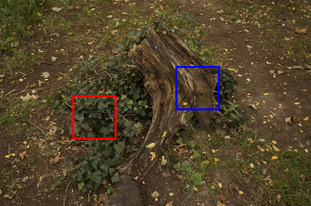}} 
&
\makecell{
\fcolorbox{red}{white}{\includegraphics[width=0.095\linewidth]{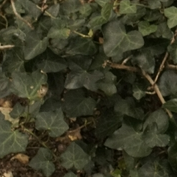}}\\[-0.25mm]
\fcolorbox{blue}{white}{\includegraphics[width=0.095\linewidth]{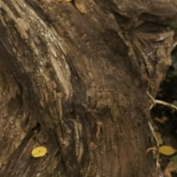}}} 
&
\makecell{
\fcolorbox{red}{white}{\includegraphics[width=0.095\linewidth]{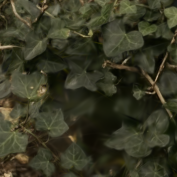}}\\[-0.25mm]
\fcolorbox{blue}{white}{\includegraphics[width=0.095\linewidth]{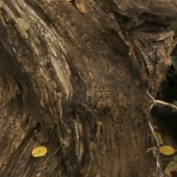}}} 
&
\makecell{
\fcolorbox{red}{white}{\includegraphics[width=0.095\linewidth]{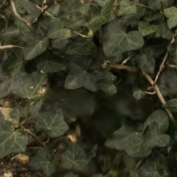}}\\[-0.25mm]
\fcolorbox{blue}{white}{\includegraphics[width=0.095\linewidth]{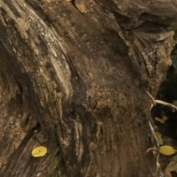}}} 
&
\makecell{
\fcolorbox{red}{white}{\includegraphics[width=0.095\linewidth]{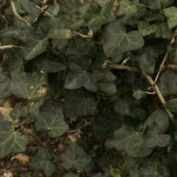}}\\[-0.25mm]
\fcolorbox{blue}{white}{\includegraphics[width=0.095\linewidth]{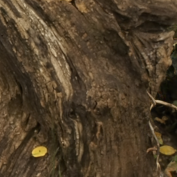}}} 
&
\makecell{
\fcolorbox{red}{white}{\includegraphics[width=0.095\linewidth]{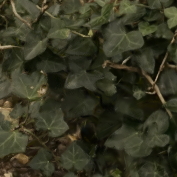}}\\[-0.25mm]
\fcolorbox{blue}{white}{\includegraphics[width=0.095\linewidth]{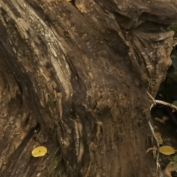}}} 
&
\makecell{
\fcolorbox{red}{white}{\includegraphics[width=0.095\linewidth]{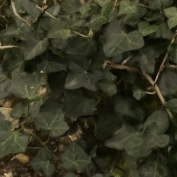}}\\[-0.25mm]
\fcolorbox{blue}{white}{\includegraphics[width=0.095\linewidth]{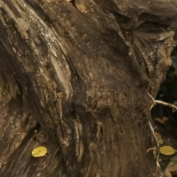}}} 
&
\makecell{
\fcolorbox{red}{white}{\includegraphics[width=0.095\linewidth]{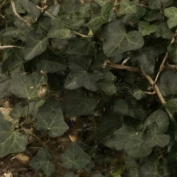}}\\[-0.25mm]
\fcolorbox{blue}{white}{\includegraphics[width=0.095\linewidth]{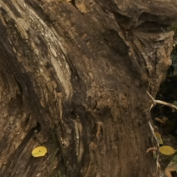}}} 
\\[-0.25mm]
PSNR\textsuperscript{$\uparrow$}/LPIPS\textsuperscript{$\downarrow$}
&& 
27.29/0.183 & 27.83/0.168 & \textbf{28.10}/\textbf{0.148} & 27.77/0.172 & 27.36/0.187 & 27.25/0.178 \\
\makecell{\includegraphics[width=0.302\linewidth]{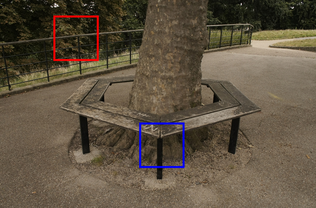}} 
&
\makecell{
\fcolorbox{red}{white}{\includegraphics[width=0.095\linewidth]{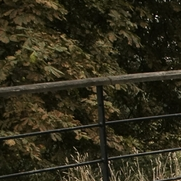}}\\[-0.25mm]
\fcolorbox{blue}{white}{\includegraphics[width=0.095\linewidth]{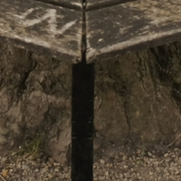}}} 
&
\makecell{
\fcolorbox{red}{white}{\includegraphics[width=0.095\linewidth]{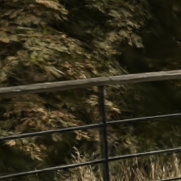}}\\[-0.25mm]
\fcolorbox{blue}{white}{\includegraphics[width=0.095\linewidth]{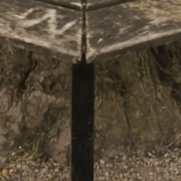}}} 
&
\makecell{
\fcolorbox{red}{white}{\includegraphics[width=0.095\linewidth]{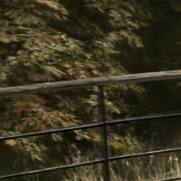}}\\[-0.25mm]
\fcolorbox{blue}{white}{\includegraphics[width=0.095\linewidth]{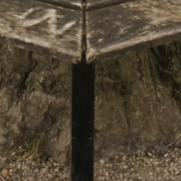}}} 
&
\makecell{
\fcolorbox{red}{white}{\includegraphics[width=0.095\linewidth]{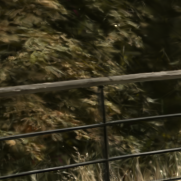}}\\[-0.25mm]
\fcolorbox{blue}{white}{\includegraphics[width=0.095\linewidth]{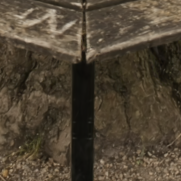}}} 
&
\makecell{
\fcolorbox{red}{white}{\includegraphics[width=0.095\linewidth]{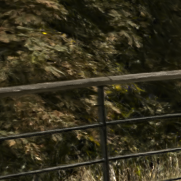}}\\[-0.25mm]
\fcolorbox{blue}{white}{\includegraphics[width=0.095\linewidth]{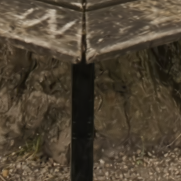}}} 
&
\makecell{
\fcolorbox{red}{white}{\includegraphics[width=0.095\linewidth]{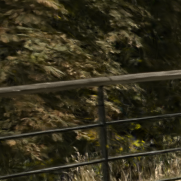}}\\[-0.25mm]
\fcolorbox{blue}{white}{\includegraphics[width=0.095\linewidth]{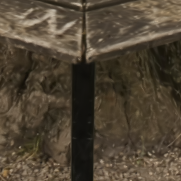}}} 
&
\makecell{
\fcolorbox{red}{white}{\includegraphics[width=0.095\linewidth]{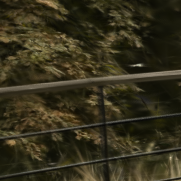}}\\[-0.25mm]
\fcolorbox{blue}{white}{\includegraphics[width=0.095\linewidth]{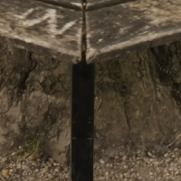}}} 
\\[-0.25mm]
PSNR\textsuperscript{$\uparrow$}/LPIPS\textsuperscript{$\downarrow$}
&& 
24.83/0.270 & 25.11/0.253 & \textbf{25.82}/\textbf{0.217} & 25.25/0.221 & 25.27/0.240 & 25.54/0.223
 \\
\makecell{\includegraphics[width=0.302\linewidth]{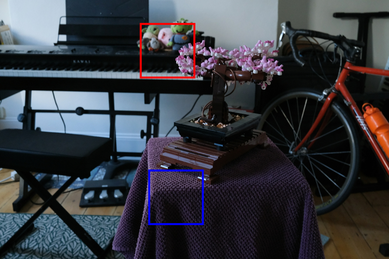}} 
&
\makecell{
\fcolorbox{red}{white}{\includegraphics[width=0.095\linewidth]{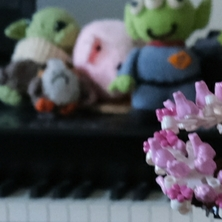}}\\[-0.25mm]
\fcolorbox{blue}{white}{\includegraphics[width=0.095\linewidth]{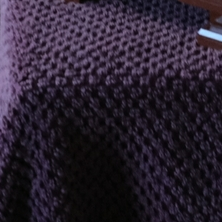}}} 
&
\makecell{
\fcolorbox{red}{white}{\includegraphics[width=0.095\linewidth]{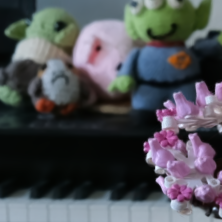}}\\[-0.25mm]
\fcolorbox{blue}{white}{\includegraphics[width=0.095\linewidth]{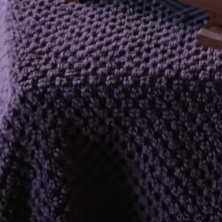}}} 
&
\makecell{
\fcolorbox{red}{white}{\includegraphics[width=0.095\linewidth]{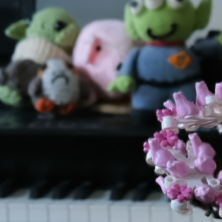}}\\[-0.25mm]
\fcolorbox{blue}{white}{\includegraphics[width=0.095\linewidth]{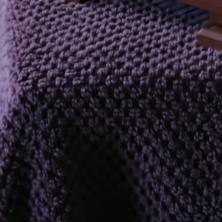}}} 
&
\makecell{
\fcolorbox{red}{white}{\includegraphics[width=0.095\linewidth]{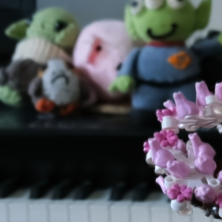}}\\[-0.25mm]
\fcolorbox{blue}{white}{\includegraphics[width=0.095\linewidth]{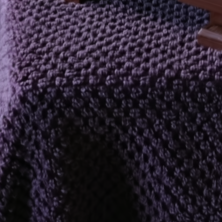}}} 
&
\makecell{
\fcolorbox{red}{white}{\includegraphics[width=0.095\linewidth]{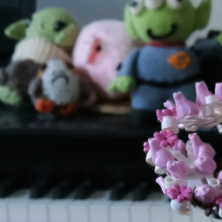}}\\[-0.25mm]
\fcolorbox{blue}{white}{\includegraphics[width=0.095\linewidth]{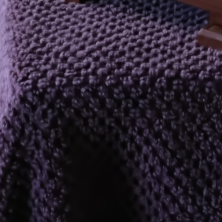}}} 
&
\makecell{
\fcolorbox{red}{white}{\includegraphics[width=0.095\linewidth]{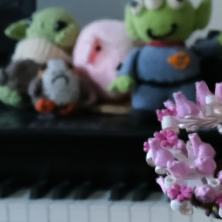}}\\[-0.25mm]
\fcolorbox{blue}{white}{\includegraphics[width=0.095\linewidth]{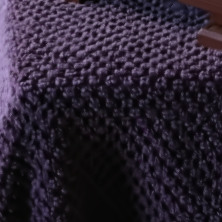}}} 
&
\makecell{
\fcolorbox{red}{white}{\includegraphics[width=0.095\linewidth]{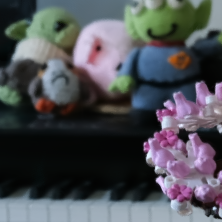}}\\[-0.25mm]
\fcolorbox{blue}{white}{\includegraphics[width=0.095\linewidth]{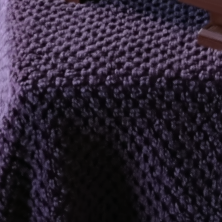}}} 
\\[-0.25mm]
PSNR\textsuperscript{$\uparrow$}/LPIPS\textsuperscript{$\downarrow$}
&& 
\textbf{32.61}/\textbf{0.180} & 32.42/0.182 & 32.05/\textbf{0.180} & 31.89/0.182 & 30.97/0.186 & 31.01/0.182
 \\
\makecell{\includegraphics[width=0.302\linewidth]{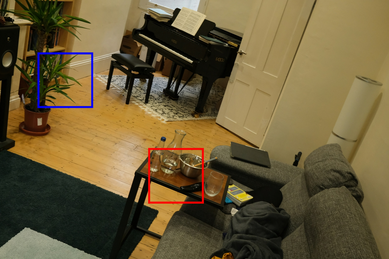}} 
&
\makecell{
\fcolorbox{red}{white}{\includegraphics[width=0.095\linewidth]{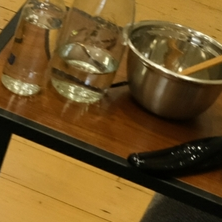}}\\[-0.25mm]
\fcolorbox{blue}{white}{\includegraphics[width=0.095\linewidth]{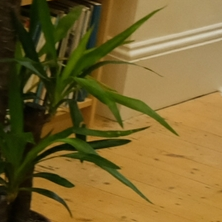}}} 
&
\makecell{
\fcolorbox{red}{white}{\includegraphics[width=0.095\linewidth]{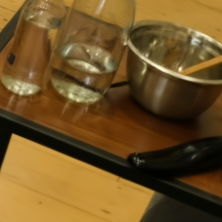}}\\[-0.25mm]
\fcolorbox{blue}{white}{\includegraphics[width=0.095\linewidth]{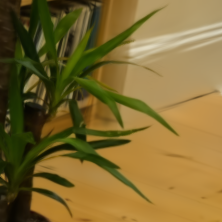}}} 
&
\makecell{
\fcolorbox{red}{white}{\includegraphics[width=0.095\linewidth]{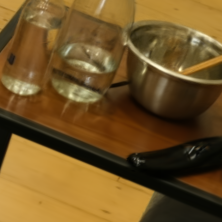}}\\[-0.25mm]
\fcolorbox{blue}{white}{\includegraphics[width=0.095\linewidth]{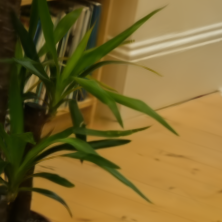}}} 
&
\makecell{
\fcolorbox{red}{white}{\includegraphics[width=0.095\linewidth]{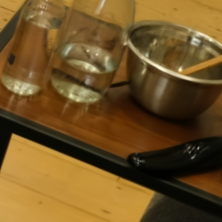}}\\[-0.25mm]
\fcolorbox{blue}{white}{\includegraphics[width=0.095\linewidth]{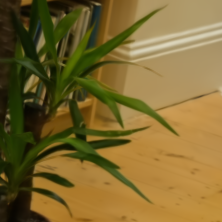}}} 
&
\makecell{
\fcolorbox{red}{white}{\includegraphics[width=0.095\linewidth]{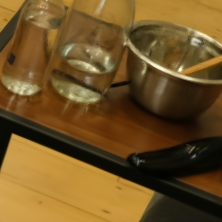}}\\[-0.25mm]
\fcolorbox{blue}{white}{\includegraphics[width=0.095\linewidth]{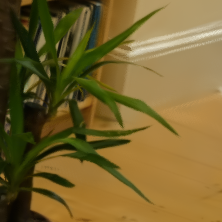}}} 
&
\makecell{
\fcolorbox{red}{white}{\includegraphics[width=0.095\linewidth]{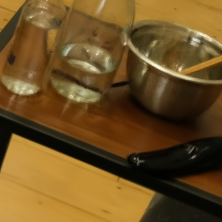}}\\[-0.25mm]
\fcolorbox{blue}{white}{\includegraphics[width=0.095\linewidth]{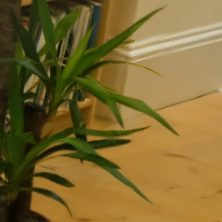}}} 
&
\makecell{
\fcolorbox{red}{white}{\includegraphics[width=0.095\linewidth]{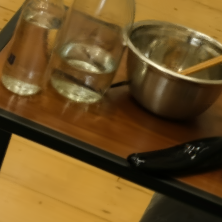}}\\[-0.25mm]
\fcolorbox{blue}{white}{\includegraphics[width=0.095\linewidth]{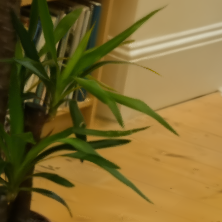}}} 
\\[-0.25mm]
PSNR\textsuperscript{$\uparrow$}/LPIPS\textsuperscript{$\downarrow$}
&& 
32.62/0.211 & 33.61/0.203 & \textbf{33.23}/\textbf{0.199} & 29.72/0.220 & 30.76/0.217 & 31.82/0.204
 \\
\end{tabular}
}
  \caption{\label{fig:renders}%
\textbf{Qualitative Comparison for Novel View Synthesis}.
We show small zoom-ins of various scenes from the Mip-NeRF 360 dataset \cite{barron2022mipnerf360}.
PSNR/LPIPS scores for this view are inset.
Overall, 3DGS-MCMC \cite{kheradmand2024mcmc} achieves the best results, while our results are on par with \gof.
  }
\end{figure*}
\begin{figure*}
\footnotesize\sffamily
\setlength{\tabcolsep}{1pt}%
\setlength{\fboxsep}{0pt}%
\setlength{\fboxrule}{0.5pt}%
\renewcommand{\arraystretch}{1.1}%
\resizebox{.99\linewidth}{!}{
\begin{tabular}{cccccccc}
Ground Truth & Ours & \gof & 2DGS & 2DGS & \gof & Ours & Ground Truth 
\\[-0.25mm]
\makecell{\includegraphics[width=0.25\linewidth]{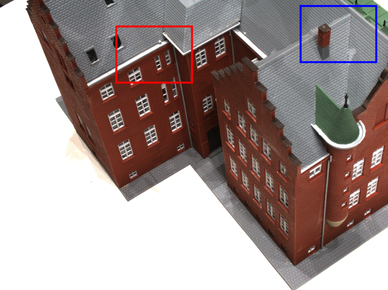}} 
&
\makecell{
\fcolorbox{red}{white}{\includegraphics[width=0.095\linewidth]{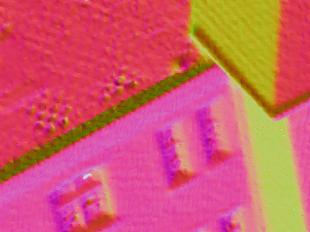}}\\[-0.25mm]
\fcolorbox{blue}{white}{\includegraphics[width=0.095\linewidth]{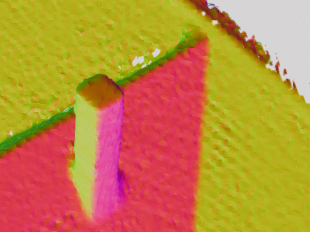}}} 
&
\makecell{
\fcolorbox{red}{white}{\includegraphics[width=0.095\linewidth]{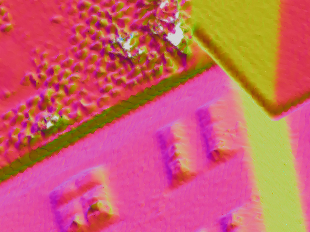}}\\[-0.25mm]
\fcolorbox{blue}{white}{\includegraphics[width=0.095\linewidth]{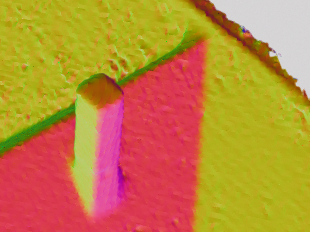}}} 
&
\makecell{
\fcolorbox{red}{white}{\includegraphics[width=0.095\linewidth]{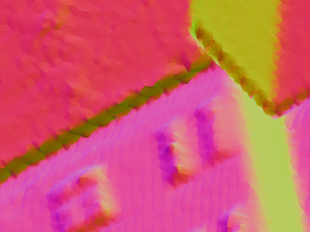}}\\[-0.25mm]
\fcolorbox{blue}{white}{\includegraphics[width=0.095\linewidth]{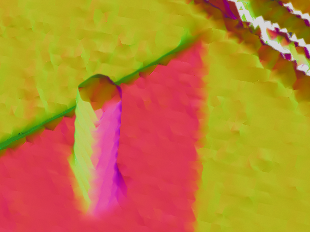}}} 
&
\makecell{
\fcolorbox{red}{white}{\includegraphics[width=0.095\linewidth]{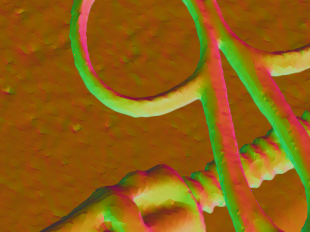}}\\[-0.25mm]
\fcolorbox{blue}{white}{\includegraphics[width=0.095\linewidth]{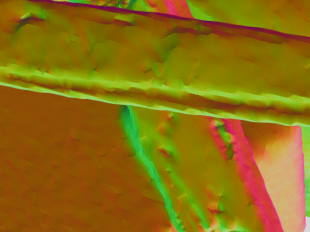}}} 
&
\makecell{
\fcolorbox{red}{white}{\includegraphics[width=0.095\linewidth]{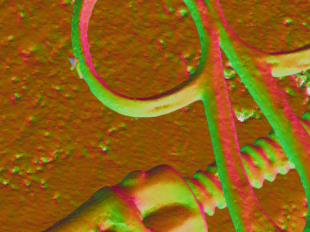}}\\[-0.25mm]
\fcolorbox{blue}{white}{\includegraphics[width=0.095\linewidth]{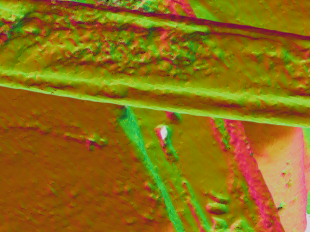}}} 
&
\makecell{
\fcolorbox{red}{white}{\includegraphics[width=0.095\linewidth]{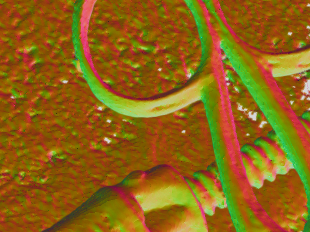}}\\[-0.25mm]
\fcolorbox{blue}{white}{\includegraphics[width=0.095\linewidth]{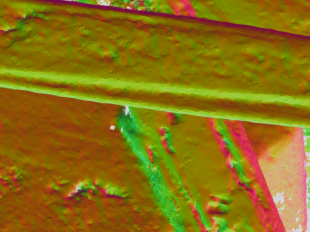}}} 
&
\makecell{\includegraphics[width=0.25\linewidth]{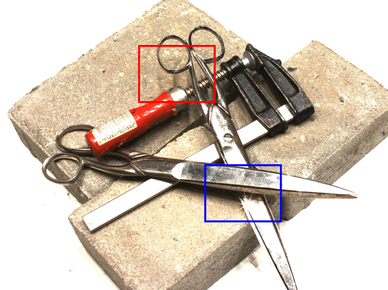}} 
\\[-0.25mm]
Chamfer Distance\textsuperscript{$\downarrow$} & 0.552 & {0.529} & \textbf{0.516} & 0.802 & 0.892 & \textbf{0.705} \\
\makecell{\includegraphics[width=0.25\linewidth]{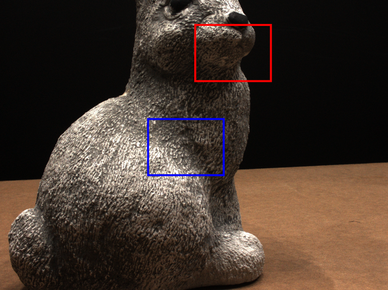}} 
&
\makecell{
\fcolorbox{red}{white}{\includegraphics[width=0.095\linewidth]{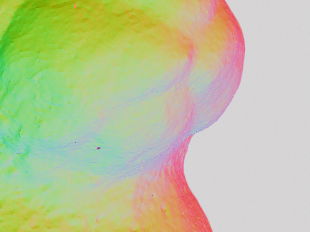}}\\[-0.25mm]
\fcolorbox{blue}{white}{\includegraphics[width=0.095\linewidth]{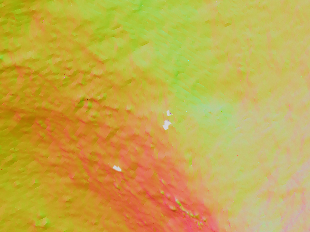}}} 
&
\makecell{
\fcolorbox{red}{white}{\includegraphics[width=0.095\linewidth]{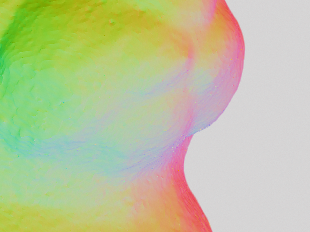}}\\[-0.25mm]
\fcolorbox{blue}{white}{\includegraphics[width=0.095\linewidth]{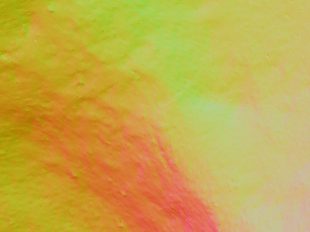}}} 
&
\makecell{
\fcolorbox{red}{white}{\includegraphics[width=0.095\linewidth]{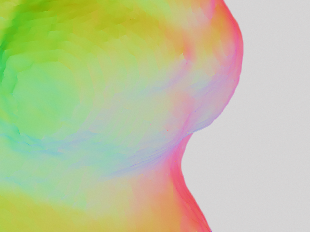}}\\[-0.25mm]
\fcolorbox{blue}{white}{\includegraphics[width=0.095\linewidth]{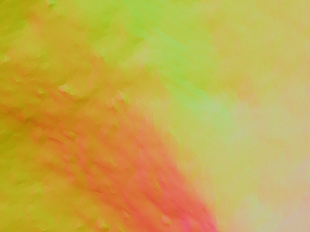}}} 
&
\makecell{
\fcolorbox{red}{white}{\includegraphics[width=0.095\linewidth]{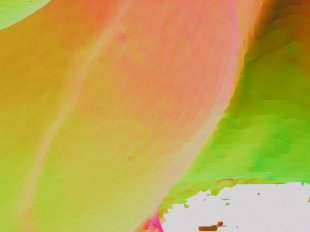}}\\[-0.25mm]
\fcolorbox{blue}{white}{\includegraphics[width=0.095\linewidth]{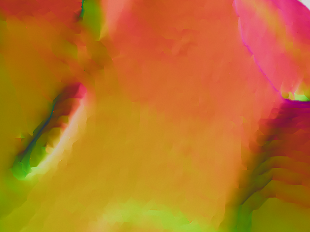}}} 
&
\makecell{
\fcolorbox{red}{white}{\includegraphics[width=0.095\linewidth]{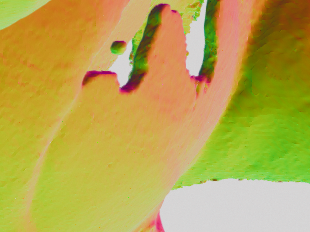}}\\[-0.25mm]
\fcolorbox{blue}{white}{\includegraphics[width=0.095\linewidth]{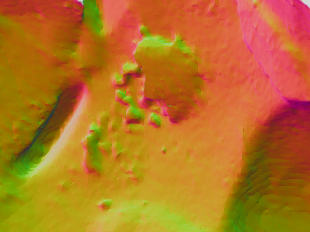}}} 
&
\makecell{
\fcolorbox{red}{white}{\includegraphics[width=0.095\linewidth]{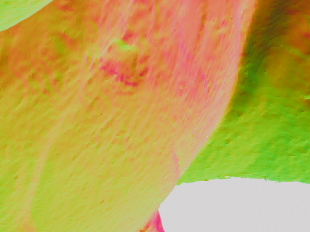}}\\[-0.25mm]
\fcolorbox{blue}{white}{\includegraphics[width=0.095\linewidth]{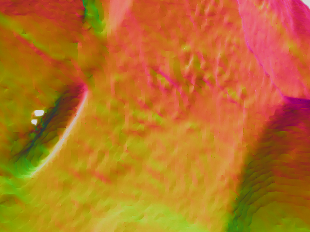}}} 
&
\makecell{\includegraphics[width=0.25\linewidth]{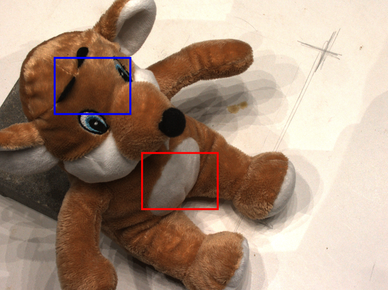}} 
\\[-0.25mm]
Chamfer Distance\textsuperscript{$\downarrow$} & {0.395} & \textbf{0.380} & {0.422} & 0.631 & 0.789 & \textbf{0.597} \\
\makecell{\includegraphics[width=0.25\linewidth]{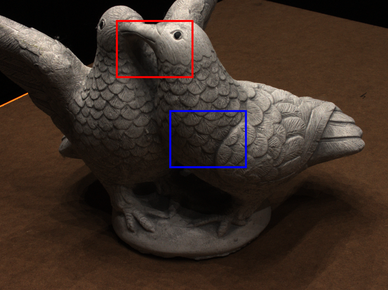}} 
&
\makecell{
\fcolorbox{red}{white}{\includegraphics[width=0.095\linewidth]{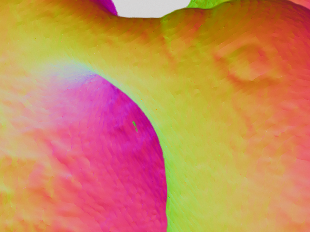}}\\[-0.25mm]
\fcolorbox{blue}{white}{\includegraphics[width=0.095\linewidth]{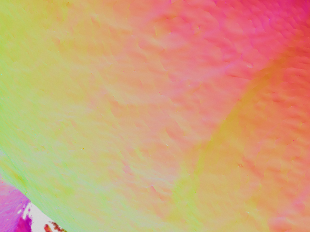}}} 
&
\makecell{
\fcolorbox{red}{white}{\includegraphics[width=0.095\linewidth]{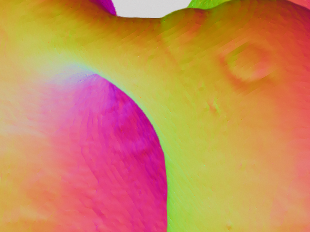}}\\[-0.25mm]
\fcolorbox{blue}{white}{\includegraphics[width=0.095\linewidth]{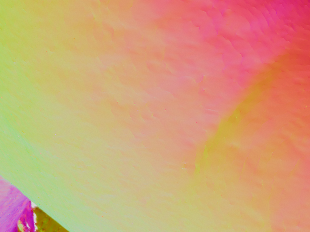}}} 
&
\makecell{
\fcolorbox{red}{white}{\includegraphics[width=0.095\linewidth]{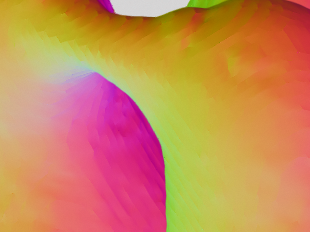}}\\[-0.25mm]
\fcolorbox{blue}{white}{\includegraphics[width=0.095\linewidth]{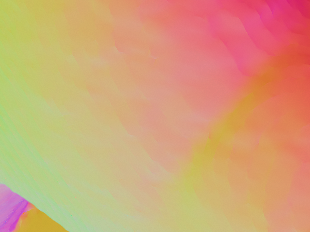}}} 
&
\makecell{
\fcolorbox{red}{white}{\includegraphics[width=0.095\linewidth]{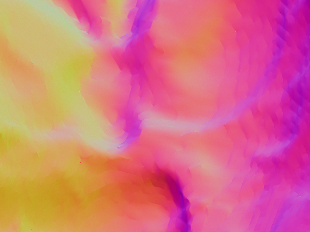}}\\[-0.25mm]
\fcolorbox{blue}{white}{\includegraphics[width=0.095\linewidth]{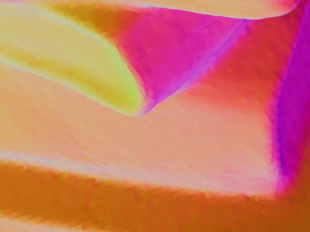}}} 
&
\makecell{
\fcolorbox{red}{white}{\includegraphics[width=0.095\linewidth]{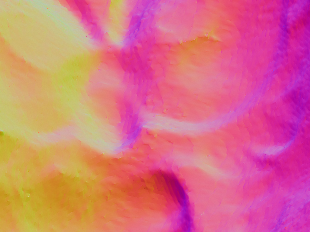}}\\[-0.25mm]
\fcolorbox{blue}{white}{\includegraphics[width=0.095\linewidth]{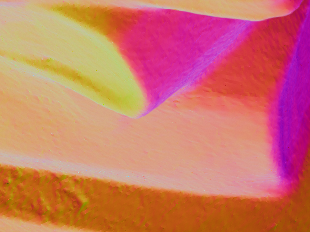}}} 
&
\makecell{
\fcolorbox{red}{white}{\includegraphics[width=0.095\linewidth]{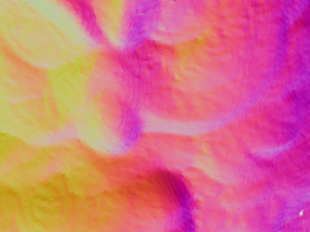}}\\[-0.25mm]
\fcolorbox{blue}{white}{\includegraphics[width=0.095\linewidth]{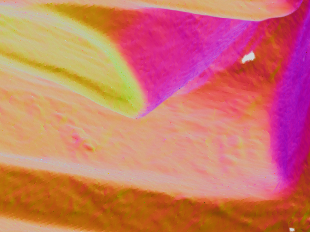}}} 
&
\makecell{\includegraphics[width=0.25\linewidth]{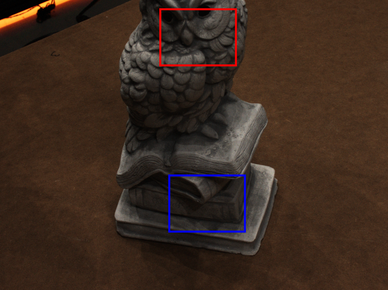}} 
\\[-0.25mm]
Chamfer Distance\textsuperscript{$\downarrow$} & \textbf{0.634} & {0.654} & {0.789} & 0.486 & 0.540 & \textbf{0.477} \\
\end{tabular}
}
  \caption{\label{fig:dtu_ablation}%
    \textbf{Qualitative Comparison for Bounded Meshes}:
We compare our bounded meshes with those from \gof \cite{yu2024gof} and 2DGS \cite{huang20242dgs}.
Compared to \gof, our meshes have fewer artifacts, while meshes extracted using 2DGS are overly smooth and lack intricate details.
However, all methods still exhibit holes and cavities due to the limitations of current view-dependent appearance.
  }
\end{figure*}
\begin{table*}[ht!]
    \centering
        \caption{
    \textbf{Ablation study for bounded meshes on the DTU dataset} \cite{jensen2014large}. 
Our method achieves better results compared to both 2DGS \cite{huang20242dgs} and \gof \cite{yu2024gof}.
Adding the losses designed for unbounded mesh extraction reduces the overall surface reconstruction quality.
    }
\resizebox{.99\linewidth}{!}{
\begin{tabular}{lrrrrrrrrrrrrrrrr}
\toprule
Method & 24 & 37 & 40 & 55 & 63 & 65 & 69 & 83 & 97 & 105 & 106 & 110 & 114 & 118 & 122 & Avg \\
\midrule
GOF & \cellcolor{tab_color!32} 0.529 & 0.892 & 0.434 & \cellcolor{tab_color!32} 0.380 & 1.332 & 0.866 & 0.770 & 1.284 & 1.289 & 0.789 & 0.767 & \cellcolor{tab_color!32} 1.148 & \cellcolor{tab_color!32} 0.457 & 0.696 & 0.540 & 0.812 \\
2DGS & \cellcolor{tab_color!49} 0.516 & 0.802 & \cellcolor{tab_color!49} 0.344 & 0.422 & \cellcolor{tab_color!49} 0.972 & 0.890 & 0.823 & 1.235 & \cellcolor{tab_color!49} 1.246 & 0.631 & \cellcolor{tab_color!15} 0.654 & 1.970 & \cellcolor{tab_color!49} 0.416 & 0.690 & \cellcolor{tab_color!15} 0.486 & 0.806 \\
Ours & \cellcolor{tab_color!15} 0.552 & \cellcolor{tab_color!32} 0.705 & \cellcolor{tab_color!15} 0.408 & 0.395 & \cellcolor{tab_color!32} 1.120 & 0.735 & \cellcolor{tab_color!32} 0.657 & \cellcolor{tab_color!32} 1.112 & 1.467 & \cellcolor{tab_color!32} 0.597 & \cellcolor{tab_color!32} 0.634 & \cellcolor{tab_color!49} 1.046 & 0.581 & \cellcolor{tab_color!32} 0.615 & \cellcolor{tab_color!32} 0.477 & \cellcolor{tab_color!49} 0.740 \\
(A) w/o Exact Depth & 0.585 & 0.882 & 0.469 & 0.423 & 1.269 & 0.973 & 0.858 & 1.188 & 1.365 & 1.062 & 0.686 & 1.272 & 0.718 & 1.009 & 0.546 & 0.887 \\
(B) w/ $\mathcal{L}_{\text{smooth}}$ & 0.585 & \cellcolor{tab_color!15} 0.716 & 0.410 & 0.399 & 1.310 & 0.774 & 0.795 & 1.176 & \cellcolor{tab_color!32} 1.253 & \cellcolor{tab_color!15} 0.627 & 0.837 & 1.188 & 0.635 & \cellcolor{tab_color!15} 0.667 & 0.553 & 0.795 \\
(C) w/ $\mathcal{L}_{\text{opa}}$ & 0.579 & 0.749 & 0.449 & 0.414 & \cellcolor{tab_color!15} 1.178 & \cellcolor{tab_color!32} 0.669 & 0.769 & \cellcolor{tab_color!15} 1.151 & \cellcolor{tab_color!15} 1.271 & 0.643 & 0.678 & 1.335 & 0.631 & 0.683 & 0.531 & \cellcolor{tab_color!32} 0.782 \\
(D) w/ $\mathcal{L}_{\text{extent}}$ & 0.590 & 0.773 & 0.486 & \cellcolor{tab_color!15} 0.381 & 1.253 & 0.779 & 0.766 & 1.190 & 1.325 & 0.755 & \cellcolor{tab_color!15} 0.654 & 1.227 & \cellcolor{tab_color!15} 0.535 & \cellcolor{tab_color!15} 0.667 & 0.524 & \cellcolor{tab_color!15} 0.794 \\
(E) w/ Attached Grad & 0.646 & \cellcolor{tab_color!49} 0.675 & 0.838 & \cellcolor{tab_color!49} 0.279 & 1.429 & \cellcolor{tab_color!49} 0.555 & \cellcolor{tab_color!49} 0.619 & \cellcolor{tab_color!49} 0.878 & 1.659 & \cellcolor{tab_color!49} 0.517 & \cellcolor{tab_color!49} 0.437 & \cellcolor{tab_color!15} 1.160 & 1.545 & \cellcolor{tab_color!49} 0.428 & \cellcolor{tab_color!49} 0.422 & 0.806 \\
(F) w/ $F=1000$ & 0.557 & 0.738 & \cellcolor{tab_color!32} 0.400 & 0.393 & 1.263 & \cellcolor{tab_color!15} 0.712 & \cellcolor{tab_color!15} 0.742 & 1.188 & 1.298 & 0.810 & 0.706 & 1.311 & 0.549 & 0.698 & 0.579 & 0.796 \\
(G) w/ PGSR Appearance & 0.585 & 0.854 & 0.577 & 0.413 & 1.300 & 0.966 & 0.799 & 1.313 & 1.556 & 1.238 & 0.750 & 1.333 & 0.863 & 1.097 & 0.532 & 0.945 \\
\bottomrule
\end{tabular}
}
    \label{tab:app_dtu}
\end{table*}
\begin{table*}[ht!]
    \centering
        \caption{
\textbf{Full per-scene results} for Novel View Synthesis for the Mip-NeRF 360 dataset \cite{barron2021mipnerf}.
    }
\resizebox{.98\textwidth}{!}{
\begin{tabular}{lrrrrrrrrrr}
\toprule
Method & bicycle & bonsai & counter & flowers & garden & stump & treehill & kitchen & room & Average \\\midrule
 &  \multicolumn{9}{c}{PSNR\textsuperscript{$\uparrow$}} \\
\cmidrule(lr){2-11}
3DGS \cite{kerbl20233dgs} & \cellcolor{tab_color!0} 25.183 & \cellcolor{tab_color!0} 32.102 & \cellcolor{tab_color!0} 28.982 & \cellcolor{tab_color!0} 21.478 & \cellcolor{tab_color!0} 27.240 & \cellcolor{tab_color!0} 26.620 & \cellcolor{tab_color!0} 22.452 & \cellcolor{tab_color!15} 31.352 & \cellcolor{tab_color!0} 31.494 & \cellcolor{tab_color!0} 27.434 \\
Mip-Splatting \cite{Yu2024MipSplatting} & \cellcolor{tab_color!0} 25.320 & \cellcolor{tab_color!15} 32.134 & \cellcolor{tab_color!15} 29.003 & \cellcolor{tab_color!0} 21.635 & \cellcolor{tab_color!15} 27.484 & \cellcolor{tab_color!0} 26.581 & \cellcolor{tab_color!0} 22.583 & \cellcolor{tab_color!0} 31.343 & \cellcolor{tab_color!15} 31.779 & \cellcolor{tab_color!15} 27.540 \\
StopThePop \cite{radl2024stopthepop} & \cellcolor{tab_color!0} 25.206 & \cellcolor{tab_color!0} 31.904 & \cellcolor{tab_color!0} 28.703 & \cellcolor{tab_color!0} 21.454 & \cellcolor{tab_color!0} 27.174 & \cellcolor{tab_color!0} 26.677 & \cellcolor{tab_color!0} 22.472 & \cellcolor{tab_color!0} 31.230 & \cellcolor{tab_color!0} 30.844 & \cellcolor{tab_color!0} 27.296 \\
Taming-3DGS \cite{mallick2024taming} & \cellcolor{tab_color!32} 25.474 & \cellcolor{tab_color!32} 32.474 & \cellcolor{tab_color!32} 29.057 & \cellcolor{tab_color!32} 21.869 & \cellcolor{tab_color!32} 27.756 & \cellcolor{tab_color!32} 27.052 & \cellcolor{tab_color!32} 22.906 & \cellcolor{tab_color!32} 31.762 & \cellcolor{tab_color!32} 32.087 & \cellcolor{tab_color!32} 27.826 \\
3DGS-MCMC \cite{kheradmand2024mcmc} & \cellcolor{tab_color!49} 25.686 & \cellcolor{tab_color!49} 32.654 & \cellcolor{tab_color!49} 29.376 & \cellcolor{tab_color!49} 22.014 & \cellcolor{tab_color!49} 27.867 & \cellcolor{tab_color!49} 27.358 & \cellcolor{tab_color!49} 22.944 & \cellcolor{tab_color!49} 32.089 & \cellcolor{tab_color!49} 32.252 & \cellcolor{tab_color!49} 28.027 \\
2DGS \cite{huang20242dgs} & \cellcolor{tab_color!0} 24.739 & \cellcolor{tab_color!0} 30.723 & \cellcolor{tab_color!0} 28.115 & \cellcolor{tab_color!0} 21.135 & \cellcolor{tab_color!0} 26.702 & \cellcolor{tab_color!0} 26.165 & \cellcolor{tab_color!0} 22.355 & \cellcolor{tab_color!0} 30.325 & \cellcolor{tab_color!0} 31.246 & \cellcolor{tab_color!0} 26.834 \\
GOF \cite{yu2024gof} & \cellcolor{tab_color!15} 25.467 & \cellcolor{tab_color!0} 31.598 & \cellcolor{tab_color!0} 28.681 & \cellcolor{tab_color!0} 21.644 & \cellcolor{tab_color!0} 27.426 & \cellcolor{tab_color!15} 26.989 & \cellcolor{tab_color!0} 22.406 & \cellcolor{tab_color!0} 30.787 & \cellcolor{tab_color!0} 30.812 & \cellcolor{tab_color!0} 27.312 \\
Ours & \cellcolor{tab_color!0} 25.457 & \cellcolor{tab_color!0} 31.246 & \cellcolor{tab_color!0} 28.419 & \cellcolor{tab_color!15} 21.780 & \cellcolor{tab_color!0} 27.245 & \cellcolor{tab_color!0} 26.925 & \cellcolor{tab_color!0} 22.508 & \cellcolor{tab_color!0} 30.502 & \cellcolor{tab_color!0} 30.345 & \cellcolor{tab_color!0} 27.159 \\
Ours (MCMC) & \cellcolor{tab_color!0} 25.232 & \cellcolor{tab_color!0} 31.388 & \cellcolor{tab_color!0} 28.585 & \cellcolor{tab_color!0} 21.622 & \cellcolor{tab_color!0} 27.182 & \cellcolor{tab_color!0} 26.600 & \cellcolor{tab_color!15} 22.843 & \cellcolor{tab_color!0} 30.492 & \cellcolor{tab_color!0} 31.174 & \cellcolor{tab_color!0} 27.235 \\
\bottomrule
\\[-3mm]
Method & bicycle & bonsai & counter & flowers & garden & stump & treehill & kitchen & room & Average \\\midrule
 &  \multicolumn{9}{c}{SSIM\textsuperscript{$\uparrow$}} \\
\cmidrule(lr){2-11}
3DGS \cite{kerbl20233dgs} & \cellcolor{tab_color!0} 0.763 & \cellcolor{tab_color!0} 0.939 & \cellcolor{tab_color!0} 0.906 & \cellcolor{tab_color!0} 0.603 & \cellcolor{tab_color!0} 0.862 & \cellcolor{tab_color!0} 0.772 & \cellcolor{tab_color!0} 0.632 & \cellcolor{tab_color!0} 0.925 & \cellcolor{tab_color!0} 0.917 & \cellcolor{tab_color!0} 0.813 \\
Mip-Splatting \cite{Yu2024MipSplatting} & \cellcolor{tab_color!0} 0.768 & \cellcolor{tab_color!15} 0.942 & \cellcolor{tab_color!15} 0.909 & \cellcolor{tab_color!0} 0.608 & \cellcolor{tab_color!15} 0.869 & \cellcolor{tab_color!0} 0.773 & \cellcolor{tab_color!0} 0.638 & \cellcolor{tab_color!15} 0.928 & \cellcolor{tab_color!0} 0.920 & \cellcolor{tab_color!0} 0.817 \\
StopThePop \cite{radl2024stopthepop} & \cellcolor{tab_color!0} 0.767 & \cellcolor{tab_color!0} 0.939 & \cellcolor{tab_color!0} 0.904 & \cellcolor{tab_color!0} 0.603 & \cellcolor{tab_color!0} 0.862 & \cellcolor{tab_color!0} 0.775 & \cellcolor{tab_color!0} 0.635 & \cellcolor{tab_color!0} 0.925 & \cellcolor{tab_color!0} 0.917 & \cellcolor{tab_color!0} 0.814 \\
Taming-3DGS \cite{mallick2024taming} & \cellcolor{tab_color!0} 0.780 & \cellcolor{tab_color!32} 0.944 & \cellcolor{tab_color!32} 0.910 & \cellcolor{tab_color!0} 0.614 & \cellcolor{tab_color!32} 0.873 & \cellcolor{tab_color!0} 0.788 & \cellcolor{tab_color!32} 0.646 & \cellcolor{tab_color!32} 0.931 & \cellcolor{tab_color!15} 0.924 & \cellcolor{tab_color!32} 0.823 \\
3DGS-MCMC \cite{kheradmand2024mcmc} & \cellcolor{tab_color!49} 0.799 & \cellcolor{tab_color!49} 0.948 & \cellcolor{tab_color!49} 0.917 & \cellcolor{tab_color!49} 0.645 & \cellcolor{tab_color!49} 0.878 & \cellcolor{tab_color!49} 0.811 & \cellcolor{tab_color!49} 0.659 & \cellcolor{tab_color!49} 0.934 & \cellcolor{tab_color!32} 0.930 & \cellcolor{tab_color!49} 0.836 \\
2DGS \cite{huang20242dgs} & \cellcolor{tab_color!0} 0.733 & \cellcolor{tab_color!0} 0.907 & \cellcolor{tab_color!0} 0.893 & \cellcolor{tab_color!0} 0.576 & \cellcolor{tab_color!0} 0.842 & \cellcolor{tab_color!0} 0.757 & \cellcolor{tab_color!0} 0.616 & \cellcolor{tab_color!0} 0.916 & \cellcolor{tab_color!49} 0.931 & \cellcolor{tab_color!0} 0.797 \\
GOF \cite{yu2024gof} & \cellcolor{tab_color!32} 0.787 & \cellcolor{tab_color!0} 0.937 & \cellcolor{tab_color!0} 0.902 & \cellcolor{tab_color!15} 0.637 & \cellcolor{tab_color!0} 0.868 & \cellcolor{tab_color!32} 0.794 & \cellcolor{tab_color!0} 0.641 & \cellcolor{tab_color!0} 0.917 & \cellcolor{tab_color!0} 0.916 & \cellcolor{tab_color!15} 0.822 \\
Ours & \cellcolor{tab_color!15} 0.786 & \cellcolor{tab_color!0} 0.936 & \cellcolor{tab_color!0} 0.900 & \cellcolor{tab_color!32} 0.638 & \cellcolor{tab_color!0} 0.866 & \cellcolor{tab_color!15} 0.791 & \cellcolor{tab_color!15} 0.645 & \cellcolor{tab_color!0} 0.914 & \cellcolor{tab_color!0} 0.915 & \cellcolor{tab_color!0} 0.821 \\
Ours (MCMC) & \cellcolor{tab_color!0} 0.779 & \cellcolor{tab_color!0} 0.940 & \cellcolor{tab_color!0} 0.905 & \cellcolor{tab_color!0} 0.625 & \cellcolor{tab_color!0} 0.863 & \cellcolor{tab_color!0} 0.781 & \cellcolor{tab_color!0} 0.643 & \cellcolor{tab_color!0} 0.917 & \cellcolor{tab_color!0} 0.922 & \cellcolor{tab_color!0} 0.819 \\
\bottomrule
\\[-3mm]
Method & bicycle & bonsai & counter & flowers & garden & stump & treehill & kitchen & room & Average \\\midrule
 &  \multicolumn{9}{c}{LPIPS\textsuperscript{$\downarrow$}} \\
\cmidrule(lr){2-11}
3DGS \cite{kerbl20233dgs} & \cellcolor{tab_color!0} 0.213 & \cellcolor{tab_color!0} 0.206 & \cellcolor{tab_color!0} 0.202 & \cellcolor{tab_color!0} 0.338 & \cellcolor{tab_color!0} 0.109 & \cellcolor{tab_color!0} 0.216 & \cellcolor{tab_color!0} 0.327 & \cellcolor{tab_color!0} 0.127 & \cellcolor{tab_color!0} 0.221 & \cellcolor{tab_color!0} 0.218 \\
Mip-Splatting \cite{Yu2024MipSplatting} & \cellcolor{tab_color!0} 0.212 & \cellcolor{tab_color!0} 0.204 & \cellcolor{tab_color!0} 0.200 & \cellcolor{tab_color!0} 0.339 & \cellcolor{tab_color!0} 0.108 & \cellcolor{tab_color!0} 0.216 & \cellcolor{tab_color!0} 0.326 & \cellcolor{tab_color!15} 0.126 & \cellcolor{tab_color!0} 0.218 & \cellcolor{tab_color!0} 0.216 \\
StopThePop \cite{radl2024stopthepop} & \cellcolor{tab_color!0} 0.205 & \cellcolor{tab_color!0} 0.203 & \cellcolor{tab_color!0} 0.199 & \cellcolor{tab_color!0} 0.335 & \cellcolor{tab_color!15} 0.107 & \cellcolor{tab_color!0} 0.210 & \cellcolor{tab_color!0} 0.319 & \cellcolor{tab_color!15} 0.126 & \cellcolor{tab_color!0} 0.216 & \cellcolor{tab_color!0} 0.213 \\
Taming-3DGS \cite{mallick2024taming} & \cellcolor{tab_color!0} 0.192 & \cellcolor{tab_color!0} 0.201 & \cellcolor{tab_color!15} 0.198 & \cellcolor{tab_color!0} 0.332 & \cellcolor{tab_color!32} 0.100 & \cellcolor{tab_color!15} 0.196 & \cellcolor{tab_color!0} 0.313 & \cellcolor{tab_color!32} 0.122 & \cellcolor{tab_color!15} 0.210 & \cellcolor{tab_color!0} 0.207 \\
3DGS-MCMC \cite{kheradmand2024mcmc} & \cellcolor{tab_color!49} 0.168 & \cellcolor{tab_color!49} 0.191 & \cellcolor{tab_color!49} 0.185 & \cellcolor{tab_color!15} 0.284 & \cellcolor{tab_color!49} 0.094 & \cellcolor{tab_color!49} 0.171 & \cellcolor{tab_color!49} 0.272 & \cellcolor{tab_color!49} 0.121 & \cellcolor{tab_color!49} 0.198 & \cellcolor{tab_color!49} 0.187 \\
2DGS \cite{huang20242dgs} & \cellcolor{tab_color!0} 0.269 & \cellcolor{tab_color!0} 0.243 & \cellcolor{tab_color!0} 0.230 & \cellcolor{tab_color!0} 0.373 & \cellcolor{tab_color!0} 0.147 & \cellcolor{tab_color!0} 0.258 & \cellcolor{tab_color!0} 0.377 & \cellcolor{tab_color!0} 0.147 & \cellcolor{tab_color!0} 0.227 & \cellcolor{tab_color!0} 0.252 \\
GOF \cite{yu2024gof} & \cellcolor{tab_color!32} 0.180 & \cellcolor{tab_color!0} 0.198 & \cellcolor{tab_color!0} 0.204 & \cellcolor{tab_color!32} 0.279 & \cellcolor{tab_color!15} 0.107 & \cellcolor{tab_color!32} 0.195 & \cellcolor{tab_color!15} 0.279 & \cellcolor{tab_color!0} 0.137 & \cellcolor{tab_color!0} 0.217 & \cellcolor{tab_color!15} 0.200 \\
Ours & \cellcolor{tab_color!0} 0.183 & \cellcolor{tab_color!32} 0.195 & \cellcolor{tab_color!0} 0.202 & \cellcolor{tab_color!49} 0.276 & \cellcolor{tab_color!15} 0.107 & \cellcolor{tab_color!0} 0.197 & \cellcolor{tab_color!32} 0.276 & \cellcolor{tab_color!0} 0.139 & \cellcolor{tab_color!0} 0.214 & \cellcolor{tab_color!32} 0.199 \\
Ours (MCMC) & \cellcolor{tab_color!15} 0.182 & \cellcolor{tab_color!32} 0.195 & \cellcolor{tab_color!32} 0.196 & \cellcolor{tab_color!0} 0.295 & \cellcolor{tab_color!0} 0.111 & \cellcolor{tab_color!0} 0.200 & \cellcolor{tab_color!0} 0.282 & \cellcolor{tab_color!0} 0.141 & \cellcolor{tab_color!32} 0.204 & \cellcolor{tab_color!0} 0.201 \\
\bottomrule
\\[-3mm]
Method & bicycle & bonsai & counter & flowers & garden & stump & treehill & kitchen & room & Average \\\midrule
 &  \multicolumn{9}{c}{\FLIP\textsuperscript{$\downarrow$}} \\
\cmidrule(lr){2-11}
3DGS \cite{kerbl20233dgs} & \cellcolor{tab_color!0} 0.158 & \cellcolor{tab_color!15} 0.082 & \cellcolor{tab_color!0} 0.105 & \cellcolor{tab_color!0} 0.225 & \cellcolor{tab_color!15} 0.118 & \cellcolor{tab_color!0} 0.150 & \cellcolor{tab_color!0} 0.186 & \cellcolor{tab_color!15} 0.096 & \cellcolor{tab_color!0} 0.093 & \cellcolor{tab_color!15} 0.135 \\
Mip-Splatting \cite{Yu2024MipSplatting} & \cellcolor{tab_color!0} 0.160 & \cellcolor{tab_color!0} 0.083 & \cellcolor{tab_color!0} 0.104 & \cellcolor{tab_color!0} 0.225 & \cellcolor{tab_color!0} 0.119 & \cellcolor{tab_color!0} 0.152 & \cellcolor{tab_color!0} 0.184 & \cellcolor{tab_color!15} 0.096 & \cellcolor{tab_color!0} 0.092 & \cellcolor{tab_color!15} 0.135 \\
StopThePop \cite{radl2024stopthepop} & \cellcolor{tab_color!0} 0.160 & \cellcolor{tab_color!0} 0.085 & \cellcolor{tab_color!0} 0.110 & \cellcolor{tab_color!0} 0.225 & \cellcolor{tab_color!0} 0.120 & \cellcolor{tab_color!0} 0.149 & \cellcolor{tab_color!0} 0.183 & \cellcolor{tab_color!0} 0.099 & \cellcolor{tab_color!0} 0.101 & \cellcolor{tab_color!0} 0.137 \\
Taming-3DGS \cite{mallick2024taming} & \cellcolor{tab_color!32} 0.157 & \cellcolor{tab_color!49} 0.079 & \cellcolor{tab_color!15} 0.103 & \cellcolor{tab_color!32} 0.214 & \cellcolor{tab_color!49} 0.113 & \cellcolor{tab_color!32} 0.145 & \cellcolor{tab_color!0} 0.186 & \cellcolor{tab_color!49} 0.091 & \cellcolor{tab_color!49} 0.089 & \cellcolor{tab_color!32} 0.131 \\
3DGS-MCMC \cite{kheradmand2024mcmc} & \cellcolor{tab_color!49} 0.156 & \cellcolor{tab_color!49} 0.079 & \cellcolor{tab_color!32} 0.101 & \cellcolor{tab_color!49} 0.209 & \cellcolor{tab_color!32} 0.115 & \cellcolor{tab_color!49} 0.142 & \cellcolor{tab_color!32} 0.176 & \cellcolor{tab_color!32} 0.093 & \cellcolor{tab_color!32} 0.090 & \cellcolor{tab_color!49} 0.129 \\
2DGS \cite{huang20242dgs} & \cellcolor{tab_color!0} 0.167 & \cellcolor{tab_color!0} 0.184 & \cellcolor{tab_color!49} 0.101 & \cellcolor{tab_color!0} 0.236 & \cellcolor{tab_color!0} 0.127 & \cellcolor{tab_color!0} 0.159 & \cellcolor{tab_color!49} 0.114 & \cellcolor{tab_color!0} 0.106 & \cellcolor{tab_color!15} 0.090 & \cellcolor{tab_color!0} 0.143 \\
GOF \cite{yu2024gof} & \cellcolor{tab_color!32} 0.157 & \cellcolor{tab_color!0} 0.086 & \cellcolor{tab_color!0} 0.109 & \cellcolor{tab_color!0} 0.217 & \cellcolor{tab_color!0} 0.121 & \cellcolor{tab_color!0} 0.147 & \cellcolor{tab_color!0} 0.184 & \cellcolor{tab_color!0} 0.104 & \cellcolor{tab_color!0} 0.102 & \cellcolor{tab_color!0} 0.136 \\
Ours & \cellcolor{tab_color!32} 0.157 & \cellcolor{tab_color!0} 0.091 & \cellcolor{tab_color!0} 0.111 & \cellcolor{tab_color!32} 0.214 & \cellcolor{tab_color!0} 0.124 & \cellcolor{tab_color!15} 0.146 & \cellcolor{tab_color!0} 0.183 & \cellcolor{tab_color!0} 0.107 & \cellcolor{tab_color!0} 0.106 & \cellcolor{tab_color!0} 0.138 \\
Ours (MCMC) & \cellcolor{tab_color!0} 0.161 & \cellcolor{tab_color!0} 0.092 & \cellcolor{tab_color!0} 0.110 & \cellcolor{tab_color!0} 0.219 & \cellcolor{tab_color!0} 0.125 & \cellcolor{tab_color!0} 0.151 & \cellcolor{tab_color!15} 0.181 & \cellcolor{tab_color!0} 0.107 & \cellcolor{tab_color!0} 0.101 & \cellcolor{tab_color!0} 0.139 \\
\bottomrule
\end{tabular}
}
    \label{tab:m360_perscene}
\end{table*}

\end{document}